\pdfoutput=1
\documentclass[%
 reprint,
 superscriptaddress,
 amsmath,amssymb,
 aps
]{revtex4-2}

\usepackage{array}
\usepackage{braket}
\usepackage{amsmath}
\usepackage{mathtools}
\usepackage{here}
\mathtoolsset{showonlyrefs} % 参照した式のみ番号を表示
\usepackage{latexsym}
\usepackage{amsfonts}
\usepackage{amsthm}
\usepackage{mathrsfs}
\usepackage{graphicx}
\usepackage{xcolor}
\usepackage{comment}
\usepackage{bm}
\usepackage{physics}
\usepackage{hyperref}
\hypersetup{
    colorlinks=true,
    linkcolor=blue,
    filecolor=magenta,   
    citecolor=magenta,
    urlcolor=cyan
    }
\usepackage{url}
\usepackage{amssymb}
\usepackage{afterpage}
\usepackage{qcircuit}
\usepackage[ruled,linesnumbered]{algorithm2e}
\theoremstyle{plain}
\newtheorem{thm}{Theorem}

\newcommand{\average}[1]{\ensuremath{\left< #1 \right> }}

\newcommand{\beb}{\begin{itembox}}
\newcommand{\enb}{\end{itembox}}

\begin{document}

\title{Decoder Switching: Breaking the Speed-Accuracy Tradeoff in Real-Time Quantum Error Correction}
% Decoder Switching: Towards Real-Time, High-Precision Quantum Error Correction
% Decoder Switching: Bridging the Gap Between Speed and Accuracy in Quantum Decoding

\author{Riki Toshio}
\thanks{These authors contributed equally to this work.\\
Email: toshio.riki@fujitsu.com} 
%\email{toshio.riki@fujitsu.com}

\affiliation{
Quantum Laboratory, Fujitsu Research, Fujitsu Limited,
4-1-1 Kawasaki, Kanagawa 211-8588, Japan
}
\affiliation{
Fujitsu Quantum Computing Joint Research Division,
Center for Quantum Information and Quantum Biology, Osaka University, 1-2 Machikaneyama, Toyonaka, Osaka, 565-8531, Japan
}

\author{Kaito Kishi}
\thanks{These authors contributed equally to this work.\\
Email: toshio.riki@fujitsu.com} 
\affiliation{
Quantum Laboratory, Fujitsu Research, Fujitsu Limited,
4-1-1 Kawasaki, Kanagawa 211-8588, Japan
}
\affiliation{
Fujitsu Quantum Computing Joint Research Division,
Center for Quantum Information and Quantum Biology, Osaka University, 1-2 Machikaneyama, Toyonaka, Osaka, 565-8531, Japan
}

\author{Jun Fujisaki}
\affiliation{
Quantum Laboratory, Fujitsu Research, Fujitsu Limited,
4-1-1 Kawasaki, Kanagawa 211-8588, Japan
}
\affiliation{
Fujitsu Quantum Computing Joint Research Division,
Center for Quantum Information and Quantum Biology, Osaka University, 1-2 Machikaneyama, Toyonaka, Osaka, 565-8531, Japan
}

\author{Hirotaka Oshima}
\affiliation{
Quantum Laboratory, Fujitsu Research, Fujitsu Limited,
4-1-1 Kawasaki, Kanagawa 211-8588, Japan
}
\affiliation{
Fujitsu Quantum Computing Joint Research Division,
Center for Quantum Information and Quantum Biology, Osaka University, 1-2 Machikaneyama, Toyonaka, Osaka, 565-8531, Japan
}

\author{Shintaro Sato}
\affiliation{
Quantum Laboratory, Fujitsu Research, Fujitsu Limited,
4-1-1 Kawasaki, Kanagawa 211-8588, Japan
}
\affiliation{
Fujitsu Quantum Computing Joint Research Division,
Center for Quantum Information and Quantum Biology, Osaka University, 1-2 Machikaneyama, Toyonaka, Osaka, 565-8531, Japan
}

\author{Keisuke Fujii}
\affiliation{
Fujitsu Quantum Computing Joint Research Division,
Center for Quantum Information and Quantum Biology, Osaka University, 1-2 Machikaneyama, Toyonaka, Osaka, 565-8531, Japan
}

\affiliation{
Graduate School of Engineering Science, Osaka University,
1-3 Machikaneyama, Toyonaka, Osaka, 560-8531, Japan
}
\affiliation{
Center for Quantum Information and Quantum Biology, Osaka University, 560-0043, Japan
}
\affiliation{
RIKEN Center for Quantum Computing (RQC), Wako Saitama 351-0198, Japan
}

\date{\today}

\begin{abstract}
The realization of fault-tolerant quantum computers hinges on the construction of high-speed, high-accuracy, real-time decoding systems. The persistent challenge lies in the fundamental trade-off between speed and accuracy: efforts to improve the decoder's accuracy often lead to unacceptable increases in decoding time and hardware complexity, while attempts to accelerate decoding result in a significant degradation in logical error rate. To overcome this challenge, we propose a novel framework, {\it decoder switching}, which balances these competing demands by combining a faster, soft-output decoder (``weak decoder") with a slower, high-accuracy decoder (``strong decoder").  In usual rounds, the weak decoder processes error syndromes and simultaneously evaluates its reliability via soft information. Only when encountering a decoding window with low reliability do we switch to the strong decoder to achieve more accurate decoding. Numerical simulations suggest that this framework can achieve accuracy comparable to, or even surpassing, that of the strong decoder, while maintaining an average decoding time on par with the weak decoder. We also develop an online decoding scheme tailored to our framework, named {\it double window decoding}, and elucidate the criteria for preventing an exponential slowdown of quantum computation. These findings break the long-standing speed-accuracy trade-off, paving the way for scalable real-time decoding devices.
\end{abstract}

\maketitle

%\tableofcontents

\section{Introduction}

Quantum computer has a great potential to offer a revolutionary impact on a broad range of applications, including cryptography~\cite{Shor1994,Shor1999, Gidney2021RSA, Gidney2025RSA}, quantum simulation of molecules~\cite{Lloyd1996,Abrams1999,Aspuru-Guzik2005, Lee2021, Low2025} and solids~\cite{Babbush2018qubitization,Kivlichan2020improved, Yoshioka2022hunting,Toshio2024,Akahoshi2024}, and linear algebraic analyses~\cite{Harrow2009}. However, in actual quantum devices, interactions with the environment always perturb the state of qubits, thereby hindering the realization of these quantum advantages.
To overcome such difficulties, fault-tolerant quantum computing (FTQC) architectures are designed to employ the so-called quantum error correction (QEC) protocol~\cite{Shor1996, Aharonov1997, Kitaev1997}. 
According to the well-established theories~\cite{Nielsen2000}, the QEC technique enables us to exponentially suppress the logical error rate in FTQC by encoding quantum information with a polynomial amount of physical qubits.

A key element for developing FTQC systems is the real-time decoder~\cite{Battistel2023review, iOlius2024review}. It infers the most likely errors that may occur in the QEC codes, given a stream of syndrome data generated continuously from quantum devices.
To achieve a reliable and scalable FTQC, decoding systems are desired to satisfy two conflicting requirements simultaneously: high accuracy and low latency. High accuracy is essential for minimizing the spatial overhead (i.e., the number of physical qubits) required to achieve a target logical error rate on FTQC. 
On the other hand, low latency is crucial for achieving a high clock rate of logical operations and avoiding the so-called backlog problem~\cite{Terhal2015Review, Skoric2023}, where syndrome data accumulate faster than the decoder can process it. In particular, to prevent an exponential increase in the backlog, we have to achieve a decoding time shorter than the syndrome extraction time of the quantum device~\cite{Terhal2015Review}.

Over the past few decades, a major challenge in decoder design has been the inherent trade-off between speed and accuracy~\cite{Battistel2023review}.  
Many existing decoders struggle to simultaneously achieve both of these requirements. For example, the so-called AlphaQubit~\cite{Bausch2024alphaqubit}, a transformer-based neural network decoder, has recently demonstrated a remarkably high accuracy compared to other standard decoders, such as Minimum Weight Perfect Matching (MWPM) decoder~\cite{Dennis2002, Fowler2015mwpm, Higgott2023sparse, Wu2023fusion} and tensor network decoders~\cite{Ferris2014, Bravyi2014, iOlius2024review}, in the experiments on Google's superconducting quantum processors~\cite{Bausch2024alphaqubit, Acharya2025google}. However, its decoding time per round, even with a current Tensor Processing Unit (TPU), exceeds 10 $\mu$s even for the smallest surface code with $d=3$. This is significantly longer than the syndrome generation time of current superconducting devices (on the order of $\sim$ 1 $\mu$s)~\cite{Jeffrey2014, Arute2019,Google2023suppressing}.  
In contrast, the Union-Find (UF) decoder, which is based on an efficient and parallelizable 
algorithm~\cite{Delfosse2021UF,Heer2023,Chan2023actis}, has recently been implemented on Field-Programmable Gate Arrays (FPGAs) or Application-Specific Integrated Circuits (ASICs). This enables us to leverage parallel computing resources for speedup, thereby achieving a decoding time shorter than 1 $\mu$s for moderate code distances up to around $d\lesssim 21$~\cite{Liyanage2024FPGA, Liyanage2025, Barber2025}. However, on the downside, the UF decoder is known to exhibit an error threshold several tens of percent lower than that of MWPM, leading to a higher logical error rate~\cite{Huang2020weightedUF, Chan2023actis}. In general, implementing decoders on dedicated classical hardware, such as FPGAs or ASICs, can achieve lower latency and higher throughput by exploiting hardware parallelism.
However, this approach often sacrifices the decoder's accuracy~\cite{Overwater2022, Liao2023, Liyanage2024FPGA, Barber2025} or scalability~\cite{Das2022, Vittal2023Astrea} due to the limited hardware resources. These fundamental dilemmas pose a significant barrier to the development of practical real-time decoders.

Recently, some researchers have shed new light on a class of decoders known as ``soft-output decoders"~\cite{Gidney2025yoked, Meister2024, Lee2025soft, Kishi2025}. These decoders not only provide an estimate of the logical error, but also quantify the reliability of the decoder's estimate, which is referred to as {\it soft information}. Conventionally, such soft information has been discussed in the context of tensor network decoders~\cite{Ferris2014, Bravyi2014, iOlius2024review}. These decoders can directly evaluate the logical error rate conditioned by the generated syndrome data $\sigma$, but they generally exhibit a significant decoding time. To avoid such high time complexity, recent seminal works~\cite{Hutter2014, Bombin2024, Gidney2025yoked, Meister2024, Lee2025soft, Kishi2025} have developed a distinct type of soft information that can be evaluated with an efficient decoder, like the MWPM or UF decoders, and utilized to evaluate a kind of conditional logical error rate.
This information can be leveraged to adaptively control subsequent classical or quantum operations, enabling the construction of more sophisticated quantum computing systems.  For instance, in the past few years, soft information has been successfully applied to post-selection in magic state preparation~\cite{Bombin2024, Gidney2024cultivation}, error mitigation in quantum computing~\cite{Smith2024mitigation}, and soft-decision decoding for concatenated surface codes~\cite{Gidney2025yoked, Meister2024}.

\begin{figure}
    \centering
    \includegraphics[width=0.9\linewidth]{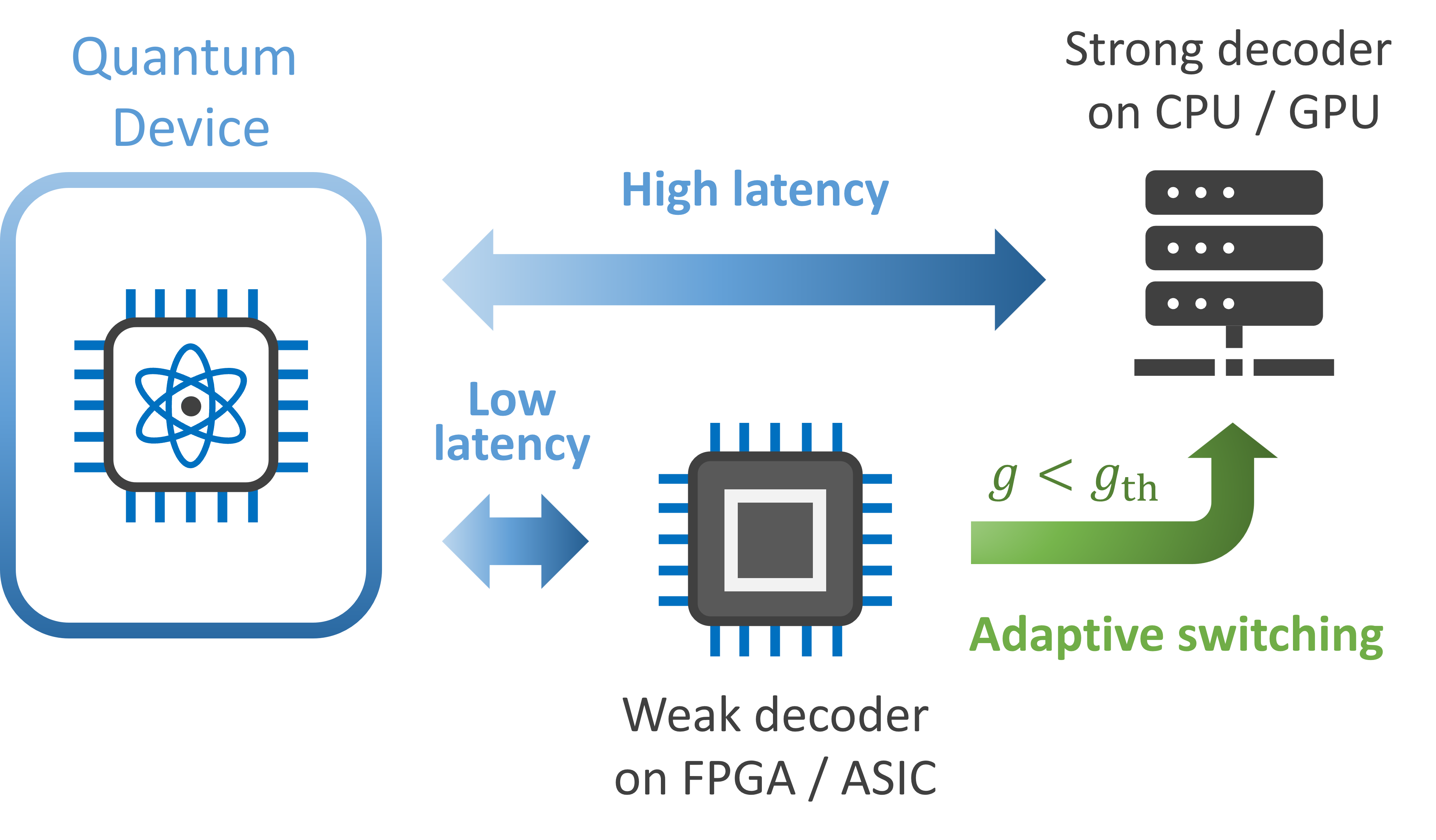}
    \caption{A possible device configuration for real-time decoding systems based on decoder switching. In this example, the {\it weak decoder} is implemented with FPGA or ASIC chips to realize a low response time, based on a fast algorithm like the UF decoder. This decoder also outputs a kind of soft information $g$, for example, by using the method proposed in Ref.~\cite{Meister2024}. On the other hand, the {\it strong decoder} is implemented with CPU or GPU-based devices to perform highly accurate decoding schemes like a transformer-based neural network decoder~\cite{Bausch2024alphaqubit}. In decoder switching, only when we obtain a smaller value of the soft information than a given threshold $g_{\text{th}}$, we wait for the outcome from the strong decoder. Otherwise, we use the decoding result from the weak decoder to maintain the average decoding speed sufficiently low.}
    \label{fig:decoder_switching}
\end{figure}

In this work, we introduce a new framework, {\it decoder switching}, which addresses the speed-accuracy trade-off in QEC decoding problems by leveraging hybrid decoding systems and soft information (see also Fig.~\ref{fig:decoder_switching}).
The core idea is to judiciously combine a fast, soft-output decoder (``weak decoder") with a slow, accurate decoder (``strong decoder"). Specifically, we primarily use the weak decoder for syndrome decoding while concurrently assessing the reliability of the decoding outcome based on its soft output $g$. Then, only when the confidence level falls below a predefined threshold $g_{\text{th}}$, we switch the decoding process to the strong decoder for more accurate decoding. 
To ensure the overall decoding accuracy becomes comparable to that achievable by the strong decoder alone, we formulate a methodology for determining the (near-)optimal value of $g_{\text{th}}$ from the perspective of the $g$-resolved logical error rate.

However, decoder switching tends to destabilize the behavior of the backlog during real-time decoding, leading to an accidental exponential slowdown of logical operations. To resolve this issue, we also propose an online decoding scheme tailored to decoder switching systems, termed the {\it double window decoding} scheme.
Based on it, we prove that satisfying $\tau_{\text{dec}}^{\text{weak}}\leq \tau_{\text{gen}}$ and $\tau_{\text{dec}}^{\text{strong}} \leq C\gamma_{\text{switch}}^{-1}\tau_{\text{gen}}$ is sufficient to avoid the backlog problem under decoder switching. Here, $\tau_{\text{dec}}^{\text{weak}}$ and $\tau_{\text{dec}}^{\text{strong}}$ represent the decoding time per round of the weak and the strong decoder respectively, while $\tau_{\text{gen}}$ denotes the syndrome generation time per round. $\gamma_{\text{switch}}$ is the switching rate per $d$-round, and $C$ is a constant factor dependent on the decoding window size and the number of logical qubits. As shown later, the value of $\gamma_{\text{switch}}$ becomes exponentially small for well-designed decoding systems as the code distance increases. This means that decoder switching systems can permit a potentially large latency of the strong decoder even for practical-sized quantum circuits.

A notable advantage of our proposal is that it enables the independent development of the weak and strong decoders. Within our framework, strong decoders can exclusively pursue higher accuracy, while weak decoders can prioritize lower latency and higher scalability without the burden of conflicting demands.
For example, we can design the weak decoder as an FPGA-based decoding system that implements the UF decoder with soft outputs proposed in Ref.~\cite{Meister2024}, and the strong decoder as a CPU- or GPU-based decoding system that implements tensor network decoders~\cite{Ferris2014, Bravyi2014, iOlius2024review} or state-of-the-art neural network decoders~\cite{Bausch2024alphaqubit, Varbanov2025, Hu2025, Blue2025}.
This strategic division of decoders' roles will dramatically simplify and accelerate the future development of real-time decoding devices.

Finally, to demonstrate the effectiveness of our framework, we have conducted several numerical simulations where we employ the MWPM decoder (or UF decoder) as a weak decoder and the belief-matching decoder~\cite{Higgott2023belief_matching} as a strong decoder. Our results elucidate that decoder switching can achieve high decoding accuracy comparable to, or even superior to, that of the strong decoder, while maintaining an average decoding speed on par with the weak decoder. 
Remarkably, we confirm that the switching rate $\gamma_{\text{switch}}$ exhibits a threshold behavior against physical error rates, and it declines exponentially below the threshold as the code distance $d$ increases. 
This behavior ensures that the requirement for $\tau_{\text{dec}}^{\text{strong}}$ is dramatically relaxed for large-scale computational problems.
These findings effectively address the long-standing trade-off inherent in existing decoding methods and pave a clear path toward the construction of high-speed, high-accuracy, real-time decoding devices in a flexible and scalable manner.

% \textcolor{red}{Here it should be noted that several hybrid decoder schemes have been proposed to date in several previous works~\cite{Battistel2023review, Delfosse2020, Smith2023, Higgott2023belief_matching, Caune2023, Shutty2024, Jones2024}. 
% These include the design of local predecoders~\cite{Delfosse2020, Meinerz2022, Ravi2023, Smith2023, Chamberland2023NN, Gicev2023}, the proposal of ensembling-based decoding approaches~\cite{Shutty2024, Jones2024}, and the switching based on the complementary gap~\cite{Jones2024}. 
% These methods are either too specific to a particular decoding method or not scalable from the perspective of the switching rate. It is also unclear whether they can truly avoid the speed-accuracy trade-off.
% In contrast, our work presents a more scalable and comprehensive framework for these topics, clearly demonstrating how it resolves the long-standing tradeoff.
% %However, our work presents a more scalable and comprehensive framework for these topics, which have primarily been explored in relation to specific decoding schemes or soft information. 
% For these aspects, we provide a detailed discussion in Sec.~\ref{sec:related works}.}

This paper is organized as follows: In Sec.~\ref{sec:Preliminary}, we introduce the fundamental concepts of real-time decoders in FTQC and provide a brief overview of recent developments in efficient soft-output decoders, which are crucial for understanding our proposal. Sec.~\ref{sec:decoder switching} details the protocol of decoder switching, and gives its theoretical interpretation and the analysis of its temporal overhead. Especially in Sec.~\ref{sec:decoding time}, we formulate an online decoding scheme, termed double window decoding scheme, to minimize the demands on the speed of the strong decoder.
Finally, we compare our formulation with related previous works in Sec.~\ref{sec:related works}. In Sec.~\ref{sec:numerical simulation}, we present numerical simulations to demonstrate the effectiveness of the proposed scheme, employing MWPM (or UF) as a weak decoder and belief-matching as a strong decoder. We first analyze the logical error rate and the switching rate around the error threshold, highlighting the significant improvement of the decoder's accuracy and the exponential decay of the switching rate. Then, by analyzing the tradeoff between switching rate and decoding accuracy, we identify the optimal values of $g_{\text{th}}$ for various setups and suggest that our scheme will work robustly even in practical-sized quantum circuits.
Finally, Sec.~\ref{sec:conclusion} concludes this paper by summarizing our findings and discussing future research directions.

\section{Preliminary}
\label{sec:Preliminary}

A decoder is a key element of a fault-tolerant quantum computer.
It processes a stream of error syndrome data and infers the most likely logical errors.
In this section, we first introduce the fundamental concepts related to real-time decoders in FTQC. Then, we give a brief overview of recent developments in efficient soft-output decoders, which is crucial to understanding our proposal.

\subsection{Basic concepts in decoding problem}
\label{sec:decoder}

The decoding problem in FTQC is a task of estimating logical errors that might occur in quantum states encoded on QEC codes, given a stream of syndrome data obtained from stabilizer measurements. Each syndrome provides partial information about the errors, but due to the inherent degeneracy of quantum codes, multiple error configurations often produce the same syndromes. The role of the decoder is to efficiently and accurately determine the most likely errors, or a suitable correction operation, that satisfy the syndrome conditions and are expected to restore the quantum state. 

In principle, any QEC code can be optimally decoded by using a {\it degenerate quantum maximum-likelihood (DQML)} decoder~\cite{Iyer2015,Poulin2006,iOlius2024review}. It adds up the probabilities of any possible error patterns that reproduce a given syndrome, with these errors grouped into equivalence classes based on their impact on logical degrees of freedom. Then, the class with the largest total likelihood is taken as the
decoding result of the DQML decoder. Although this method offers an optimal solution to decoding problems, its complexity generally belongs to \#P-complete problem~\cite{Iyer2015}. Tensor network decoders~\cite{Ferris2014, Bravyi2014, iOlius2024review} can simplify the DQML decoding problem through the use of tensor network techniques, yet they still demand significant computational resources. A more scalable and familiar approach is to find the most likely physical error pattern that produces a given syndrome. We refer to such a type of decoder as {\it most likely error (MLE)} decoder. 
For some class of QEC codes, such as surface codes~\cite{iOlius2024review}, we can map the MLE decoding problem into a matching problem on a decoding graph, enabling us to efficiently solve the task by employing classical algorithms developed in the context of graph theory, like Edmonds’ blossom algorithm~\cite{Edmonds1965_1,Edmonds1965_2}.
This approach is currently known as minimum-weight perfect matching (MWPM) decoder~\cite{Dennis2002, Fowler2015mwpm, Higgott2023sparse, Wu2023fusion}. While the MWPM decoder enables much faster decoding than tensor network decoders, the accuracy of the decoding results is substantially reduced since it is not always optimal.

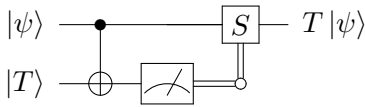
\begin{figure}[tbp]
\hspace{-10mm}
\centering
\fontsize{11pt}{11pt}\selectfont
  \mbox{
  \Qcircuit @C=1em @R=.7em {
    \lstick{\ket{\psi}}      & \ctrl{1}  &   \qw  & \gate{S}  \qw  & \rstick{T \ket{\psi} } \qw    \\
    \lstick{\ket{T}} & \targ &  \meter & \cctrlo{-1}
    }
  }
  \caption{A gate-teleportation circuit to execute a $T$ gate by utilizing a magic state denoted as $\ket{T}:=T\ket{+}$. This circuit incorporates a classically controlled $S$ gate, whose operation is contingent on the measurement outcome. In fault-tolerant implementations based on QEC codes, it is necessary to decode the logical $Z$ measurement outcome prior to the adaptive application of the $S$ correction. The response time is predominantly determined by the decoding time, but is also affected by communication and control latency.}
  \label{fig:GT_circ}
\end{figure}

In FTQC, adaptive logical operations conditioned on measurement outcomes are frequently required throughout the entire circuit, especially for executing non-Clifford gates. Consequently, decoding systems must process all previously generated syndrome data to reliably determine these measurement outcomes~\cite{Riesebos2017}. For example, to perform the gate teleportation circuit for applying a $T$-gate (as shown in Fig.~\ref{fig:GT_circ}), the outcome of a logical $Z$ measurement must be determined to decide whether a logical $S$ correction needs to be applied before performing the subsequent logical operations.

Unfortunately, realistic classical computing environments possess finite computational power and communication throughput. Consequently, decoders inevitably introduce a finite latency in FTQC. The following terminologies are commonly used when discussing the temporal aspects of decoder's performance~\cite{Delfosse2023tradeoff, Skoric2023}:
\begin{itemize}
    \item \textbf{Decoding Time:} The time taken by the decoding algorithm to process the syndrome and determine the appropriate correction. This is a critical factor to determine the overall response time (see Ref.~\cite{Battistel2023review} for review).
    \item \textbf{Syndrome Generation Time:} The time required to perform the syndrome measurements and generate the syndrome data. This is determined by the hardware and control system. For example, on a current superconducting qubit chip~\cite{Jeffrey2014, Arute2019,Google2023suppressing}, a single round of syndrome extraction takes around 1 $\mu$s for the surface code.
    \item \textbf{Response Time:} The total time elapsed from the generation of the final syndrome to the application of the correction. This includes the decoding time and any communication and control latency.
    \item \textbf{(Decoding) Window:} In online decoding schemes~\cite{Dennis2002, Skoric2023,Tan2023, Viszlai2024}, the decoder processes a subset of the syndrome data within a specific time window. The size of the decoding window affects both the accuracy and the latency of the decoder. 
\end{itemize}

In particular, the so-called {\it backlog problem} is crucially important in the context of decoding time.
This arises when the decoding time per round, $\tau_{\text{dec}}$, exceeds the syndrome generation time $\tau_{\text{gen}}$, namely, $\tau_{\text{dec}}>\tau_{\text{gen}}$. Such a situation leads to an accumulation of unprocessed syndrome data, which can exponentially slow down the fault-tolerant quantum computation~\cite{Terhal2015Review}. Although this issue might be relaxed to some extent with the use of parallelization techniques~\cite{Skoric2023,Tan2023}, the implementation of fast real-time decoders remains a crucial challenge for developing FTQC architectures~\cite{Battistel2023review}. In Sec.~\ref{sec:decoding time}, we will discuss how our proposal alleviates this crucial issue efficiently.

\subsection{Soft information in decoding problem}
\label{sec:soft_information}

In our context, soft information refers to some analog parameters that quantify the reliability of the decoder’s
estimate, rather than a hard decision on the most likely logical error. 
This concept should be distinguished from the analog information related to the signals in physical-level measurement, discussed in Refs.~\cite{Pattison2021,Hanisch2025, Majaniemi2025}, which is also referred to as ``soft information" in the context of decoding problems.

\begin{figure*}
    \centering
    \includegraphics[width=0.8\linewidth]{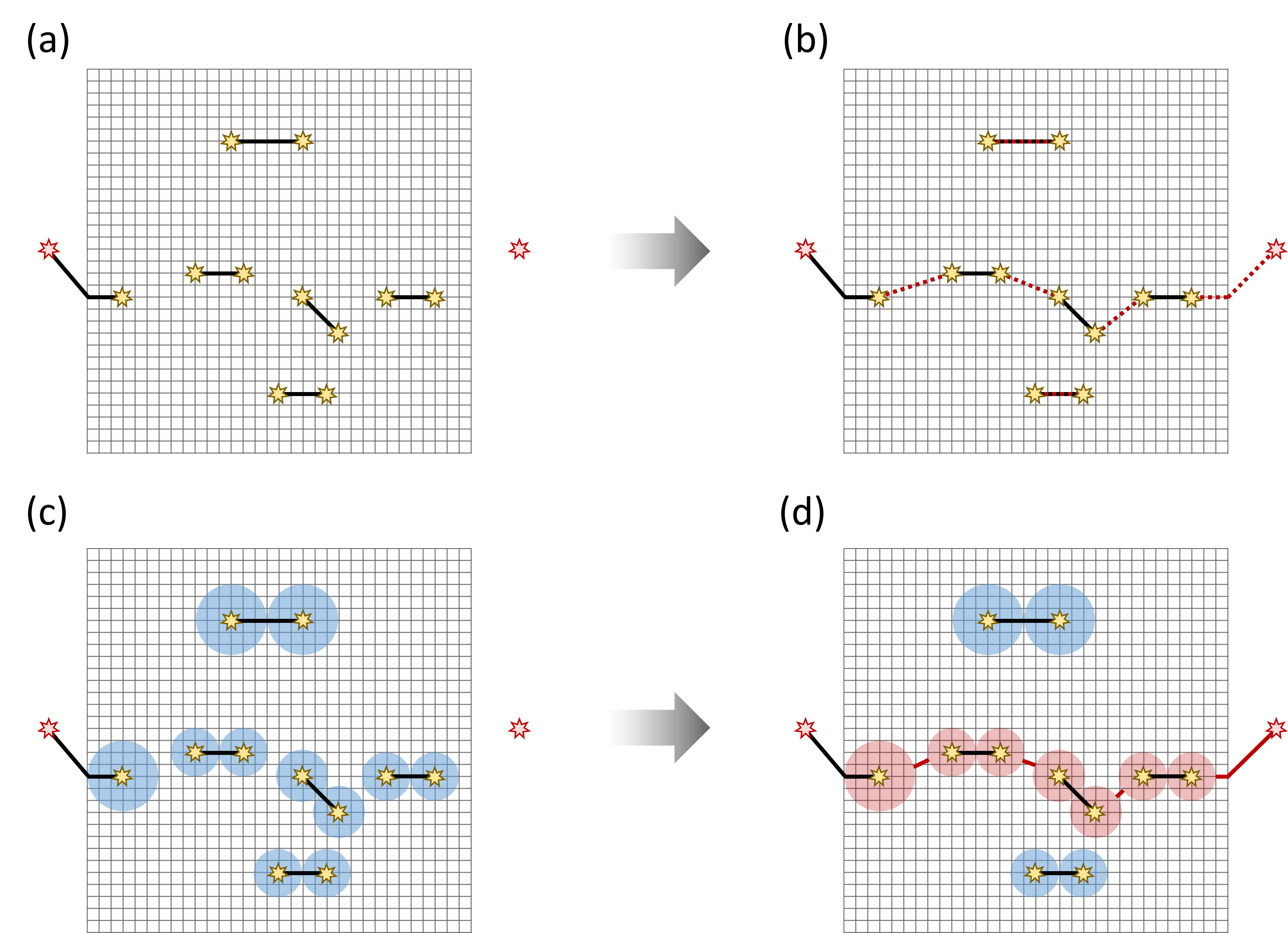}
    \caption{Schematic diagrams to describe how to calculate two types of soft outputs. ({\bf a,b}) {\bf Complementary gap}~\cite{Bombin2024, Gidney2025yoked}: First, we find the minimum-weight perfect matching (black solid lines) by using some matching-based decoder, given a syndrome configuration (yellow stars). By flipping the boundary conditions (red stars), we then find the so-called complementary matching (red dotted lines), which has a minimum weight conditioned on the new boundary conditions~\cite{Hutter2014}. The complementary gap is defined as the difference between the weights of these two matchings. ({\bf c,d}) {\bf Cluster gap}~\cite{Meister2024}: First, we find a perfect matching (black solid lines) by using some clustering-based decoder, such as the Sparse-Blossom decoder or the UF decoder. Then we find the shortest path (red solid lines) between boundary nodes on the quotient graph $G/C$, where $G$ is the decoding graph (background lattice) and $C$ is the set of clusters formed by the decoder (blue circles). The cluster gap is defined as the weights of the shortest path.}
    \label{fig:soft-output decoder}
\end{figure*}

%This analog information can be used to improve the overall performance of QEC systems, particularly in decoding of concatenated codes or when performing circuit sampling.
Here, let us consider a syndrome $\sigma$ obtained from a quantum error-correcting code. In this case, a hard-output decoder uniquely determines the most likely logical error $e$ among all possible logical errors.
In contrast, a soft-output decoder provides a probability distribution $P(e|\sigma)$ over the possible logical errors, conditioned on the observed syndrome $\sigma$. This distribution represents the decoder's confidence in different error scenarios.

Conventionally, such soft information is directly calculated by solving the DQML decoding problem with tensor network decoders~\cite{Ferris2014, Bravyi2014, iOlius2024review}.
However, these decoders require a considerably large decoding time, which is much higher than the typical value of the syndrome generation time. To avoid such a substantial time overhead, recent seminal works~\cite{Hutter2014, Bombin2024, Gidney2025yoked, Meister2024, Lee2025soft} have developed another type of soft information that can be evaluated with efficient decoders, like MWPM and UF decoders.
These soft-output decoders generate an analog parameter $g(\sigma)$, instead of directly estimating $P(e|\sigma)$.
Then, by collecting the statistics of $g$ for randomly generated syndromes, we can establish the conditional probability $P(e|g)$ as an empirical distribution. 
In what follows, we collectively refer to any conditional logical probabilities, such as $P(e|g)$ and $P(e|\sigma)$, as soft information. Similarly, we define soft outputs as any analog parameters $g(\sigma)$ that are calculated by decoders and utilized to estimate the conditional logical error rate $P(e|g)$.

The utility of soft information stems from its ability to capture the decoder's uncertainty, thereby providing a way to make adaptive decisions in subsequent logical operations.
For example, in the context of concatenated surface code, such as yoked surface code~\cite{Gidney2025yoked} and hierarchical codes~\cite{Pattison2023hierarchical, Meister2024}, soft information from the inner code is utilized as a prior error distribution in the decoding of the outer code. This breaks the degeneracy of error probability and enhances the performance of the concatenated codes~\cite{Poulin2006}. Furthermore, soft-output decoders can improve the reliability of the output states from various quantum circuits by flagging and discarding runs with a high probability of logical errors~\cite{Bombin2024, Gidney2024cultivation, Smith2024mitigation}.

In what follows, when mentioning the soft output $g$, we assume it to satisfy the following requirements: 
\begin{itemize}
    \item The soft output $g$ is normalized so that it has a non-negative value, and a smaller value of $g$ indicates that the decoder has a lower confidence.
    \item We can efficiently estimate the $g$-conditioned logical error rate $P(e|g)$ heuristically.
\end{itemize}
The first requirement means that $P(e|g)$ can be regarded as a monotonically increasing function of $g$. This does not seem to be necessarily needed to formulate our proposal. But, it is satisfied for the soft outputs treated in this work, and makes the discussion below easier to understand.
Furthermore, for simplicity, we assume that the decoder separates the decoding problem into $Z$ and $X$ subproblems, as is often the case in usual MWPM, and that the parameter $e$ always denotes the opposite class of the logical-$X$ or $Z$ error predicted by the decoder.

\subsection{Examples of efficient soft-output decoder}

In this section, we provide several examples of efficient soft-output decoders that will be applicable to our proposal. As discussed in some previous works~\cite{Gidney2025yoked, Meister2024, Shutty2024, Lee2025soft}, various types of existing decoding algorithms can be modified to provide soft outputs. 

A familiar approach is the {\it complementary gap} method~\cite{Bombin2024, Gidney2025yoked}, which evaluates the difference between the weights of minimum weight perfect matching for the two possible homology classes (see also Fig.~\ref{fig:soft-output decoder} (a-b)).
In this method, we first perform a usual MWPM decoder, and thereby, obtain the minimum weight perfect matching for some homology class. Here we denote the total weight as $w_{\text{min}}$.
Then, we perform the MWPM decoding again after constructing a decoding graph with modified boundary syndromes conditioned on the complementary logical outcome~\cite{Hutter2014}. This procedure leads to the complementary matching for the other homology class, whose weight is denoted as $w_{\text{comp}}$.
Finally, by calculating the difference between the weights of these matchings, we obtain the soft output $g_{\text{comp}}\equiv |w_{\text{comp}}-w_{\text{min}}|$ called {\it (unsigned) complementary gap}. This definition indicates that a small complementary gap (close to zero) implies the decoder lacks confidence in its decision, while a large one suggests high confidence.
In principle, this method would be applicable to any type of MLE decoders, including the MWPM decoder for any QEC codes with a matchable decoding graph.

The downside of the above method is that we have to apply a MWPM decoder to modified syndrome configurations, which totally differ from the typically sampled ones, to obtain the complementary perfect matching. This procedure tends to deteriorate the time overhead of the fast implementations of MWPM, known as the Sparse-Blossom decoder~\cite{Higgott2023sparse}. Another drawback is that the complementary gap method becomes further complex when there exist multiple logical operators on a patch, as in the case of lattice surgery~\cite{Horsman2012, Litinski2019}, since we have to consider various boundary syndrome conditions to cover any types of possible logical errors.

To alleviate these issues, Meister {\it et al}.~\cite{Meister2024} proposed an alternative approach that extracts soft output from the geometry of the clusters formed by clustering-based decoders like Sparse-Blossom~\cite{Higgott2023sparse} and UF decoder~\cite{Delfosse2021UF,Heer2023,Chan2023actis}. 
The soft output is defined as the total weights of the shortest path that covers a logical operator in a modified decoding graph where edges within the clusters have zero weight. 
More specifically, the soft output is computed by performing the following procedures (Fig.~\ref{fig:soft-output decoder} (c-d)):
\begin{enumerate}
    \item Run clustering-based decoders, such as Sparse-Blossom or UF decoder, to generate a set of clusters $C$ based on the syndrome $\sigma$. Here, the cluster set $C$ is defined as a union of balls formed around syndrome vertices on the decoding graph $G=(V,E,\omega)$.
    \item Create a quotient graph $G'\equiv G/C=(V,E,\omega')$ where the weight $\omega'(\mathcal{E})$ of each edge $\mathcal{E}$ ($\in E$) is set to zero if $\mathcal{E}$ is within a cluster in $C$, and to $\omega(\mathcal{E})$ otherwise.
    \item Run Dijkstra's algorithm~\cite{Dijkstra1959} on $G'$ to find the shortest path between inequivalent boundaries. The length of this path is provided as the soft output of the decoder.
\end{enumerate}
This algorithm can estimate the soft output quite efficiently because the UF decoder and Dijkstra's algorithm have almost linear time complexity. In addition, we can further improve the decoding time to almost constant overhead by leveraging the techniques of early stopping and parallelization~\cite{Kishi2025}.

In what follows, to distinguish the above soft output from the complementary gap, we refer to it as {\it cluster gap} and denote it as $g_{\text{cluster}}(\sigma)$.
Interestingly, according to Ref.~\cite{Meister2024}, it is shown that the cluster gap always provides a lower bound on the complementary gap, namely, $g_{\text{comp}}(\sigma) \geq g_{\text{cluster}}(\sigma)$ for any syndromes $\sigma$.

Another promising approach to define a soft output might be to quantify the degree of consensus among the ensemble of independent decoders with weakly perturbed priors, as demonstrated in Ref.~\cite{Shutty2024}.
This approach will be applicable to a fairly broad class of decoders, and it enables us to fully benefit from hardware parallelism since each decoder in the ensemble can be executed in a completely independent manner.

% \subsection{Definition of soft information}

% In this paper, we only assume that 
% \begin{enumerate}
%     \item We can estimate the $g$-conditioned logical error rate $P_L(g)$ numerically or theoretically.
%     \item $P_L(g)$ can be regarded as a monotonically increasing (or monotonically decreasing) function of $g$.
% \end{enumerate}

% Namely, given a decoding window and the related soft information, we can 

% \subsection{Examples of soft information}

% \begin{itemize}
%     \item complementary gap
%     \item collision gap 
%     \item collision pseudogap 
% \end{itemize}

% Interestingly, according to Ref.~\cite{}, these soft information satisfy the following simple inequalities:  
% \begin{equation}
%     g_{\text{comp}}(\sigma)\  \leq  \ g_{\text{col}}(\sigma) \ \leq \  g_{\text{pseudo}}(\sigma)
% \end{equation}

\section{Decoder switching}
\label{sec:decoder switching}

In this section, we propose a new decoding framework termed {\it decoder switching}. This framework addresses the fundamental trade-off between accuracy and speed in QEC decoding problems by leveraging hybrid decoding systems and their soft information. First, in Sec.~\ref{sec:protocol}, we introduce the protocol of decoder switching and highlight its significance in the hardware implementation of real-time decoders. Then, in Sec.~\ref{sec:thresholded logical error rate}, we provide a theoretical interpretation of decoder switching through the lens of the $g$-spectral decomposition of logical error rate. This theoretical analysis demonstrates how to determine the optimal threshold value for soft output. In Sec.~\ref{sec:decoding time}, we analyze the intricate relationship between the backlog problem and the switching rate under decoder switching. Specifically, we formulate the {\it double window decoding} scheme, an online decoding strategy tailored to decoder switching systems, and clarify straightforward criteria to prevent an exponential growth of the backlog.
While other types of hybrid decoder schemes have been proposed in the past few years~\cite{Battistel2023review, Delfosse2020, Smith2023, Higgott2023belief_matching, Caune2023, Shutty2024, Jones2024}, our work presents a more scalable and comprehensive framework for these topics, clearly demonstrating how it resolves the long-standing
tradeoff. To highlight these novelties, we provide a comprehensive comparison with these related works in Sec.~\ref{sec:related works}.

\subsection{Protocol}
\label{sec:protocol}

The decoding system based on decoder switching is composed of a fast soft-output decoder and a highly accurate decoder with a relatively high latency.
In what follows, we refer to these paired decoders as ``weak decoder" and ``strong decoder", respectively.
The core principle of decoder switching is to dynamically adopt the appropriate decoder based on the confidence of the weak decoder. The primitive procedures unfold as follows:
\begin{enumerate}
\item {\bf Parallel Decoding}: For a given decoding window, a sequence of syndrome data $\sigma$ is simultaneously fed to both the weak and strong decoders, initiating their decoding processes independently.
\item {\bf Early Assessment with Weak Decoder}: The weak decoder rapidly estimates the most likely logical errors and computes a soft output $g(\sigma)$, which quantifies the decoder's confidence in the estimate.
\item {\bf Conditional Early Exit}: If the soft output $g$ exceeds a preset threshold $g_{\text{th}}$, indicating high confidence in the weak decoder's result, its estimation is immediately adopted as the output of the hybrid decoding system. Simultaneously, the ongoing computation on the strong decoder is halted to save computational resources.
\item {\bf Switching to Strong Decoder}: Conversely, if $g$ falls below the threshold, signifying low confidence, the weak decoder's outcome is discarded. The decoding system then waits for the more accurate result from the strong decoder, which is subsequently used as the final estimate for the hybrid decoding system.
\end{enumerate}
This adaptive strategy ensures that the strong decoder is invoked only when truly necessary, allowing the system to maintain a sufficiently small decoding time on average by predominantly relying on the fast weak decoder. 
Increasing the threshold $g_{\text{th}}$ normally improves the decoding accuracy at the cost of the average decoding time. To balance this tradeoff, we need to properly determine the threshold value of the soft output $g_{\text{th}}$. This point will be discussed in detail in Sec.~\ref{sec:thresholded logical error rate}. 

Additionally, if one aims to conserve classical computational resources as much as possible, the simultaneous execution of the weak and strong decoders in Step 1 can be forgone. In such a scenario, the protocol is modified to execute the strong decoder only when necessary, after the early assessment by the weak decoder (Step 2) is complete. 
As seen in Sec.~\ref{sec:numerical simulation}, the switching rate to the strong decoder decays exponentially with increasing code distance. Therefore, such a modification is expected to have a negligible impact on the overall decoding time. 
Moreover, even in scenarios where numerous code patches are processed concurrently by multiple weak decoders, it is anticipated that the decoder switching system will invoke at most one strong decoder at each timeslice, given a sufficiently large code distance.
%Moreover, in a situation where many code patches are processed in parallel by multiple weak decoders, we expect that the number of strong decoders invoked at a specific window will be at most one when assuming a sufficiently large code distance.
In this sense, the above approach will significantly improve the scalability of classical resources without compromising the decoder's accuracy. This is particularly important when developing a real-time decoding system for large FTQC devices at the scale of thousands of logical qubits. We will delve into these perspectives in more detail in Sec.~\ref{sec:decoding time}.

A significant advantage of our framework is its broad applicability across various decoding algorithms and QEC codes. To ensure the validity of decoder switching, we only assume that each decoder satisfies the following requirements:
\begin{itemize}
    \item The weak decoder efficiently processes syndrome data with a decoding time per round, $\tau_{\text{weak}}$, that is less than the syndrome generation time $\tau_{\text{gen}}$ of the quantum devices. Furthermore, it simultaneously generates a soft output $g$ for each decoding window, which measures the reliability of the decoding outcome (as illustrated in Sec.~\ref{sec:soft_information}).
    \item The strong decoder offers higher accuracy than the weak decoder, achieving an error threshold higher than the physical error rate of the quantum devices.
    \item The value of $g_{\text{th}}$ is determined to guarantee that the switching rate remains below a certain threshold value, given by Eq.~\eqref{eq:condition for switching rate} later, to avoid the backlog problem.
\end{itemize}
Crucially, our framework differs from traditional single-decoder approaches in that it does not impose the stringent requirement that the strong decoder's decoding time $\tau_{\text{strong}}$ be shorter than the syndrome generation time $\tau_{\text{gen}}$. Similarly, it does not require the reaction time to be comparable to the required clock time of logical operations.
In fact, as shown later in Theorem~\ref{thm: sufficient condition}, the upper bound on $\tau_{\text{strong}}$ to avoid the backlog problem is lifted inversely with the switching rate.
This significant relaxation of latency requirements opens up unprecedented opportunities: it enables us to implement highly complex and accurate decoders on powerful, yet potentially high-latency and energy-intensive, computing environments such as High-Performance Computing (HPC) clusters. This is vital for pushing the boundaries of real-time decoding systems, where the response time between quantum devices and classical processors might otherwise be a bottleneck.

% Finally, \textcolor{red}{it should be noted that the third requirement often necessitates the error threshold of the weak decoder to be comparable to or less than the physical error rate of quantum devices. At least, this holds for the examples examined in this work}, where we utilize well-established decoders like the MWPM decoder or the UF decoder as the weak decoder.
% Specifically, in Sec.~\ref{sec:numerical simulation}, we see that the switching rate to the strong decoder varies in close correlation with the logical error rate of the weak decoder. This means that the weak decoder needs to have a certain level of accuracy to suppress the switching rate substantially.
Finally, we note that the third requirement typically necessitates the weak decoder to have a certain level of accuracy to suppress the switching rate in a scalable manner. In fact, in Sec.~\ref{sec:numerical simulation}, we see that the switching rate varies almost proportionally to the logical error rate of the weak decoder. This indicates that the threshold of the weak decoder should be comparable to or larger than the physical error rate of quantum devices to suppress the switching rate exponentially.
Nonetheless, if we do not demand exponential decay in switching rate, decoder switching might work well even for some weak decoders that do not show an error threshold.
It might be an interesting open problem to determine to what extent the performance of the weak decoder can be reduced to ease hardware requirements.

\subsection{Theory}
\label{sec:thresholded logical error rate}

In this subsection, we give a more theoretical viewpoint of decoder switching by considering the relation between the threshold $g_{\text{th}}$ and the logical error rate. Then, we discuss how to estimate the near-optimal value of the threshold $g_{\text{th}}$ by introducing the concept of a {\it thresholded logical error rate}.

%As explained below, the near-optimal value of the threshold $g_{\text{th}}$ can be estimated by introducing the concept of a {\it thresholded logical error rate}.

First, let us begin to interpret the logical error rate under a usual decoding scheme from the viewpoint of the soft output. Generally, the average logical error rate $P_L$ for a given soft-output decoder can be expressed in a spectrally decomposed form using the soft output $g$ as follows:
\begin{equation}
\label{eq:average logical error rate}
P_L = \int_0^{\infty} dg  \  P(e|g)\cdot p(g)
\end{equation}
where $p(g)$ is the probability distribution of the soft output $g$, and $P(e|g)$ is the $g$-conditional logical error rate.
In principle, the probability distribution $p(g)$ and $P(e|g)$ are uniquely determined by defining the noise model for the syndrome extraction circuit. Here, we assume that we can empirically determine the form of these functions from the statistical data obtained via the stabilizer circuit simulations, as demonstrated below.

Next, consider the case where we use a hybrid decoding system under the decoder switching scheme. In this case, we can evaluate the logical error rate under decoder switching as 
\begin{equation}
\label{eq:error rate under switching}
\begin{aligned}
P_{L,\text{switch}}(g_{\text{th}}) &= \int_0^{g_{\text{th}}} dg\   P_{\text{strong}}(e|g)\cdot p(g) \\ 
& \qquad \qquad + \int_{g_{\text{th}}}^\infty dg \  P_{\text{weak}}(e|g)\cdot p(g),\\
\end{aligned}
\end{equation}
or equivalently, 
\begin{equation}
\label{eq:comparison of error rate}
\begin{aligned}
&P_{L,\text{switch}}(g_{\text{th}}) - P_{L, \text{strong}}\\  &\qquad\quad =\int_{g_{\text{th}}}^\infty dg \ \left[P_{\text{weak}}(e|g)-P_{\text{strong}}(e|g)\right]\cdot p(g),
\end{aligned}
\end{equation}
where $P_{L, \text{strong}}$ is the average logical error rate of the strong decoder, and $P_{\text{strong}}(e|g)$ and $P_{\text{weak}}(e|g)$ are the conditional logical error rate of the weak and strong decoder, respectively, under the constraint that the value of soft output is evaluated as $g$ by the weak decoder. These equations suggest that the threshold value $g_{\text{th}}$ should be determined to make the difference between $P_{\text{strong}}(e|g)$ and $P_{\text{weak}}(e|g)$ sufficiently small for $g>g_{\text{th}}$.
Furthermore, if the inequality $P_{\text{weak}}(e|g)<P_{\text{strong}}(e|g)$ holds in a specific range of soft output, we can achieve an even lower error rate than that of the strong decoder. Interestingly, we later observe this kind of behavior in a decoder switching system utilizing the MWPM (or UF) and belief-matching decoders. These findings may provide insight into a new aspect of decoders: the $g$-resolved performance of decoding algorithms.

\begin{figure}
    \centering
    \includegraphics[width=0.95\linewidth]{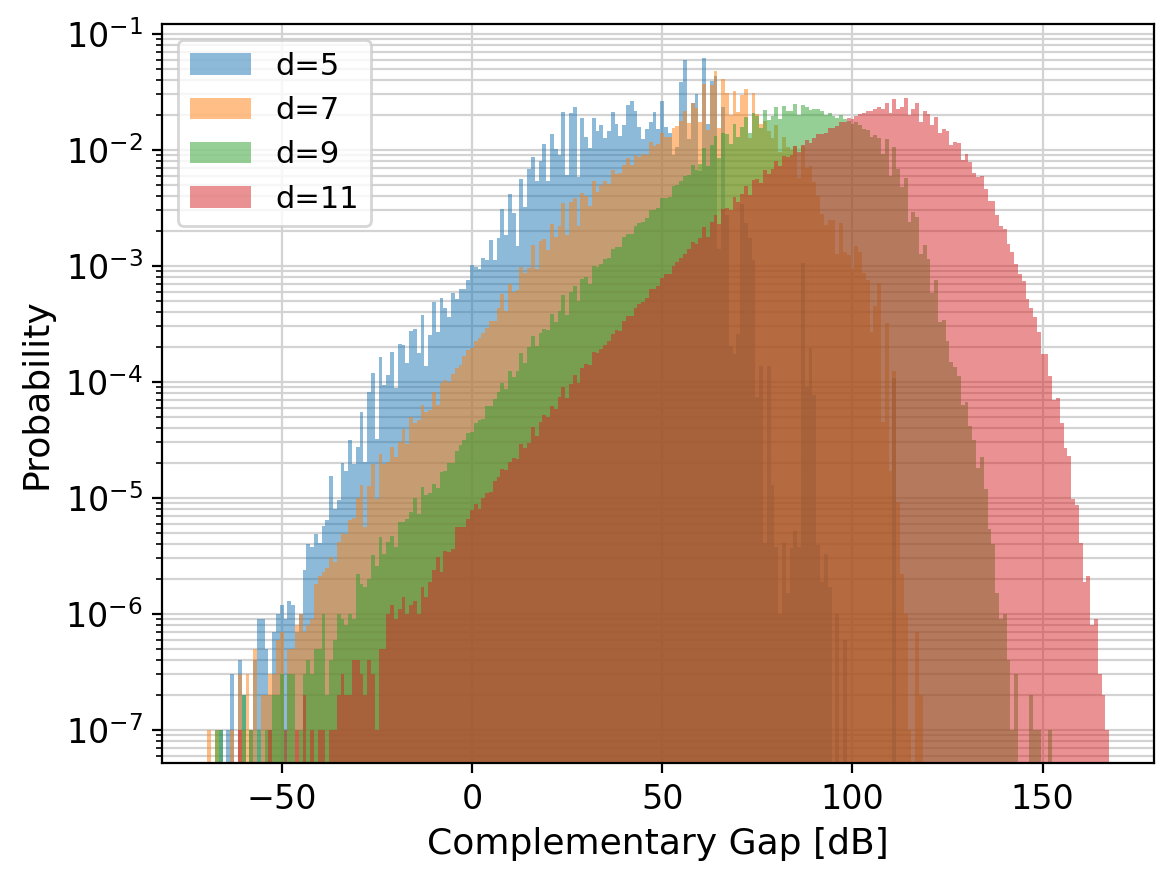}
    \caption{Histograms of (signed) complementary gaps $\tilde{g}_{\text{comp}}$ for rotated surface code, sampled from $10d$-rounds memory experiments with perfect terminal time boundaries. which is reproduced based on the approach in Ref.~\cite{Gidney2025yoked}. This data was gathered with $N=10^7$ shots under the uniform circuit-level noise model at a physical error rate of $p_{\text{ph}}=10^{-3}$. Here we denote the value of complementary gaps in the unit of dB.}
    \label{fig:histogram}
\end{figure}

\begin{figure}
    \centering
    \includegraphics[width=0.95\linewidth]{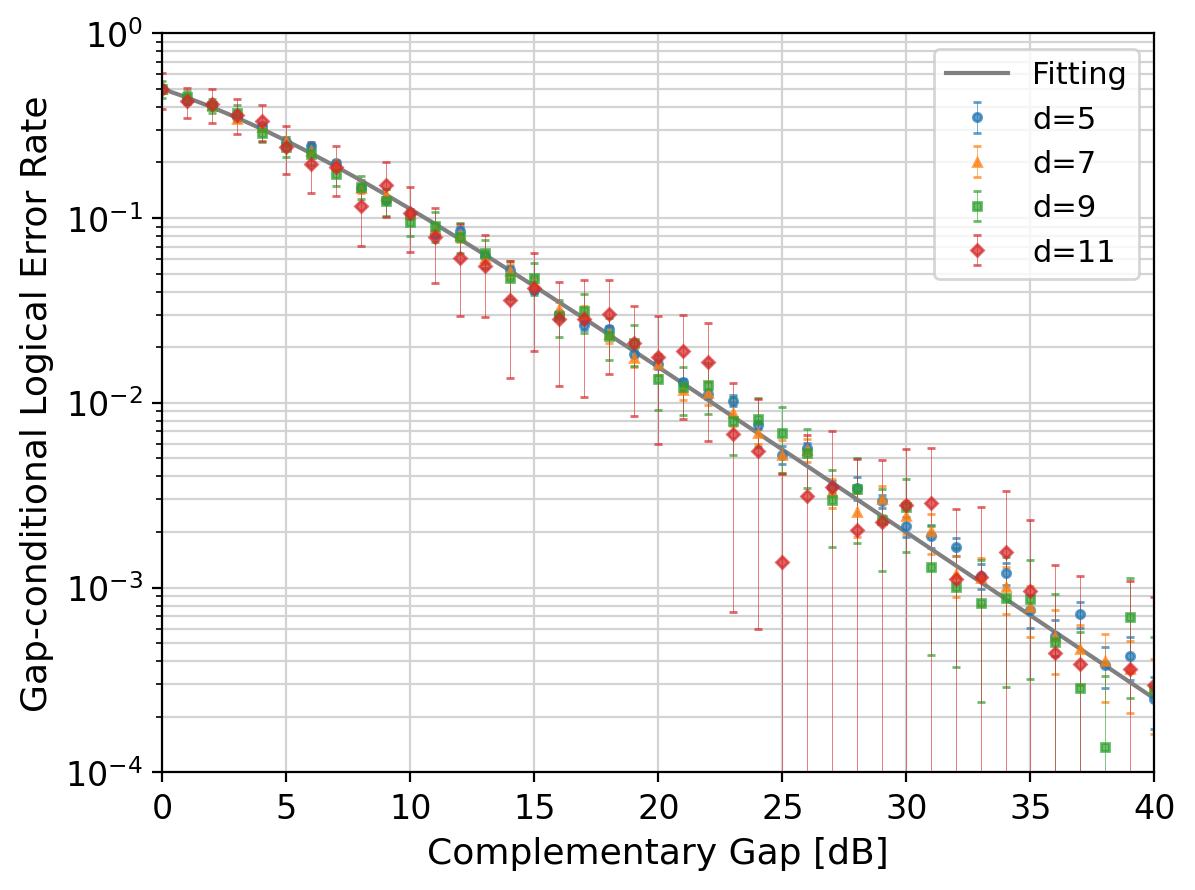}
    \caption{Gap-conditional logical error rate $P_{\text{weak}}(e|g_{\text{comp}})$ for MWPM decoder, which is produced from the histogram data in Fig.~\ref{fig:histogram}. As demonstrated in Ref.~\cite{Gidney2025yoked}, each plot is fitted well with an empirical function $f(g_{\text{comp}})=(1+10^{0.09g_{\text{comp}}})^{-1}$ (gray solid line), where the complementary gap $g_{\text{comp}}$ is also denoted in the unit of dB.}
    \label{fig:conditional_error_rate}
\end{figure}

\begin{figure}
    \centering
    \includegraphics[width=0.95\linewidth]{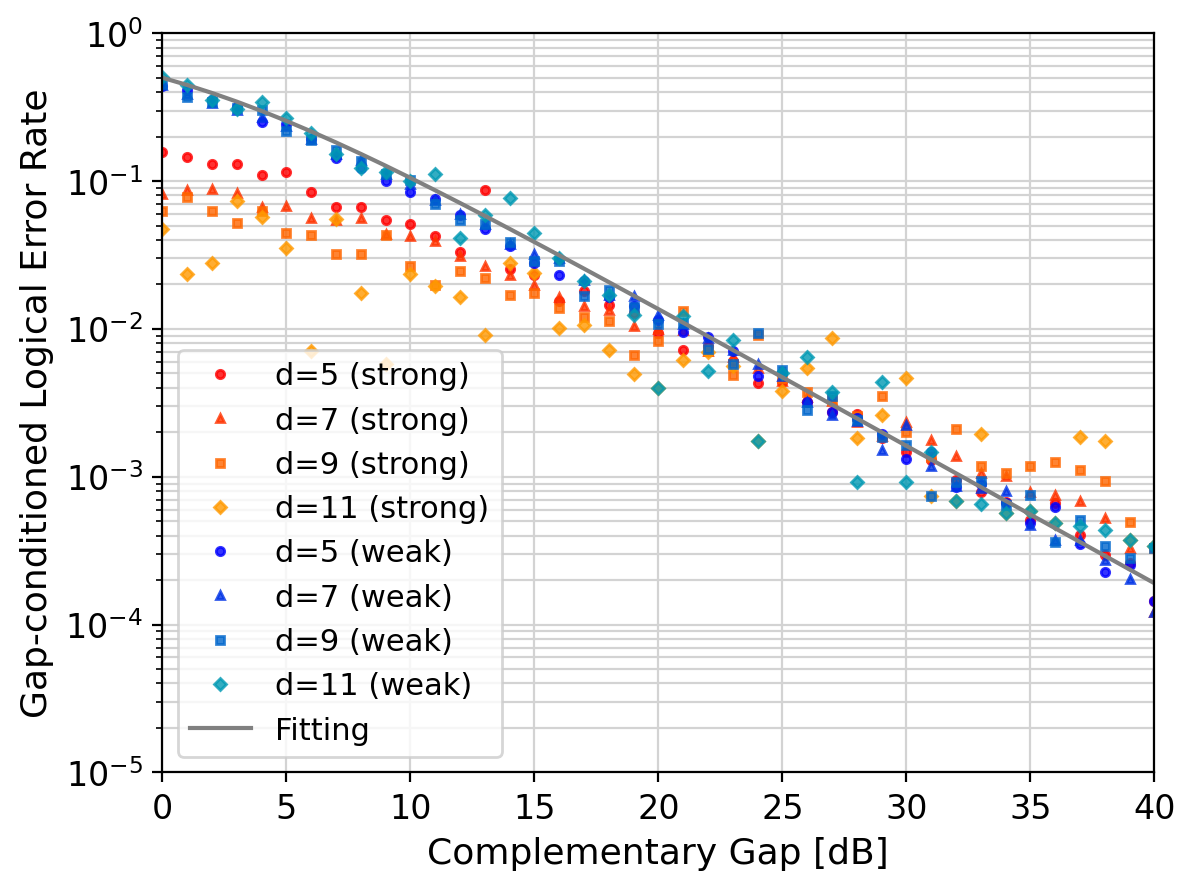}
    \caption{Comparison of gap-conditional logical error rate for weak and strong decoder, namely, $P_{\text{weak}}(e|g_{\text{comp}})$ and $P_{\text{strong}}(e|g_{\text{comp}})$. Here, we select the MWPM decoder as the weak decoder and the belief-matching decoder as the strong decoder. This data was gathered with $N=10^8$ shots of $d$-rounds memory experiments, rather than $10d$-rounds memory experiments, to suppress the computation costs on the strong decoder.}
    \label{fig:conditional_error_rate_for_strong}
\end{figure}

In what follows, we numerically illustrate how the above parameters, such as $p(g)$ and $P(e|g)$, actually behave, taking the complementary gap of rotated surface codes~\cite{Bombin2007, Horsman2012} as an example. We employ the MWPM decoder as the weak decoder and the belief-matching decoder as the strong decoder.
In Fig.~\ref{fig:histogram}, we show the histograms of the {\it signed} complementary gap $\tilde{g}_{\text{comp}}$ for rotated surface codes with a code distance $d\in \{5,7,9,11\}$. Here we define the {\it signed} complementary gap for convenience as follows:
\begin{equation}
    \tilde{g}_{\text{comp}}\equiv \left\{
\begin{array}{ll}
   w_{\text{comp}}-w_{\text{min}} & \ \ (\text{If MWPM is correct}) \\
   w_{\text{min}}-w_{\text{comp}} & \ \ (otherwise)
\end{array}
\right.
\end{equation}
In this simulation, we have executed $10d$-rounds memory experiments with perfect terminal time boundaries, according to the approach used in Ref.~\cite{Gidney2025yoked}. 
The data was gathered with $N=10^7$ shots using PyMatching~\cite{Higgott2023sparse, Higgott2022PyMatching} under the uniform circuit-level noise model at a physical error rate of $p_{\text{ph}}=10^{-3}$. 
The figure shows that the peak position of the histogram shifts to the right as the code distance $d$ becomes larger. In other words, the probability that the gap's value falls below some constant decreases exponentially as the code distance increases.
By converting the signed complementary gap into an unsigned one, we can obtain the distribution function $p(g_{\text{comp}})$ in Eq.~\eqref{eq:average logical error rate}.

In Fig.~\ref{fig:conditional_error_rate}, we plot the gap-conditional logical error rate $P_{\text{weak}}(e|g_{\text{comp}})$, which is generated by using the statistical data in Fig.~\ref{fig:histogram}.
Interestingly, this figure suggests that the curve of $P_{\text{weak}}(e|g_{\text{comp}})$ can be fitted well with a empirical function $f(g_{\text{comp}})=(1+10^{0.09g_{\text{comp}}})^{-1}$, which is the same as that in Ref.~\cite{Gidney2025yoked}, despite the difference of the noise models and the decoders used in each simulation.
Moreover, we show the comparison of gap-conditional logical error rate
for the weak and strong decoder in Fig.~\ref{fig:conditional_error_rate_for_strong}.
As might be expected, the difference between $P_{\text{weak}}(e|g_{\text{comp}})$ and $P_{\text{strong}}(e|g_{\text{comp}})$ is most pronounced near $g_{\text{comp}}=0$ and gradually diminishes as the value of $g_{\text{comp}}$ increases. The most remarkable is that the magnitude relationship between $P_{\text{weak}}(e|g_{\text{comp}})$ and $P_{\text{strong}}(e|g_{\text{comp}})$ inverts around $20$ - $30$ dB, allowing the weak decoder to achieve a lower $g$-conditional logical error rate than the strong decoder.
According to Eq.~\eqref{eq:comparison of error rate}, this inversion implies that, in a specific range of $g_{\text{comp}}$, we can achieve a minimum value of logical error rate, which is even smaller than that of the strong decoder. This point will be demonstrated more clearly in Sec.~\ref{sec:numerical simulation}.
We expect that the phenomenon mentioned above originates from the fact that the belief-matching decoder is a heuristic decoder based on the belief propagation (BP), which does not guarantee that it always produces more optimal outputs than the weak decoder (MWPM decoder). This means that similar behaviors might emerge in other heuristic strong decoders, such as neural-network decoders~\cite{Lenssen2025fooling}.

A practical issue when implementing the decoder switching is how to set the threshold value $g_{\text{th}}$ optimally. 
In principle, the optimal threshold $g_{\text{th}}$ can be determined in a brute-force manner, as demonstrated later in Sec.~\ref{sec:numerical simulation}. Namely, we initially set $g_{\text{th}}$ to a sufficiently small value and numerically calculate the logical error rate $P_L(g_{\text{th}})$ under decoder switching. Then, by gradually increasing the value of $g_{\text{th}}$, we can find the value of $g_{\text{th}}$ at which $P_L(g_{\text{th}})$ exhibit a minimal point or sufficiently saturates to $P_{L, \text{strong}}$. These procedures will give an optimal value of $g_{\text{th}}$.

However, in practice, the strong decoder requires much more computational or 
engineering cost than the weak decoder. Therefore, it is desirable to enable determining a near-optimal threshold without actually executing the strong decoder for large code distances or even without specifying the details of the strong decoder.
To this end, here we introduce the concept of the {\it thresholded logical error rate}, which is defined as 
\begin{equation}
    P_{L,\text{th}}(g_{\text{th}}) = \int_{g_{\text{th}}}^\infty dg \  P_{\text{weak}}(e|g)\cdot p(g).
\end{equation}
This factor corresponds to the second term in the right-hand side of Eq.~\eqref{eq:error rate under switching}. Then, by setting the threshold $g_{\text{th}}$ to be a minimum value that satisfies the condition
\begin{equation}
\label{eq:condition for optimal threshold}
    P_{L,\text{th}}(g_{\text{th}}) \leq \epsilon \cdot   P_{L, \text{strong}},
\end{equation}
we can guarantee that the following inequality holds:
\begin{equation}
P_{L,\text{switch}}(g_{\text{th}})\  \leq \ P_{L, \text{strong}}+ P_{L,\text{th}}(g_{\text{th}}) \ \leq\  (1+\epsilon)P_{L, \text{strong}},
\end{equation}
where $\epsilon$ is a tunable parameter to determine the upper bound of the logical error rate. In other words, provided the $g_{\text{th}}$-dependence of the function $P_{L,\text{th}}(g_{\text{th}})$ and a target error rate $P_{L, \text{strong}}$, we can determine $g_{\text{th}}$ to be small enough to realize $P_{L,\text{switch}}(g_{\text{th}})\simeq P_{L, \text{strong}}$ within a target accuracy $\epsilon$.

\begin{figure}
    \centering
    \includegraphics[width=0.95\linewidth]{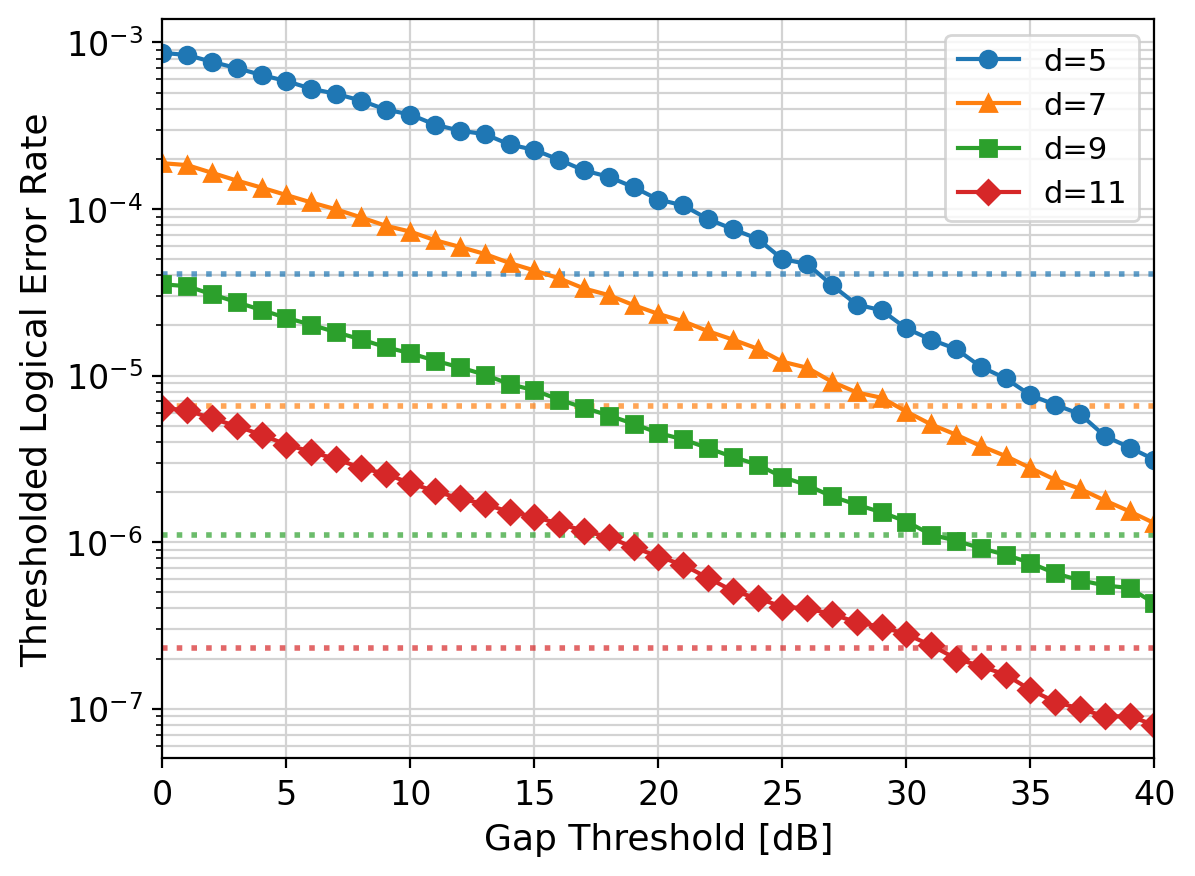}
    \caption{Thresholded logical error rate per $d$-rounds for MWPM decoder with complementary gap as a soft output. This data is generated by combining the data in Fig.~\ref{fig:histogram} and Fig.~\ref{fig:conditional_error_rate} and normalizing the error rate for $10d$-rounds to that for $d$-rounds. The parameter setup is the same as that in Fig.~\ref{fig:histogram}. For comparison, we also plot the values of $\epsilon \cdot  P_{L, \text{strong}}$ with dotted lines, where we set $\epsilon=0.1$ and $P_{L, \text{strong}}$ denotes the logical error rate of the belief-matching decoder.}
    \label{fig:thresholded_error_rate}
\end{figure}

In Fig.~\ref{fig:thresholded_error_rate}, we plot the thresholded logical error rate per $d$-rounds for the MWPM decoder with complementary gap as a soft output.
This plot clearly shows that the thresholded logical error rate decreases exponentially as the gap threshold $g_{\text{th}}$ and the code distance $d$ increase.
In addition, we plot the values of $\epsilon \cdot  P_{L, \text{strong}}$ with dotted lines for comparison. Here we set $\epsilon=0.1$ and $P_{L, \text{strong}}$ denotes the logical error rate of the belief-matching decoder. Compared to Eq.~\eqref{eq:condition for optimal threshold}, the plotted data indicate that setting $g_{\text{th}}$ to be around 30 dB is sufficient to ensure the hybrid decoding system achieves accuracy comparable to the belief-matching decoder~\cite{Higgott2023belief_matching} within a 10\% margin of deviation. 
In the next section, however, we will find that these threshold values are somewhat overestimated to achieve the target error rate when compared to the optimal values obtained by a brute-force approach. Specifically, as depicted later in Fig.~\ref{fig:tradeoff}, we will demonstrate that gap thresholds below 20 dB are sufficient to achieve a logical error rate even within 1\% of that attained by the strong decoder for $d=9,11$.

\begin{figure}
    \centering
    \includegraphics[width=0.95\linewidth]{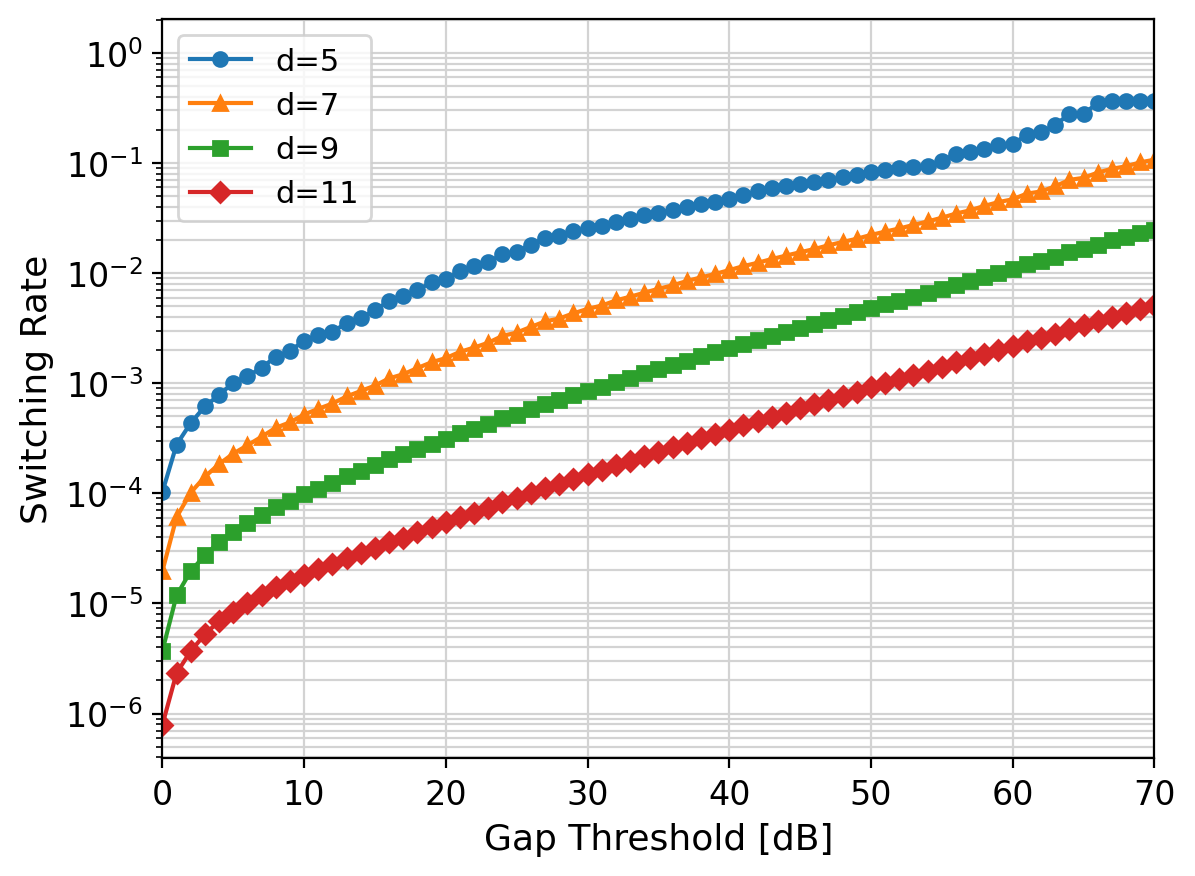}
    \caption{Switching rate per $d$-rounds for MWPM decoder with complementary gap $g_{\text{comp}}$ as a soft output. Here we estimated the switching rate per $d$-rounds from the rates per $10d$-rounds by assuming that the switching event occurs under an independent and identical distribution
in each decoding window of $d$-rounds. The parameter setup is the same as that in Fig.~\ref{fig:histogram}.}
    \label{fig:swithcing_rate_theory}
\end{figure}

\begin{figure*}
    \centering
    \includegraphics[width=0.7\linewidth]{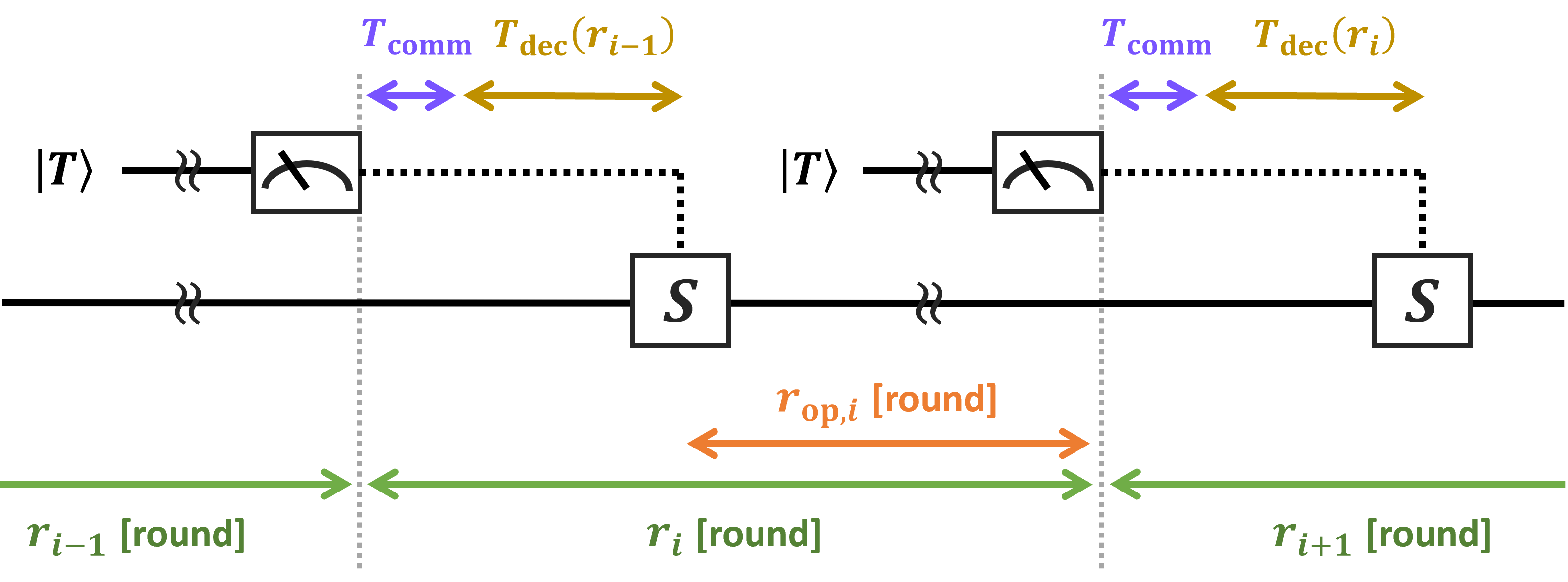}
    \caption{Time evolution of the syndrome backlog when executing a sequence of non-Clifford gates.  Key parameters are defined as follows: $r_i$, the number of accumurated rounds (i.e., syndrome backlog) between destructive measurements that determine the feedback operations for the previous and the next non-Clifford gates; $T_{\text{comm}}$, the latency time for a single round of error syndromes to be transmitted to the real-time decoder; $T_{\text{dec}}(r_i)$, the decoding time for processing the syndromes generated over $r_i$ rounds; $r_{\text{op}}$, the number of rounds between the last feedback operation and the subsequent desctructive measurement of an ancilla qubit.}
    \label{fig:backlog}
\end{figure*}

\subsection{Decoding time and backlog problem}
\label{sec:decoding time}

As mentioned in Sec.~\ref{sec:decoder}, a real-time decoding system is required to suppress its decoding and response time as much as possible. This is critical to prevent the exponential accumulation of syndrome backlog and the consequent slowdown of logical operations. In what follows, we analyze the accumulation of the syndrome backlog under decoder switching and thereby clarify the requirements for avoiding the above issues.

In decoder switching, the switching rate per $d$-rounds is evaluated based on the distribution function of soft output $p(g)$ as follows:
\begin{equation}
\gamma_{\text{switch}}(g_{\text{th}}) = \int_0^{g_{\text{th}}} dg  \  p(g),
\end{equation}
where we assume that the distribution function $p(g)$ is normalized to describe the gap distribution for a decoding window of $d$-rounds. Fig.~\ref{fig:swithcing_rate_theory} shows the switching rate per $d$-rounds for the MWPM decoder with complementary gap as its soft output. 
This figure shows that the switching rate decreases exponentially with increasing code distance $d$ and increases exponentially with increasing gap threshold $g_{\text{th}}$.

Next, following the discussion in Refs.~\cite{Chamberland2023NN, Terhal2015Review}, let us overview the backlog problem in conventional single-decoder approaches. For simplicity, here we do not consider the sliding window scheme~\cite{Dennis2002}, but assume a ``naive" online decoding scheme. More specifically, as shown in Fig.~\ref{fig:backlog}, we first collect all the syndrome data generated during $r_i$ rounds between the destructive measurements for gate teleportation. Right after receiving the data, the decoder processes it collectively to determine the feedback operations for the previous non-Clifford gate. 
We assume that the syndrome data for each round is transferred to the real-time decoder with a latency time of $T_{\text{comm}}$, and the decoder can process the syndrome data within the decoding time of $T_{\text{dec}}(r_i)=\tau_{\text{dec}} r_i$, where $\tau_{\text{dec}}$ is a constant denoting the decoding time per round.
Consequently, the backlog $r_i$ includes the rounds for awaiting the previous decoding outcome and the rounds for logical operations to execute the next non-Clifford gate, $r_{\text{op},i}$:
\begin{equation}
\label{eq:backlog}
    r_i = \frac{T_{\text{comm}} + \tau_{\text{dec}} r_{i-1}}{\tau_{\text{gen}}} + r_{\text{op},i}
\end{equation}
Strictly speaking, the first term on the right-hand side does not necessarily take an integer value, and so, should be rounded up using the ceiling function. However, for simplicity, we will ignore such rounding operations here, since it will not matter for the following discussion.
The value of $r_{\text{op},i}$ depends on the details of the quantum circuit. For example, assuming the compilation in Ref.~\cite{Litinski2019}, $r_{\text{op},i}$ has a value ranging from $d$ rounds to $9d$ rounds for a compact footprint of surface codes.

Supposing that $r_{\text{op},i}=r_{\text{op}}$, we can easily solve the recursive equation in Eq.~\eqref{eq:backlog} and the solution is given as follows:
\begin{equation}
\label{eq:solution}
    r_i = f^i r_0 + \frac{1-f^i}{1-f}\left( r_{\text{op}} + \frac{T_{\text{comm}} }{\tau_{\text{gen}}} \right),
\end{equation}
where we introduce $f \equiv \tau_{\text{dec}}/\tau_{\text{gen}}$.
It is apparent from this formula that the backlog $r_i$ diverges exponentially when $f>1$, and it converges to the following value when $f<1$:
\begin{equation}
    \lim_{i\to\infty} r_i = \frac{1}{1-f}\left( r_{\text{op}} + \frac{T_{\text{comm}} }{\tau_{\text{gen}}} \right)
\end{equation}
In particular, the deviation of the converged value from $r_{\text{op}}$ describes how decoding processes contribute to the total time overhead for executing each non-Clifford gate.

In this work, we extend the above analysis to the case of hybrid decoding systems under the decoder switching scheme.
In this case, we need to introduce the decoding time per round $\tau_{\text{dec}}^{\text{weak}}$, $\tau_{\text{dec}}^{\text{strong}}$ and the communication latency $T_{\text{comm}}^{\text{weak}}$, $T_{\text{comm}}^{\text{strong}}$ for the weak and the strong decoder, respectively.
Furthermore, the backlog $r_i$ is no longer a deterministic variable but becomes a probabilistic one. So we introduce the probability distribution of $r_i$ as $\{r_i^{(k)}, p_k\}_{k=1,2,\cdots, N_k}$. Then, in decoder switching systems, we can estimate the average value of $r_i$ by using the distribution of the previous backlog $r_{i-1}$ as follows:
\begin{equation}
\label{eq:backlog for decoder switching}
\begin{aligned}
    \average{r_i} &=\ \sum_k p_k \left(1-\frac{\gamma_{\text{switch}} r_{i-1}^{(k)}}{d}\right)\left(\frac{T_{\text{comm}}^{\text{weak}}
    + \tau_{\text{dec}}^{\text{weak}} r_{i-1}^{(k)}}{\tau_{\text{gen}}}\right) \\ & \ \ \  + \sum_k p_k\frac{\gamma_{\text{switch}}r_{i-1}^{(k)}}{d}  \left(\frac{T_{\text{comm}}^{\text{strong}}
    + \tau_{\text{dec}}^{\text{strong}} r_{i-1}^{(k)}}{\tau_{\text{gen}}}\right)
    + r_{\text{op},i}\\
    & = \frac{T_{\text{comm}}^{\text{weak}}
    + \tau_{\text{dec}}^{\text{weak}} \average{r_{i-1}}}{\tau_{\text{gen}}}
    + r_{\text{op},i}\\
    & \ \ \   + \frac{\gamma_{\text{switch}}}{d} \left(
    \frac{\Delta T_{\text{comm}}\average{r_{i-1}} + \Delta\tau_{\text{dec}}\average{r_{i-1}^2}  }{\tau_{\text{gen}}}
    \right),
\end{aligned}
\end{equation}
where 
\begin{equation}
    \Delta T_{\text{comm}} = T_{\text{comm}}^{\text{strong}}-T_{\text{comm}}^{\text{weak}},\ \ 
    \Delta\tau_{\text{dec}} = \tau_{\text{dec}}^{\text{strong}} - \tau_{\text{dec}}^{\text{weak}}.
\end{equation}
Here, we implicitly assume that the switching rate per $d$-round, $\gamma_{\text{switch}}$, is sufficiently small and so we can approximate the total switching rate within the backlog $r$ as $\gamma_{\text{switch}}r/d$, since the switching event can be regarded as independent and identical between different small windows.
Unlike Eq.~\eqref{eq:backlog}, the above equation is not closed in the linear term of $r_i$ and includes non-linear ones. This type of equation is difficult to solve analytically and tends to exhibit unstable behaviors.

\begin{figure}
    \centering
    \includegraphics[width=0.95\linewidth]{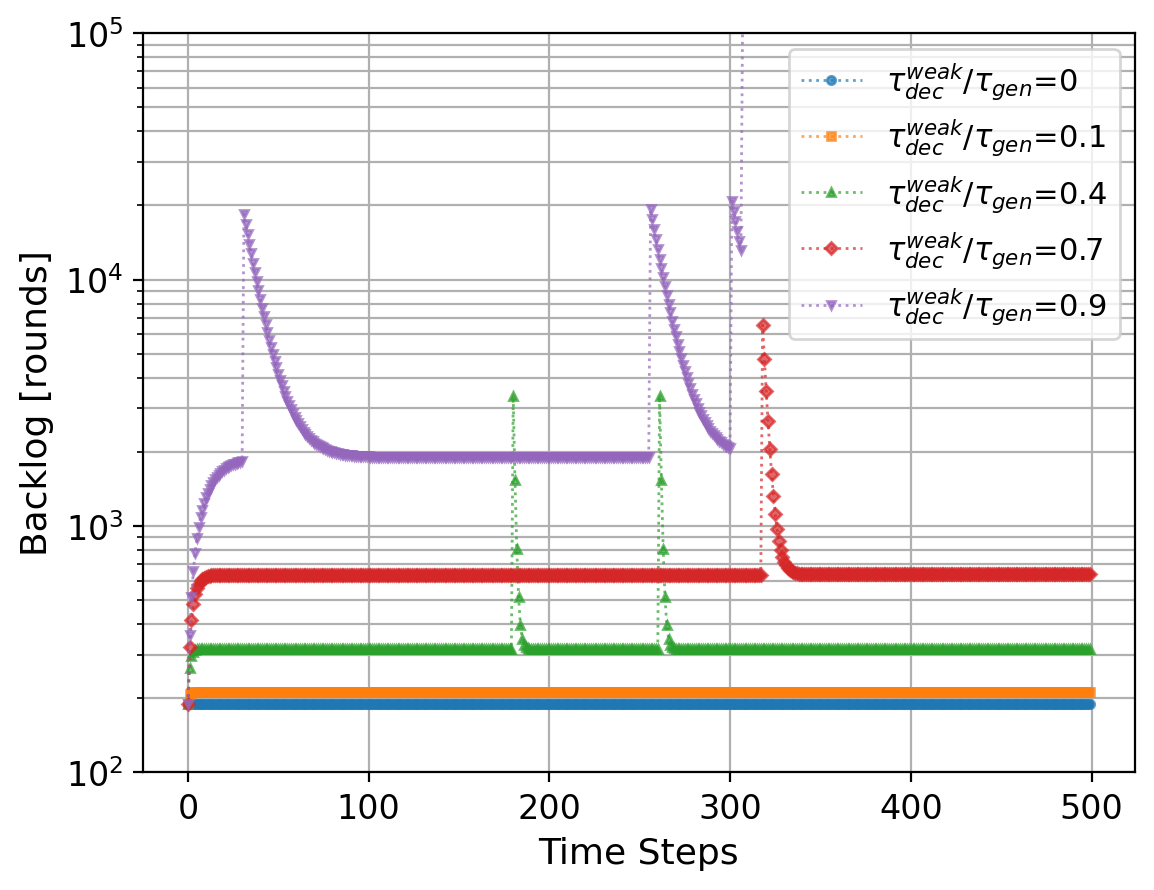}
    \caption{Sampled trajectories of backlog $r_i$ in the naive online decoding scheme with a decoder switching system. 
    Here we vary the decoding time of weak decoder as $\tau_{\text{dec}}^{\text{weak}}/\tau_{\text{gen}}=0, 0.1, 0.4, 0.7, 0.9$ and fix the other parameters with $d=21$, $r_{op,i}=9d$, $\gamma_{\text{switch}}=10^{-4}$, $T_{\text{comm}}^{\text{weak}}=\tau_{\text{gen}}$ and $T_{\text{comm}}^{\text{strong}}=\tau_{\text{dec}}^{\text{strong}}=10\tau_{\text{gen}}$. The horizontal axis denotes the integer index $i$ in $r_i$ and corresponds to the number of executed non-Clifford gates.}
    \label{fig:backlog_w_naive_decoding}
\end{figure}

To gain a deeper understanding of the backlog dynamics under decoder switching, we conduct Monte-Carlo simulations, where the switching event at each time step $i$ is treated as a random variable. In these simulations, we configure the decoders' parameters as $T_{\text{comm}}^{\text{weak}}=\tau_{\text{gen}}$ and $T_{\text{comm}}^{\text{strong}}=\tau_{\text{dec}}^{\text{strong}}=10\tau_{\text{gen}}$.
In Fig.~\ref{fig:backlog_w_naive_decoding}, we showcase illustrative examples of backlog trajectories $\{r_i\}_{i=0,1,\cdots}$ for several values of $\tau_{\text{dec}}^{\text{weak}}$, assuming $\gamma_{\text{switch}}=10^{-4}$.
This figure illustrates that an instantaneous surge in backlog occurs upon switching to the strong decoder, and the backlog is subsequently reduced by rapid decoding on the weak decoder, allowing the backlog to converge towards a stable value over time. 
However, when $\tau_{\text{dec}}^{\text{weak}}$ is not sufficiently small, a new switching event is occasionally triggered before the backlog converges to the stable values, leading to an endless accumulation of backlogs (see the case of $\tau_{\text{dec}}^{\text{weak}}/\tau_{\text{gen}}=0.9$ in Fig.~\ref{fig:backlog_w_naive_decoding}).
More precisely, for the naive online decoding scheme, $M$-consecutive switching events cause an exponential growth of the backlog and the switching rate, which increase proportionally to the factor of $(\tau_{\text{dec}}^{\text{strong}}/\tau_{\text{gen}})^{M}$.
%Then, once the backlog exceeds a certain threshold at which the switching rate exceeds $1/2$, 

\begin{figure*}
    \centering
  \begin{tabular}{lll}
{\normalsize (a)} & \hspace{0.3cm} & {\normalsize (b)} \\
   \includegraphics[width=8.2cm]{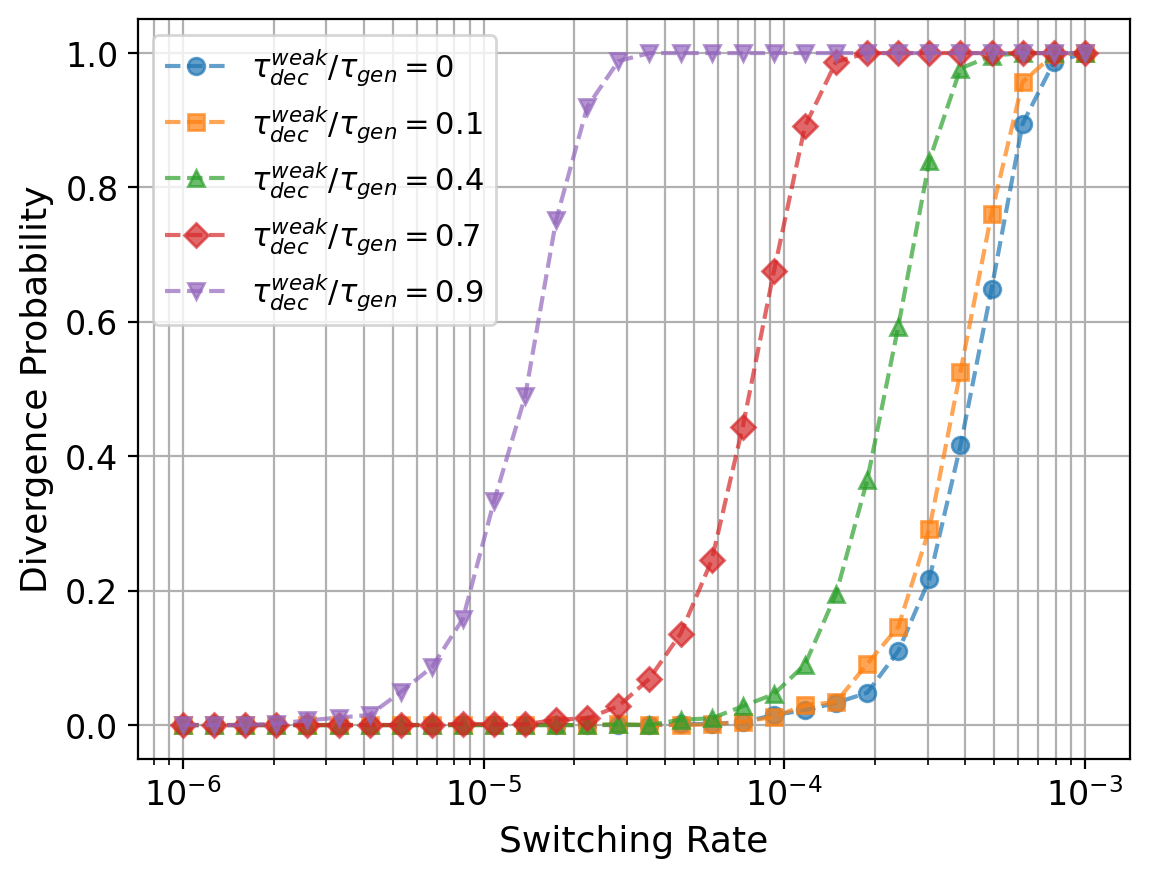}& &
   \includegraphics[width=8.6cm]{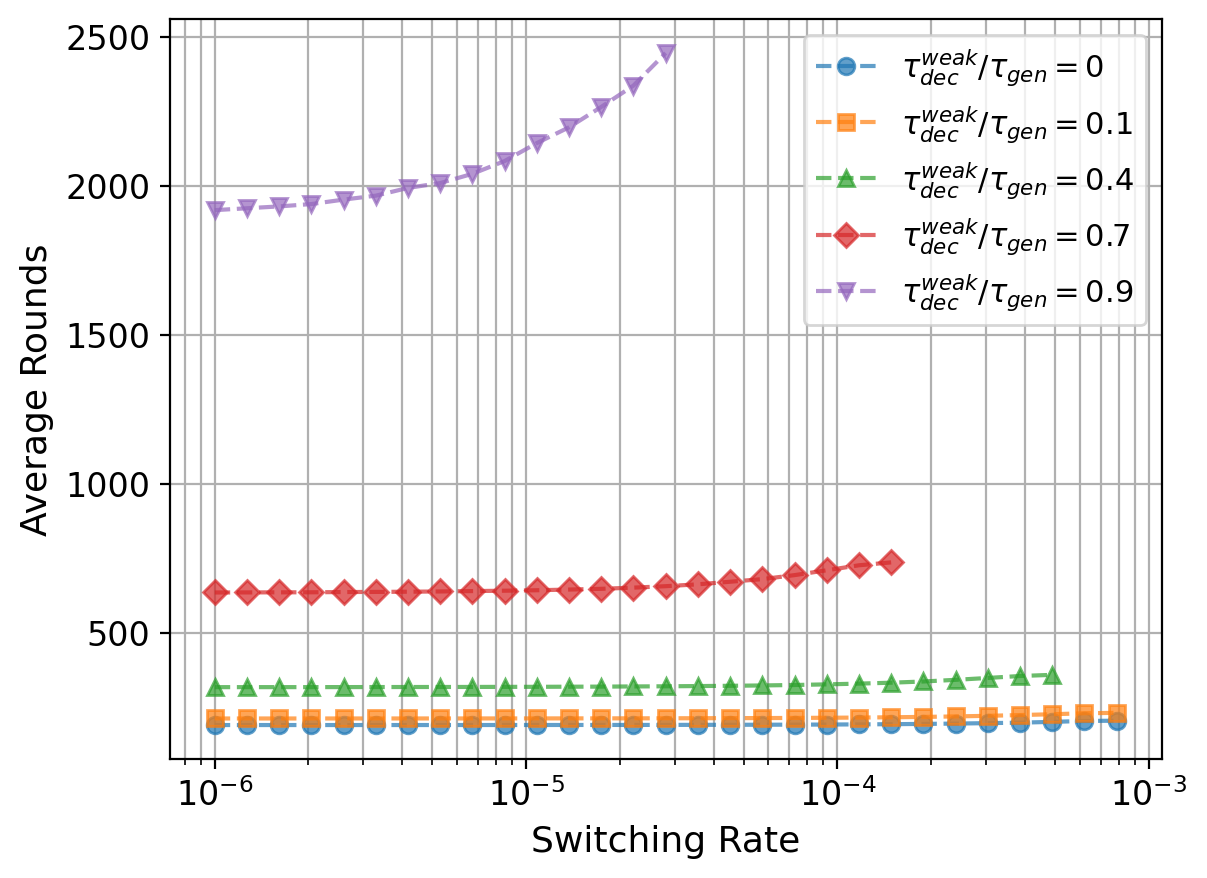}\\
  \end{tabular}
    \caption{(a) The divergence probability and (b) the average size of backlogs in the naive
online decoding scheme with a decoder switching system. We sample $10^3$ backlog trajectories, and each sample is judged to be diverging when the backlog $r_i$ exceeds $10^6$ rounds even once during $N_{\text{gate}}=10^4$ time steps. The average size of backlogs is evaluated by averaging over all time steps and backlog trajectories that have not diverged. We vary the decoding time of weak decoder as $\tau_{\text{dec}}^{\text{weak}}/\tau_{\text{gen}}=0, 0.1, 0.4, 0.7, 0.9$ and  fix the other parameters with $d=21$, $r_{op,i}=9d$, $T_{\text{comm}}^{\text{weak}}=\tau_{\text{gen}}$ and $T_{\text{comm}}^{\text{strong}}=\tau_{\text{dec}}^{\text{strong}}=10\tau_{\text{gen}}$.}
    \label{fig:divergence}
\end{figure*}

To elucidate statistical behaviors of such divergence, we simulate $10^3$ samples of backlog trajectories and evaluate the probability that the backlog diverges while executing $N_{\text{gate}}=10^4$ non-Clifford gates, as shown in Fig.~\ref{fig:divergence} (a). This result suggests that we need to reduce the switching rate to the order of $10^{-4}$ or below to avoid an accidental divergence of backlogs.
To make matters worse, we numerically confirm that the threshold value of the switching rate gradually shifts to smaller values as $N_{\text{gate}}$ increases.
In Fig.~\ref{fig:divergence} (b), we also plot the average size of backlogs over all time steps and
backlog trajectories that have not diverged. These plots suggest that the average value of backlog increases little by little as the switching rate increases, unless a divergent event occurs.

\begin{figure*}
    \centering
    \includegraphics[width=0.8\linewidth]{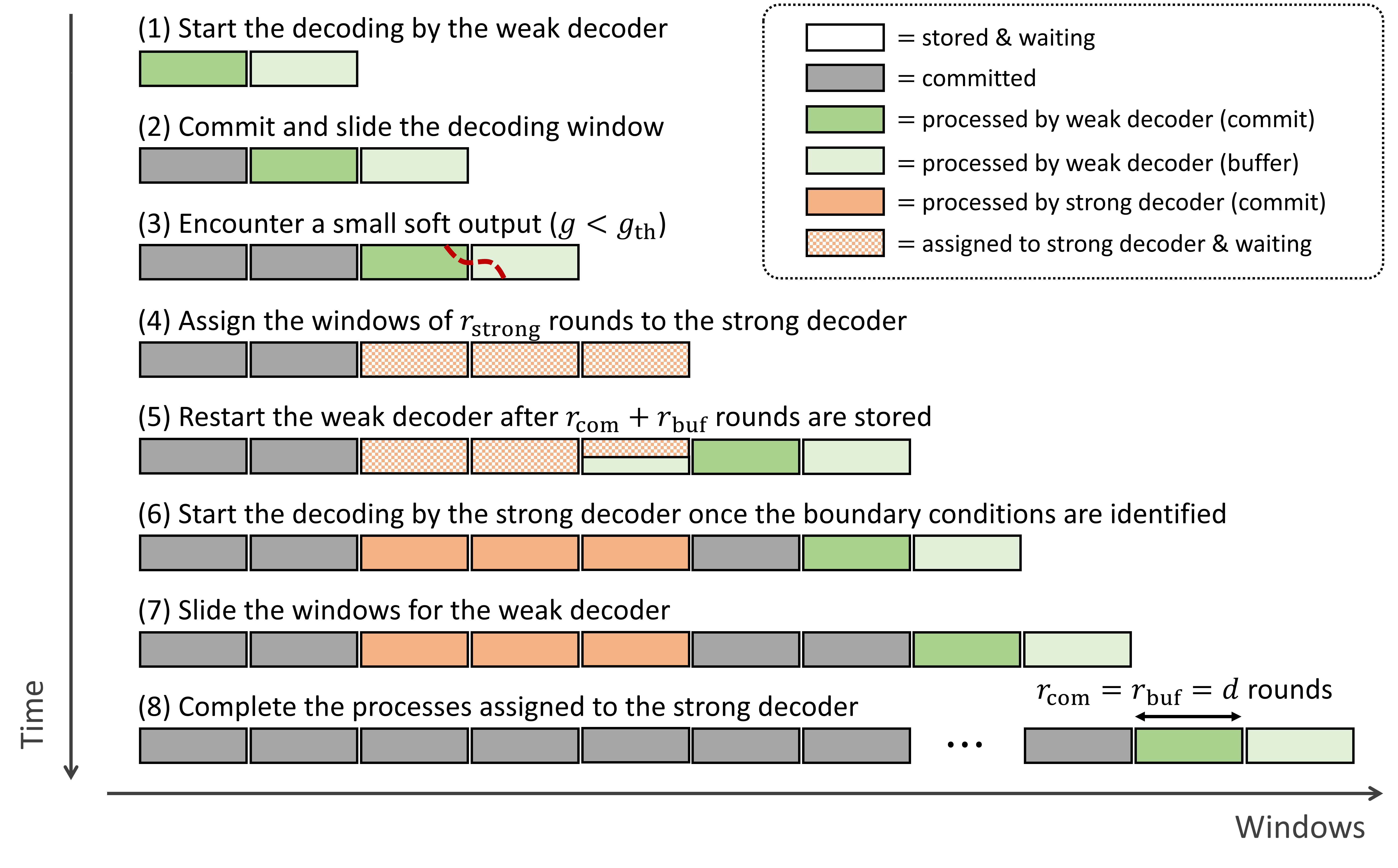}
    \caption{Double window decoding scheme. Each block represents stored syndrome data for $d$-rounds syndrome measurement. In this scheme, the weak decoder (green) and strong decoder (orange) operate in parallel, processing different decoding windows. Specifically, the weak decoder constantly tracks and processes the latest stored data, preventing the accumulation of backlog not assigned to the strong decoder. The weak decoder's decoding window consists of a commit region with $r_{\text{com}}$ rounds and a buffer region with $r_{\text{buf}}$ rounds, and is updated using a sliding window decoding scheme. Here, for simplicity, we assume that both of the commit and buffer sizes are equal to $d$ rounds: $r_{\text{com}}=r_{\text{buf}}=d$. When encountering a small soft output ($g<g_{\text{th}}$), we assign the syndrome data of $r_{\text{strong}}$ rounds, which includes the region with the small soft output, to the strong decoder. The strong decoder processes all the assigned data at once, after the boundary conditions at both ends have been determined by the weak decoder. In this paper, we assume that $r_{\text{strong}}=r_{\text{com}}+2r_{\text{buf}}$.}
    \label{fig:double_window_decoding}
\end{figure*}

\begin{figure}
    \centering
    \includegraphics[width=0.95\linewidth]{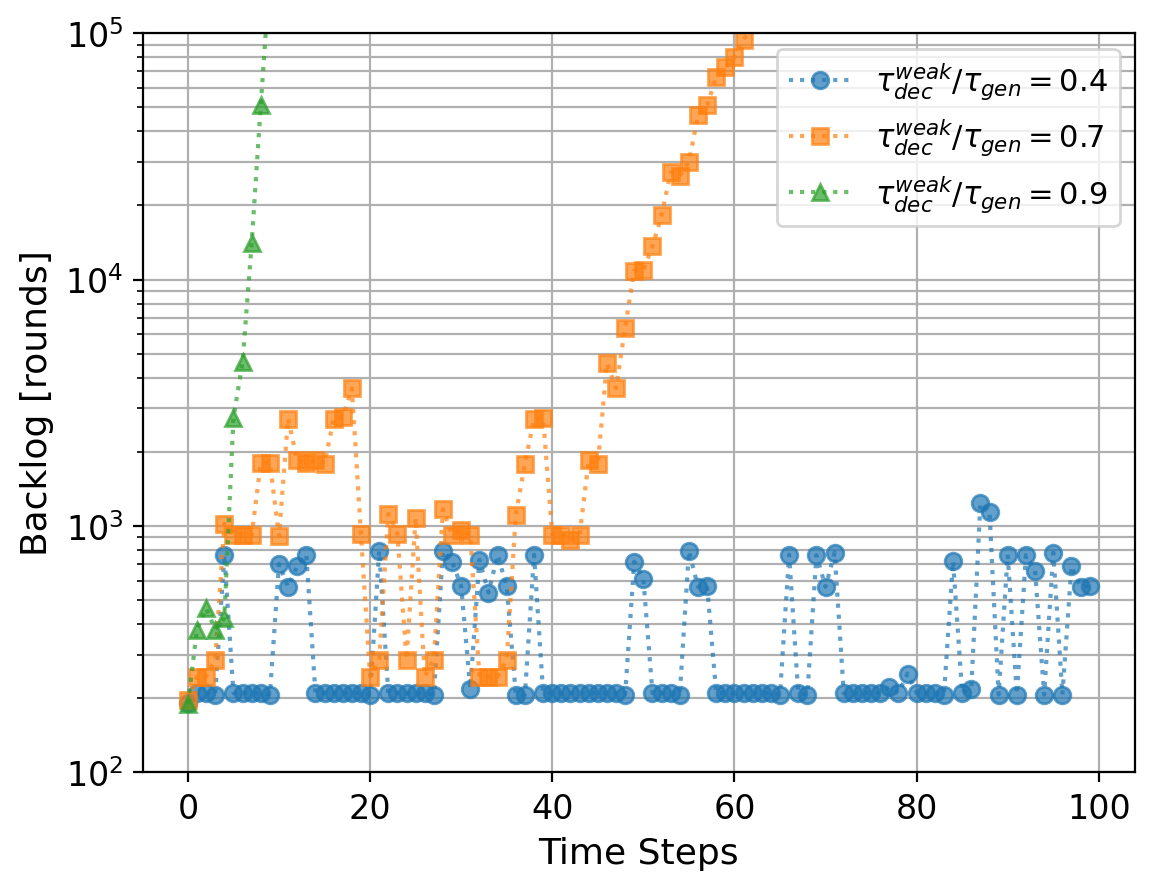}
    \caption{Sampled trajectories of backlog $r_i$ in the double window decoding scheme with a decoder switching system. 
    Here we vary the decoding time of weak decoder as $\tau_{\text{dec}}^{\text{weak}}/\tau_{\text{gen}}=0.4, 0.7, 0.9$ and fix the other parameters with $d=21$, $r_{op,i}=9d$, $\gamma_{\text{switch}}=3\times 10^{-2}$, $T_{\text{comm}}^{\text{weak}}=\tau_{\text{gen}}$ and $T_{\text{comm}}^{\text{strong}}=\tau_{\text{dec}}^{\text{strong}}=10\tau_{\text{gen}}$.}
    \label{fig:backlog_w_double_window}
\end{figure}

\begin{figure*}
    \centering
  \begin{tabular}{lll}
{\normalsize (a)} & \hspace{0.3cm} & {\normalsize (b)} \\
   \includegraphics[width=8.2cm]{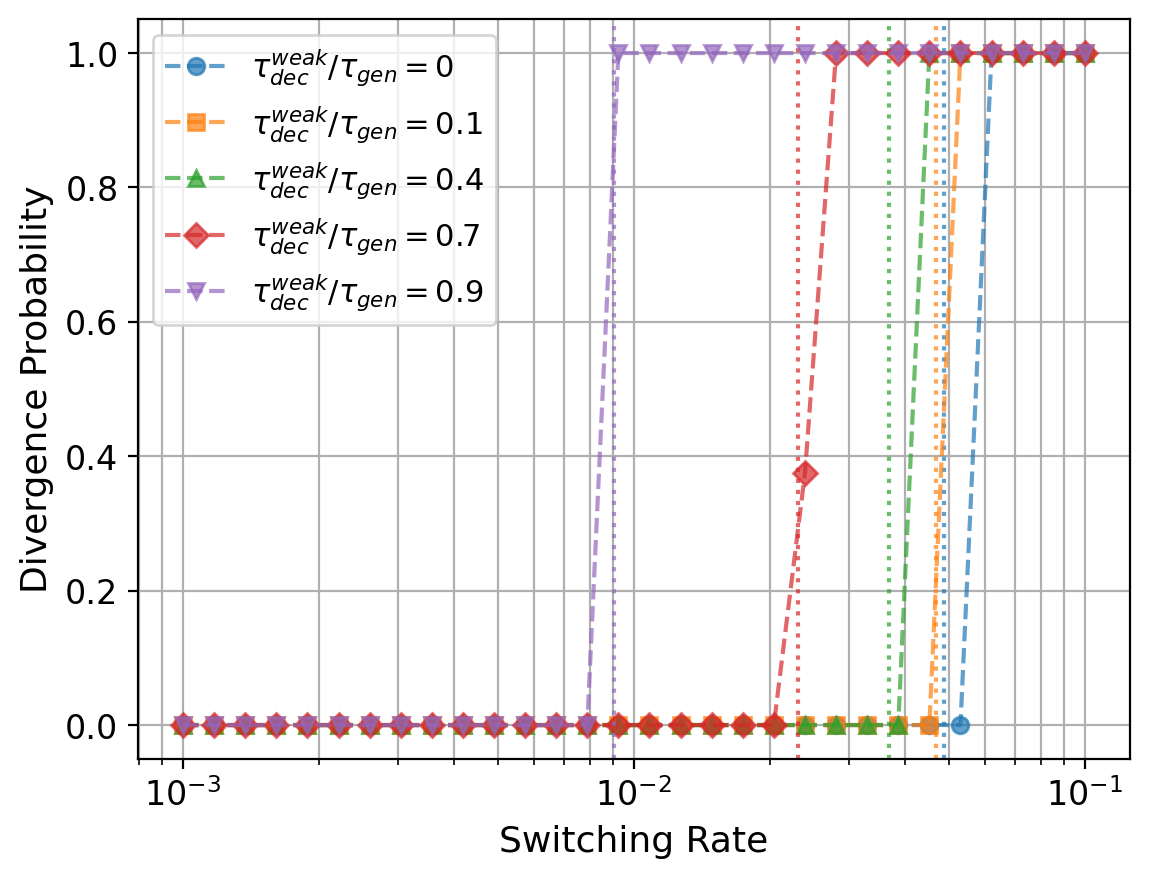}& &
   \includegraphics[width=8.6cm]{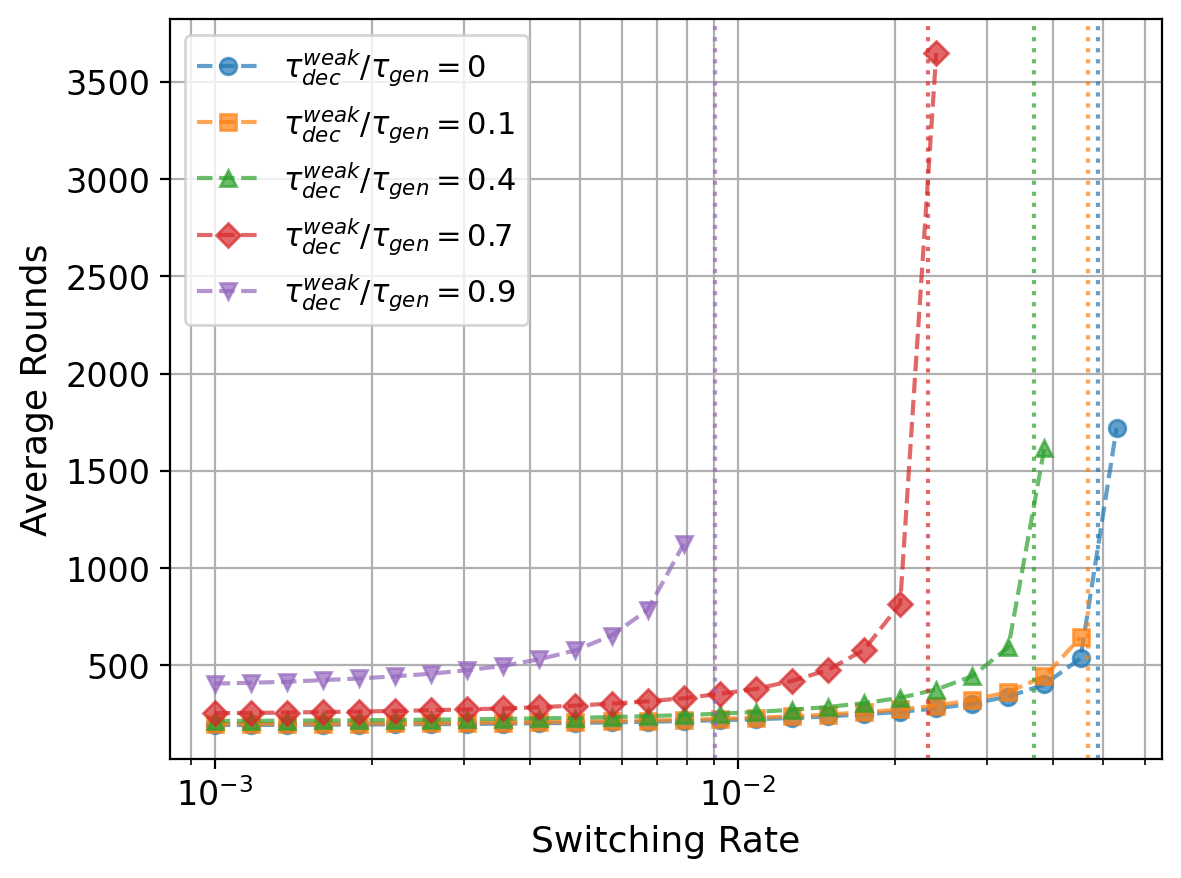}\\
  \end{tabular}
    \caption{(a) The divergence probability and (b) the average size of backlogs in the double window decoding scheme with a decoder switching system. We sample $10^3$ backlog trajectories, and each sample is judged to be diverging when the backlog $r_i$ exceeds $10^6$ rounds even once during $N_{\text{gate}}=10^4$ time steps. The average size of backlogs is evaluated by averaging over all time steps and backlog trajectories that have not diverged. We vary the decoding time of weak decoder as $\tau_{\text{dec}}^{\text{weak}}/\tau_{\text{gen}}=0, 0.1, 0.4, 0.7, 0.9$ and  fix the other parameters with $d=21$, $r_{op,i}=9d$, $T_{\text{comm}}^{\text{weak}}=\tau_{\text{gen}}$ and $T_{\text{comm}}^{\text{strong}}=\tau_{\text{dec}}^{\text{strong}}=10\tau_{\text{gen}}$. For comparison, we also plot the threshold values of the switching rate per $d$-rounds, given in Eq.~\eqref{eq:condition for switching rate}, for each value of $\tau_{\text{dec}}^{\text{weak}}$ with dotted vertical lines. We confirm that the divergence probabilities are always zero for switching rates below the threshold values.}
    \label{fig:divergence_prob_w_double_window.png}
\end{figure*}

As detailed in the analysis above, the backlog behavior sometimes becomes unstable in a hybrid decoding system based on decoder switching. Especially, to avoid a probabilistic exponential growth, the switching rate must be suppressed to at most $10^{-4}$ or below, even for $N_{\text{gate}}=10^4$. Furthermore, when considering that there are many logical qubits, as would be the case in actual quantum computation, the switching rate needs to be further reduced in inverse proportion to the number of logical qubits. Such a stringent requirement for the switching rate will significantly narrow the applicability of decoder switching, making it less practical.

Fortunately, by introducing a sliding window decoding scheme~\cite{Dennis2002}, we can greatly stabilize the backlog trajectory and substantially relax the requirements for the switching rate (for details, see also Appendix.~\ref{appendix:sliding window}).
This is because, when $M$-consecutive switching events happen, the backlog increases only linearly with $M$ in the scheme, unlike in the case of the naive online decoding scheme.
However, it should be noted that two shortcomings arise when the sliding window decoding scheme is applied naively to a decoder switching system.
The first one is that while the strong decoder slowly processes a decoding task after a switching event, the weak decoder is not assigned new tasks, leading to an accumulation of upcoming syndrome data (see also Fig.~\ref{fig: sliding window decoding} in Appendix.~\ref{appendix:sliding window}). The second one is that batch decoding without a buffer region is preferable for the strong decoder to process the syndrome data as fast as possible, assuming the total decoding time is proportional to the number of input rounds.
This is because introducing a finite size of decoding window effectively enlarges the decoding time per round due to the need for a buffer area.
Here, we note that this point requires a more delicate discussion if the total decoding time increases superlinearly with the number of rounds.

To address these two issues, we propose a novel online decoding scheme tailored to decoder switching systems, named the {\it double window decoding} scheme. As illustrated in Fig.~\ref{fig:double_window_decoding}, this scheme operates the weak and strong decoders independently in parallel. The weak decoder consistently tracks and processes the latest syndrome data, preventing the accumulation of backlog that is not assigned to the strong decoder. The decoding window for the weak decoder consists of a commit region and a buffer region, which includes $r_{\text{com}}$ and $r_{\text{buf}}$ rounds, respectively. Then, the position of the decoding window is shifted after the decoding process completes, as in the case of the sliding window decoding scheme.  
If a switching event occurs within a specific decoding window, the weak decoder resumes its process after allocating a data region of $r_{\text{strong}}$ rounds to the strong decoder, and after $r_{\text{com}}+r_{\text{buf}}$ rounds are subsequently stored. 
Here, the value of $r_{\text{strong}}$ should be determined to guarantee that the decoding region with a small soft output does not adversely affect subsequent decoding processing by the weak decoder. In this work, we set this value as $r_{\text{strong}}=r_{\text{com}}+2r_{\text{buf}}$ to make the above requirement satisfied certainly.

On the other hand, the strong decoder should process assigned syndrome data using the best method to maximize its decoding speed. In particular, when the decoding time is proportional to the number of rounds, it is preferable to execute the decoding process in bulk after the syndrome boundary conditions on both sides are determined by the weak decoder.
The detailed handling of the strong decoder could be finely tuned depending on the scaling of the decoding time and available computational resources. However, in this paper, we will not delve into these details and instead assume a linear decoding time and batch decoding.

Based on the double window decoding scheme, we can readily derive the sufficient conditions of $\tau_{\text{dec}}^{\text{weak}}$ and $\tau_{\text{dec}}^{\text{strong}}$ to avoid the backlog problem. 
They can be summarized as follows:
%Here, for the sake of generality, we assume the case where there are many logical data or ancillary qubits, and thus, the decoding system has to process at most $N_{\rm{window}}$ decoding windows simultaneously at each timeslice, based on the spatially parallel decoding scheme~\cite{Skoric2023, bombin2023modular, Lin2025}. In this setup, the required conditions can be summarized as follows:

\begin{thm}[Sufficient conditions for avoiding the backlog problem in the double window decoding scheme]
\label{thm: sufficient condition}
Assume a hybrid decoding system consisting of a single weak decoder and a single strong decoder, and focus on the decoding problem for a single code patch using the double window decoding scheme.
%Assume a hybrid decoding system that consists of $N_{\rm{window}}$ weak decoders and a single strong decoder. Each decoding window has a commit region of
Then, the following conditions
\begin{equation}
\label{eq:condition for decoding time}
    \tau_{\rm{dec}}^{\rm{weak}} < \tau_{\rm{gen}}, \ \  \tau_{\rm{dec}}^{\rm{strong}}\leq \gamma_{\rm{switch}}^{-1}\left(\frac{d}{r_{\rm{strong}}}\right) \tau_{\rm{gen}}
\end{equation}
are sufficient for avoiding the backlog problem in the decoding system. Here, the commit size of the decoding window is defined to satisfy
\begin{equation}
\label{eq:condition for commit size}
    r_{\rm{com}} \geq  \left\lceil \frac{\tau_{\rm{dec}}^{\rm{weak}}}{\tau_{\rm{gen}}-\tau_{\rm{dec}}^{\rm{weak}}}r_{\rm{buf}} \right\rceil.
\end{equation}
\end{thm}

\vspace{0.2cm}

\noindent
{\it --- Proof.} In the double window decoding scheme, to prevent the backlog problem, the weak decoder is required to have a capability to complete the latest decoding task before receiving the next $r_{\text{com}}$ rounds of syndrome data.
Since the weak decoder requires the decoding time of $\tau_{\rm{dec}}^{\rm{weak}}(r_{\text{com}}+r_{\text{buf}})$ for a single decoding window, the above requirement implies that
\begin{equation}
    \tau_{\rm{dec}}^{\rm{weak}}(r_{\text{com}}+r_{\text{buf}}) \leq \tau_{\rm{gen}} r_{\rm{com}},
\end{equation}
which leads to the inequality in Eq.~\eqref{eq:condition for commit size} and the first one in Eq.~\eqref{eq:condition for decoding time}. 

Once this condition is satisfied, our focus can shift to the temporal evolution of the backlog assigned to the strong decoder. In the double window decoding scheme, the switching event occurs with a probability of $\gamma_{\rm{switch}}r_{\rm{com}}/d$ for each decoding window of the weak decoder. Consequently, the backlog for the strong decoder is incremented by an average of $r_{\text{strong}}\times\gamma_{\rm{switch}}r_{\rm{com}}/d$ rounds every time $r_{\rm{com}}$ rounds of syndrome data are generated.
Meanwhile, the strong decoder processes the syndrome data of $r_{\rm{com}}\tau_{\rm{gen}}/\tau_{\rm{dec}}^{\rm{strong}}$ rounds for the same period.
Therefore, to prevent the backlog growth, we need to impose
\begin{equation}
    r_{\text{strong}}\times\gamma_{\rm{switch}}r_{\rm{com}}/d \ \leq \  r_{\rm{com}}\tau_{\rm{gen}}/\tau_{\rm{dec}}^{\rm{strong}}.
\end{equation}
This inequality equals to the second one in Eq.~\eqref{eq:condition for decoding time}. $\square$

\vspace{0.2cm}

The above results readily lead to the requirement for the switching rate. Setting $r_{\text{strong}}=r_{\text{com}}+\alpha r_{\text{buf}}$ and $r_{\text{buf}}=d$, Eqs.~\eqref{eq:condition for decoding time} and \eqref{eq:condition for commit size} suggest that the switching rate is required to satisfy 
\begin{equation}
\label{eq:condition for switching rate}
\begin{aligned}
    \gamma_{\rm{switch}}\ &\leq\   \frac{d}{r_{\text{com}}+\alpha r_{\text{buf}}} \times \frac{\tau_{\rm{gen}}}{\tau_{\rm{dec}}^{\rm{strong}}}\\
    &\leq\   \frac{1-f_{\text{weak}}}{\alpha+(1-\alpha)f_{\text{weak}}} \times \frac{1}{f_{\text{strong}}}.
\end{aligned}
\end{equation}
to avoid the backlog problem in the double window decoding scheme. Here we introduce the notations of $f_{\text{weak}}\equiv\tau_{\rm{dec}}^{\rm{weak}}/\tau_{\text{gen}}$ $(<1)$ and $f_{\text{strong}}\equiv\tau_{\rm{dec}}^{\rm{strong}}/\tau_{\text{gen}}$. In the second inequality, we use Eq.~\eqref{eq:condition for commit size} and neglect the ceiling function for simplicity.
For example, assuming the case where $f_{\text{weak}}=0.7$, $f_{\text{strong}}=10$, and $\alpha=2$, this inequality leads to $\gamma_{\rm{switch}}\lesssim 2.3\times 10^{-2}$. 
This result indicates that the double window decoding scheme can relax the requirement for the switching rate by more than three orders of magnitude compared to that in the naive online decoding scheme (see Fig.~\ref{fig:divergence}), and by more than one order of magnitude compared to that in the sliding decoding scheme (see Fig.~\ref{fig:divergence_prob_w_sliding_window} in Appendix.~\ref{appendix:sliding window}).

In the discussion so far, we have implicitly assumed a situation where we decode only a single QEC code.
However, in a realistic setup of FTQC, there are a large number of logical data or ancillary qubits. Then, the decoding system is required to process multiple decoding windows in parallel with an array of decoders, for example, using the spatially parallel decoding scheme~\cite{Skoric2023, Tan2023, bombin2023modular, Lin2025, Zhang2025LATTE, Liyanage2025}. We briefly comment on how to extend the above theorem to such a realistic scenario below. In this scenario, the factor $\gamma_{\rm{switch}}^{-1}$ in Eq.~\eqref{eq:condition for decoding time} will be replace with $(\gamma_{\rm{switch}}N_{\rm{window}})^{-1}$. Here $N_{\rm{window}}$ is the maximum number of spatially parallel windows at each timeslice. This is because the total probability that the switching events happen increases in proportion to the number of code blocks or the volume of decoding graphs. 
Especially, by supposing a decoding system consisting of $N_{\rm{window}}$ weak decoders and a single strong decoder, we can develop an argument nearly identical to the proof of Theorem~\ref{thm: sufficient condition}. Namely, we can focus solely on the accumulation of the backlog assigned to the strong decoder, since the weak decoder continues to track and process the latest syndrome data without causing any delay. 
On the other hand, the factor $d/r_{\text{strong}}$ in Eq.~\eqref{eq:condition for decoding time} will be replaced with another factor that does not depend on the value of $\gamma_{\rm{switch}}$ and $\tau_{\rm{dec}}^{\rm{strong}}$, but on the details of the spatially parallel window decoding scheme, including the size of the commit/buffer region and how to assign suspicious spatial regions to the strong decoder. We leave more detailed analyses to the interested reader.

Finally, we numerically demonstrate the performance of the double window decoding scheme. As is the case in Fig.~\ref{fig:backlog_w_naive_decoding} and Fig.~\ref{fig:divergence}, we set the decoders' parameters as $T_{\text{comm}}^{\text{weak}}=\tau_{\text{gen}}$ and $T_{\text{comm}}^{\text{strong}}=\tau_{\text{dec}}^{\text{strong}}=10\tau_{\text{gen}}$. In addition, the commit size is fixed to satisfy the equality in Eq.~\eqref{eq:condition for commit size}.
Fig.~\ref{fig:backlog_w_double_window} first illustrates typical trajectories of backlog $r_i$ sampled for $\gamma_{\rm{switch}}=3\times 10^{-2}$ and various values of $\tau_{\rm{dec}}^{\rm{weak}}$. In this figure, only the case of $\tau_{\rm{dec}}^{\rm{weak}}/\tau_{\text{gen}}=0.4$ satisfies the requirement in Eq.~\eqref{eq:condition for decoding time}. Acutually, we can confirm that only the corresponding trajectory exhibits stable behavior without diverging.
In Fig.~\ref{fig:divergence_prob_w_double_window.png} (a), we also show the divergence probability in the double window decoding scheme, comparing with the threshold in Eq.~\eqref{eq:condition for switching rate}. 
These plots clearly demonstrate that Eq.~\eqref{eq:condition for switching rate} well describes the sufficient (and almost necessary) conditions for avoiding the backlog problem in the double window decoding scheme. Furthermore, Fig.~\ref{fig:divergence_prob_w_double_window.png} (b) depicts the 
mean size of backlogs, which is averaged over all time steps and backlog trajectories that have not diverged. From this figure, we can see that reducing the switching rate to less than half of the value of Eq.~\eqref{eq:condition for switching rate} is preferable to prevent the effective logical operation speed from decreasing markedly.

%In the following, we will roughly discuss how to extend the theorem to such situations.
%Assume that there are at most $N_{\rm{window}}$ decoding windows simultaneously at each timeslice, and that the hybrid decoding system consists of $N_{\rm{window}}$ weak decoders and a single strong decoder.

%Here, for the sake of generality, we assume the case where there are many logical data or ancillary qubits, and thus, the decoding system has to process at most $N_{\rm{window}}$ decoding windows simultaneously at each timeslice, based on the spatially parallel decoding scheme~\cite{Skoric2023, bombin2023modular, Lin2025}. 

% As previously mentioned, the switching rate $\gamma_{\text{switch}}(g_{\text{th}})$ exponetially declines when increasing the code distance $d$ with $g_{\text{th}}$ fixed. Therefore, Eq.~\eqref{eq:average decoding time} implies that, when the code distance becomes sufficiently large, we can approximate the ratio $f_{\text{ave}}\simeq f_{\text{weak}} = \tau_{\text{dec}}^{\text{weak}}/\tau_{\text{gen}}$.
% In this limit, the decoding time of the strong decoder no longer affects the backlog problem and the clock speed of logical operations in any way.
% Remarkably, this means that the strong decoders can exclusively pursue higher accuracy within the framework of decoder switching.

\subsection{Comparison with previous works}
\label{sec:related works}

Finally, we should note that the concept of combining multiple decoders to address the speed-accuracy trade-off has been previously investigated (see also Ref.~\cite{Battistel2023review} for review).
This subsection reviews such related efforts and highlights the key distinctions of our proposal.

The most standard approach for hybrid decoding systems is to combine a computationally-simple {\it local} pre-decoder and more complex {\it global} decoder~\cite{Delfosse2020, Meinerz2022, Ravi2023, Smith2023, Chamberland2023NN, Gicev2023}. In this approach, the local pre-decoder either performs some local pre-processing to facilitate the problem on the global decoder, or solves any locally-solvable problems, otherwise passing the task to the global decoder. 
For example, Ref.~\cite{Delfosse2020} proposes a ``lazy decoder" as a local pre-decoder to handle simple error configurations before resorting to a more sophisticated global decoder.  The lazy decoder attempts to correct easy error configurations locally and, if it fails, the syndrome data is sent to a global decoder. 
Such a localized approach tends to face a rapid growth in switching rates as the code distance $d$ increases, since the number of local subproblems also increases cubically with $d$. This makes it difficult to satisfy the requirements given in Theorem.~\ref{thm: sufficient condition}, or in Eq.~\eqref{eq:condition for switching rate}, for practical-sized decoding problems. Actually, Ref.~\cite{Delfosse2020} suggests that, for $d\geq15$ and $p_{\text{ph}}=10^{-3}$, the lazy decoder almost constantly fails, and it cannot achieve any reduction of the bandwidth.

In contrast, our framework leverages soft output information from the weak decoder to determine whether the accurate decoder is needed, enabling a more scalable switching strategy. In particular, as already mentioned, the switching rate declines exponentially with the increase in the code distance, resulting in a dramatic reduction in the average decoding time.
Furthermore, our approach also offers greater flexibility in the choice of decoders.  While the lazy decoder is specifically designed for surface codes and relies on a hard-decision approach, our framework can accommodate various error-correcting codes by properly defining soft information, such as complementary or cluster gap.

Furthermore, the belief-matching decoder~\cite{Higgott2023belief_matching} and its variant~\cite{Caune2023} are also based on a somewhat similar approach to Ref.~\cite{Delfosse2020}. However, these decoders utilize the belief propagation decoder as the first pre-decoder, which operates globally rather than locally and provides the prior error probabilities to the second decoder. In these approaches, switching to a more powerful decoder depends on whether the BP converges sufficiently and its output satisfies the stabilizer constraints.
Unfortunately, the BP decoder does not have a threshold when used on its own, and therefore, the switching rate will increase as the code distance grows.
This point is a crucial difference from our proposal. 
However, Ref.~\cite{Muller2025relayBP} has recently proposed a new BP-based decoder called {\it Relay-BP}, which achieves high accuracy comparable to the MWPM decoder without compromising the speed of the BP decoder. 
By combining this decoder with ideas in Refs.~\cite{Higgott2023belief_matching, Caune2023}, it might be possible to design a good soft output for the BP-based decoder, and thereby, construct a novel instance of decoder switching. This could be an interesting future work and might become a powerful real-time decoding approach for quantum low-density parity check (qLDPC) codes~\cite{Breuckmann2021}, such as bivariate-bicycle codes~\cite{Bravyi2024}.

As another example, the authors of Ref.~\cite{Shutty2024} recently introduced an ensembling-based decoding method called {\it harmonization}, which combines multiple MWPM decoders with perturbed priors to achieve near-optimal accuracy. By querying an ensemble of decoders and pooling their results, they achieved improved logical error rates compared to individual MWPM decoders, approaching the performance of maximum-likelihood decoding. In addition, they also propose a {\it layered decoding scheme} to reduce the computational overhead for querying large ensembles by using the
degree of consensus among the ensemble as a kind of soft information. 
In the scheme, a small ensemble first processes the syndrome data and, only when its confidence is low (indicated by disagreement within the ensemble), the data is passed to a larger ensemble.
Furthermore, Ref.~\cite{Jones2024} developed a decoder named {\it Libra}, which takes the solutions of an ensemble of approximate
solvers for the minimum-weight hypergraph matching problem, and produces a new solution that combines the best local solutions. In this scheme, the author proposed to use the complementary gap~\cite{Bombin2024, Gidney2025yoked} to invoke the ensemble only on a small fraction of “hard” cases to decrease the average decoding time. 

While these two methods share with our proposal the core idea of employing a more accurate decoder adaptively based on the confidence of the initial result, our framework is formulated to encompass a broader array of hybrid decoding systems. Crucially, it remains agnostic to the specific soft information and decoding mechanisms, thereby integrating the proposals in Refs.~\cite{Shutty2024, Jones2024} as special examples.
Furthermore, we have formulated comprehensive and practically-oriented theories addressing some crucial aspects of the time overhead and the backlog problem. These include a practical method for setting (near-)optimal gap thresholds, detailed analyses of backlog dynamics under decoder switching, the proposal of the double window decoding scheme, and clarifying the requirements for the switching rate. 
%These analyses are indispensable for the practical implementation of hybrid decoding systems based on decoder switching.
In the next section, we also present several important observations, notably that decoder switching can even surpass the accuracy of the strong decoder, and that our proposal performs robustly even in practical-sized quantum circuits. We believe that these contributions lay a crucial foundation for the development of scalable real-time decoding systems based on paired complementary decoders in the near future.

\begin{figure*}
    \centering
  \begin{tabular}{lll}
{\normalsize (a)} & \hspace{0.3cm} & {\normalsize (b)} \\
   \includegraphics[width=8cm]{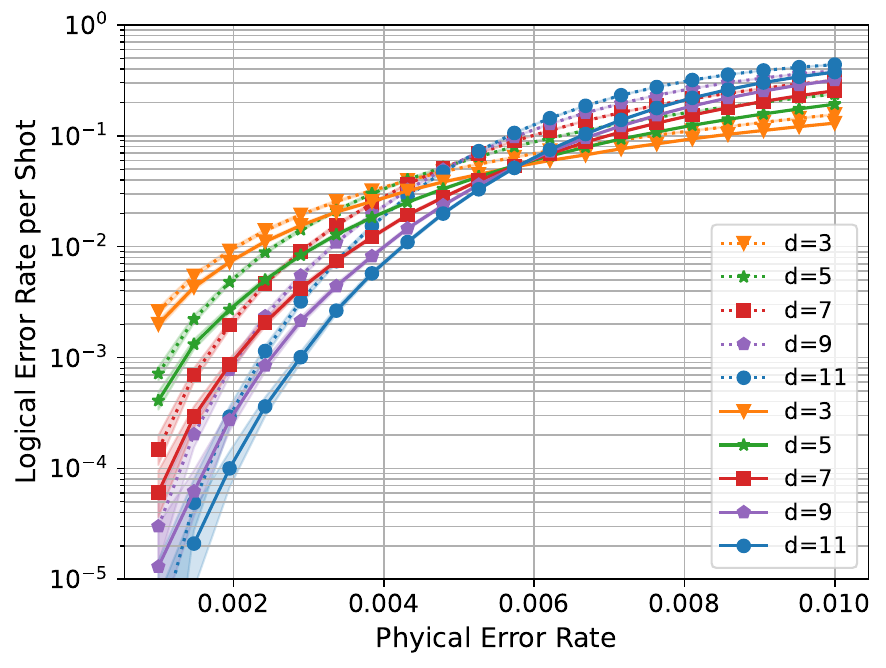}& &
   \includegraphics[width=8cm]{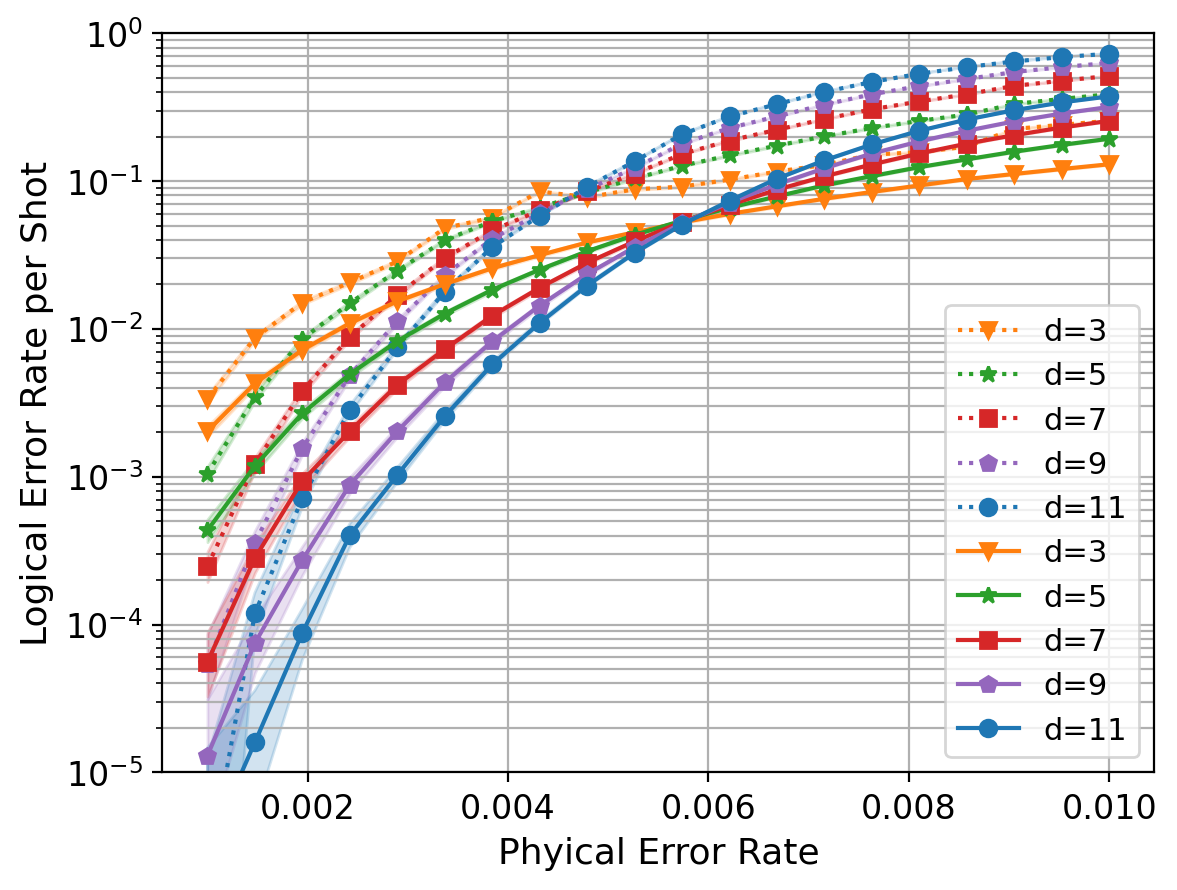}\\
  \end{tabular}
    \caption{Logical error rate around error thresholds for the decoder switching ({\bf a}) between MWPM and belief-matching decoder or ({\bf b}) between UF and belief-matching decoder. Here we fix the threshold value of complementary or cluster gap with $g_{\text{th}}=20$ [dB] and execute $N=10^6$ shots. The solid lines denote the results for the decoder switching. For comparison, we also show the results for the weak decoders with dotted lines. The value of the error threshold is improved by about 0.1\% when decoder switching is employed, compared to that without switching.}
    \label{fig:250822_error_rate_w_complementary_and_threshold=20.0_N=1000000}
\end{figure*}

\begin{figure*}
    \centering
  \begin{tabular}{lll}
{\normalsize (a)} & \hspace{0.3cm} & {\normalsize (b)} \\
   \includegraphics[width=8cm]{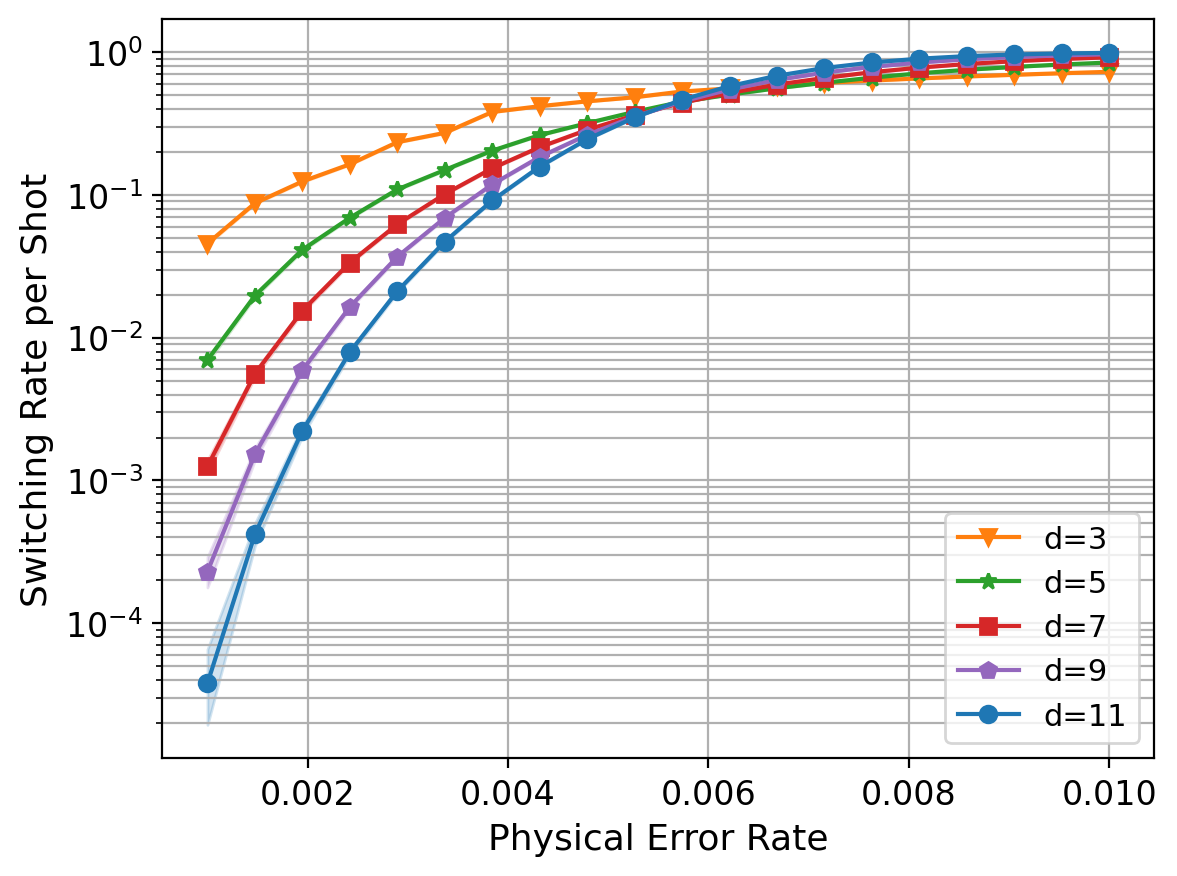}& &
   \includegraphics[width=8cm]{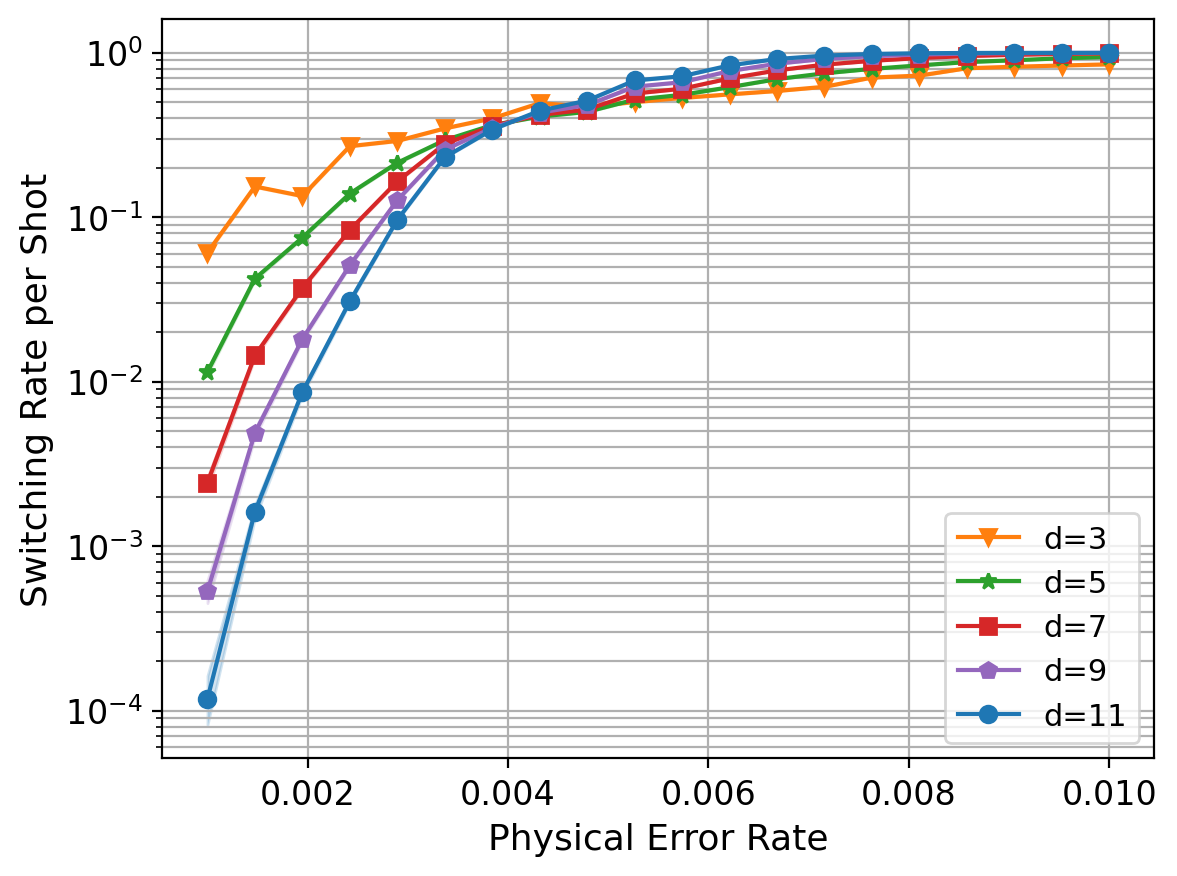}\\
  \end{tabular}
    \caption{Switching rate per shot around error thresholds for the decoder switching ({\bf a}) between MWPM and belief-matching decoder or ({\bf b}) between UF and belief-matching decoder. Here we set parameters to the same values as in Fig.~\ref{fig:250822_error_rate_w_complementary_and_threshold=20.0_N=1000000}.}
    \label{fig:250822_switching_rate_w_complementary_and_threshold=20.0_N=1000000.pdf}
\end{figure*}

\section{Numerical simulation}
\label{sec:numerical simulation}

In this section, we numerically demonstrate the effectiveness of the decoder switching scheme by considering two types of hybrid decoding systems. Specifically, we employ the MWPM decoder or the UF decoder as a weak decoder, and utilize the complementary gap or cluster gap as their soft output, respectively.
For the strong decoder, we adopt the belief-matching decoder~\cite{Higgott2023belief_matching}, a highly accurate decoder that leverages all available noise information by integrating the BP technique with the MWPM decoder. Unlike a usual MWPM decoder, which neglects $Y$ errors by decomposing the related hyperedges, the belief-matching decoder fully exploits correlations between $X$ and $Z$ decoding problems by estimating the posterior marginal probabilities by using BP. This approach results in a significant improvement of accuracy, achieving more than 10\% larger error threshold compared to that of the MWPM decoder for the surface code.

In what follows, we apply these hybrid decoding systems to the $d$-rounds memory experiments of the rotated surface code with perfect terminal time boundaries, assuming the physical error rate to be $p_{\text{ph}}=10^{-3}$. Here we note that, unlike the analysis in Fig.~\ref{fig:histogram}, we reduce the number of rounds from $10d$ to $d$ to reduce the simulation costs. Although this simplification causes finite-size effects on the distribution of complementary gap, we believe it does not alter the quantitative behavior of decoder switching significantly.

\subsection{Logical error rate around the error threshold}

We first begin with the numerical simulation of decoder switching around the error threshold. For simplicity, we fix the gap threshold to be $g_{\text{th}}=20$ dB. 
In Fig.~\ref{fig:250822_error_rate_w_complementary_and_threshold=20.0_N=1000000}~(a), we plot the logical error rate around the error thresholds for decoder switching between the MWPM and belief-matching decoder. For comparison, we also plot the results for the weak decoders (MWPM) with dotted lines.
This figure shows that the decoder switching substantially reduces the logical error rate, and the error threshold value is shifted upward by about 0.1\%.
As shown in Appendix.~\ref{appendix:belief-matching}, we also confirm that the error threshold under decoder switching is located at almost the same value as that of the strong decoder (belief-matching). Fig.~\ref{fig:250822_switching_rate_w_complementary_and_threshold=20.0_N=1000000.pdf}~(a) shows the switching rate for the same setup. Interestingly, this data clearly illustrates that the switching rate decays exponentially with respect to the code distance $d$ below a specific value of physical error rate.
This threshold behavior of the switching rate is highly beneficial for satisfying the requirements in Eq.~\eqref{eq:condition for decoding time} and reducing the average decoding time of the hybrid decoding system.
We note that similar behaviors have been reported only in the code-capacity level simulation of the surface code in some previous works~\cite{Smith2024mitigation, English2024}.
Comparing to the results in Fig.~\ref{fig:divergence_prob_w_double_window.png}, we find that the switching rates for $d\geq5$ are sufficiently small to avoid the backlog problem in the double window decoding scheme, even for the case of $\tau_{\text{dec}}^{\text{weak}}/\tau_{\text{gen}}=0.9$ and $\tau_{\text{dec}}^{\text{strong}}/\tau_{\text{gen}}=10$.
Here, it is also notable that the threshold of switching rate is slightly higher than that for the logical error rate of the weak decoder.

Similar results are obtained for the decoder switching between UF and belief-matching decoder with cluster gap $g_{\text{cluster}}$ as a soft output, as shown in Fig.~\ref{fig:250822_error_rate_w_complementary_and_threshold=20.0_N=1000000}~(b) and Fig.~\ref{fig:250822_switching_rate_w_complementary_and_threshold=20.0_N=1000000.pdf} (b).
In this case, the accuracy improvement achieved by decoder switching becomes even more pronounced, since the accuracy of the weak decoder further deteriorates compared to MWPM.
Still, we find that the logical error rate under decoder switching is almost equivalent to that in Fig.~\ref{fig:250822_error_rate_w_complementary_and_threshold=20.0_N=1000000}~(a), despite the performance degradation of the weak decoder.
These observations encourage us to seek an even faster real-time soft-output decoder with lower hardware costs by cutting its accuracy to the absolute minimum.
However, we also find that the threshold of the switching rate is clearly lower than that of the MWPM, in accordance with the decline in the accuracy of the weak decoder.
This implies that the weak decoder must also achieve a certain level of accuracy to guarantee the exponential decay of the switching rate.

\begin{figure*}
    \centering
  \begin{tabular}{lll}
{\normalsize (a)} & \hspace{0.3cm} & {\normalsize (b)} \\
   \includegraphics[width=8cm]{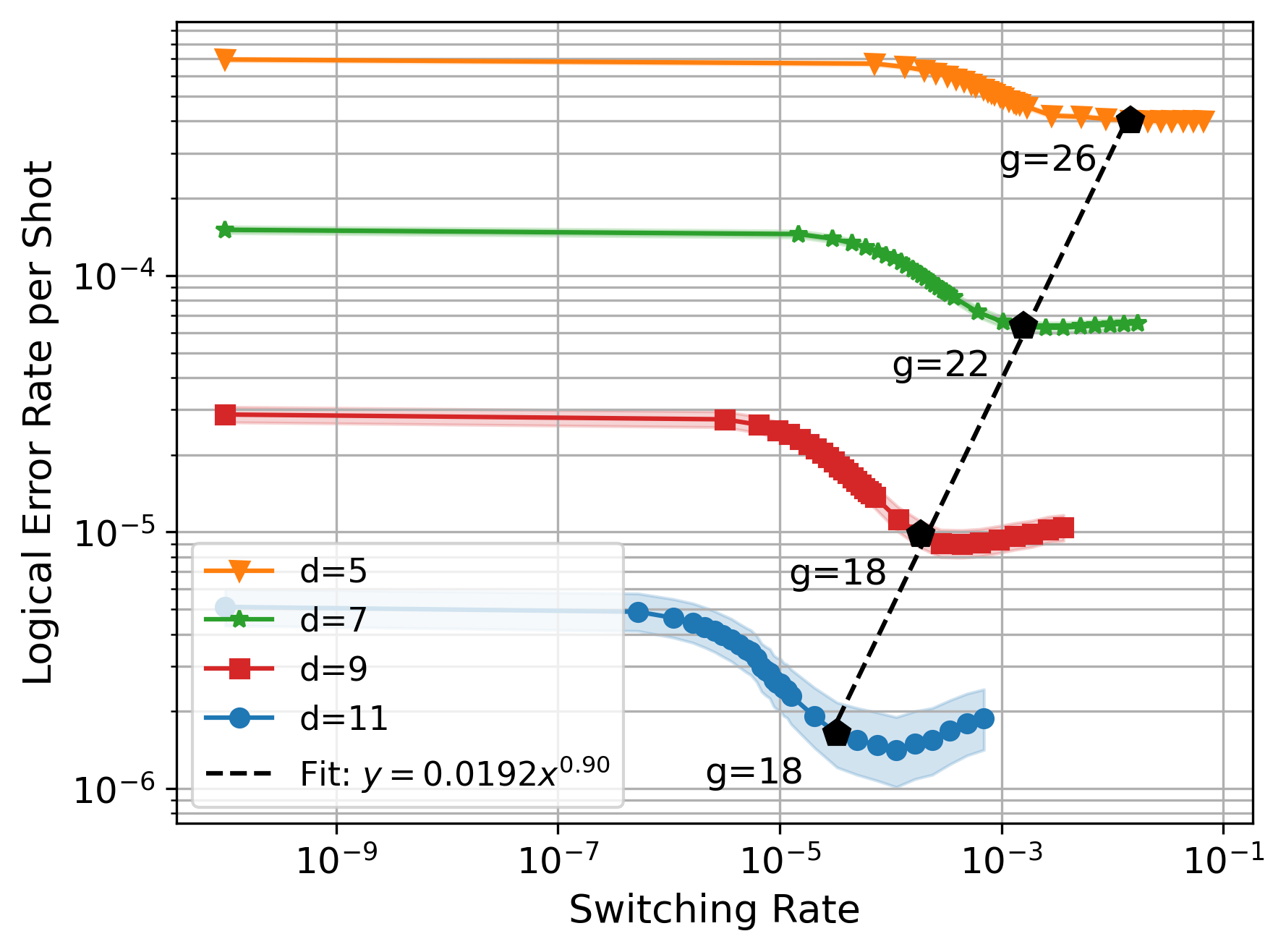}& &
   \includegraphics[width=8cm]{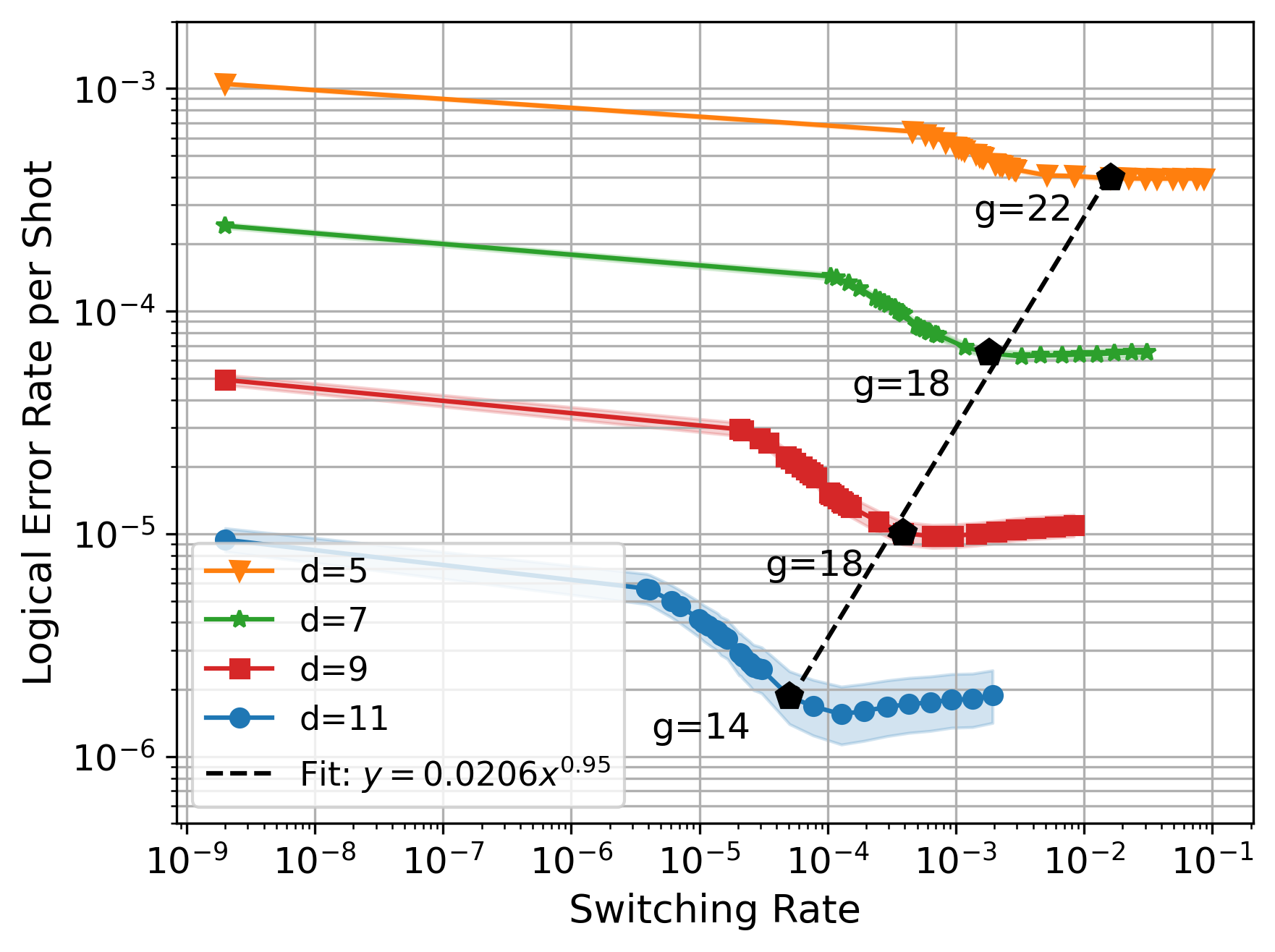}\\
  \end{tabular}
    \caption{Tradeoff between logical error rate and switching rate when we perform the decoder switching ({\bf a}) between MWPM and belief-matching decoder with complementary gaps or ({\bf b}) between UF and belief-matching decoder with cluster gaps. Here we plot the logical error rate per shot and the switching rate by varying the threshold value of complementary gap or cluster gap in the dB range of $g_{\text{th}}\in[0, 50]$. We perform $N=10^8$ shots to obtain each data point, respectively. For comparison, we plot the data point for the case without decoder switching (i.e. $g_{\text{th}}=0$) to have a switching rate of $1/N\times10^{-2}$. The black pentagonal points denote the minimum point where the difference between the error rate and the asymptotic value is within 1\%. The black dashed line denotes the one fitted to these optimal data points.}
    \label{fig:tradeoff}
\end{figure*}

\subsection{Tradeoff between logical error rate and switching rate}

In this subsection, we analyze the tradeoff between the logical error rate and the switching rate under decoder switching. This analysis is crucial for determining the optimal value of $g_{\text{th}}$.
To this end, we calculated the logical error rate and the switching rate per shot by varying the gap threshold of each soft output in the dB range of $g_{\text{th}}\in[0, 50]$ under $p_{\text{ph}}=10^{-3}$.

In Fig.~\ref{fig:tradeoff} (a), we show the results for the decoder switching between the MWPM and the belief-matching decoder.
This plot clearly shows that the logical error rate gradually decreases as the gap threshold increases, eventually approaching that of the strong decoder. However, it is noteworthy that, for $d\geq5$, the tradeoff curve forms a minimum point at a certain gap threshold before approaching the asymptotic value.
This indicates that decoder switching can achieve even lower error rate than the strong decoder by properly adjusting the threshold of soft output, which is consistent with the implication of the results in Fig.~\ref{fig:conditional_error_rate_for_strong}.
At the minimum point, the decoder switching achieves a 3.6x (or 1.3x) lower logical error rate compared to the weak (or strong) decoder on the distance-11 surface code.

Next let us consider the scaling of the switching rate at the optimal value of $g_{\text{th}}$.
Here we define the optimal value of the gap threshold, $g_{\text{op}}$, as the minimum value where the difference between the logical error rate and the asymptotic value is within 1\%. We plot the positions of $g_{\text{op}}$ with black pentagonal points.
Remarkably, these points can be fitted well to a straight line on the double logarithmic graph, which indicates a close relation between logical error rate and switching rate as follows:
\begin{equation}
    P_{L,\text{switch}}(g_{\text{op}}) \ \simeq \ 0.0192 \times \left[\gamma_{\text{switch}}(g_{\text{op}})\right]^{0.90}
\end{equation}
This relation suggests that the switching rate at the optimal threshold $g_{\text{th}}=g_{\text{op}}$ decays in a nearly linear fashion with respect to the target logical error rate.
For example, when applying this formula to a TeraQuop decoder (i.e, $P_{L,\text{switch}}\simeq10^{-12}$)~\cite{Riverlane2023}, which is typically required to solve practical problems including prime factoring~\cite{Gidney2021RSA, Gidney2025RSA} and materials simulation~\cite{Babbush2018qubitization, Kivlichan2020improved, Lee2021, Yoshioka2022hunting, Low2025}, the corresponding value of the switching rate at $g_{\text{th}}=g_{\text{op}}$ is expected to be $\gamma_{\text{switch}}\simeq 3.75\times10^{-12}$ [shot${}^{-1}$]. This means that only a few switches to the strong decoder occur even during the execution of a large-scale circuit with a depth of $10^{9}$-$10^{12}$ gates, making the decoding time of the strong decoder almost negligible on average.
Such a rapid decline in switching rate might even enable discarding the output of the entire circuit and retrying it rather than switching to a strong decoder.
This approach is essentially equivalent to the {\it exclusive decoders} in Ref.~\cite{Smith2024mitigation}, and we can regard the exclusive decoder as the most extreme form of the strong decoder.
Of course, the above analysis is based on extrapolation from small-scale problems up to $d=11$, and so, its quantitative reliability is fairly limited. However, these observations strongly indicate that the contribution of the strong decoder to the average decoding time will be almost negligible in the TeraQuop region.

Similarly, in Fig.~\ref{fig:tradeoff} (b), we show the numerical results for the decoder switching between the UF and the belief-matching decoder with cluster gap as soft outputs. 

As in the case of Fig.~\ref{fig:tradeoff} (a), these plots also form a minimum point at a certain gap threshold before approaching the asymptotic value.
At the minimum point for $d=11$, decoder switching achieves a 6.1x lower logical error rate than that of the weak decoder (UF).
Additionaly, in this case, we can obtain the fitting curve to the optimal values of $g_{\text{th}}$ as follows:
\begin{equation}
    P_{L,\text{switch}}(g_{\text{op}}) \ \simeq \ 0.0206 \times \left[\gamma_{\text{switch}}(g_{\text{op}})\right]^{0.95}
\end{equation}
This result also supports the effectiveness of the decoder switching for large-scale quantum tasks.

\section{Conclusion}
\label{sec:conclusion}

In this work, we have proposed a general framework termed {\it decoder switching}, which efficiently harmonizes the demands for accuracy and speed in decoding systems by combining a faster soft-output decoder (weak decoder) and a slower, high-accuracy decoder (strong decoder). 
As a demonstration, we have analyzed the case where the MWPM (or UF) decoder is used as a weak decoder and the belief-matching decoder is used as a strong decoder. Regarding the soft output, we have leveraged the complementary gap (or the cluster gap) for the MWPM (or UF) decoder.
Our numerical simulations confirm that decoder switching systems can achieve an accuracy comparable to that of the strong decoder, while maintaining an average decoding time on par with the weak decoder.
This framework can be readily extended to a more general type of soft outputs and QEC codes other than surface codes.
Moreover, to efficiently manage the backlog in real-time decoding, we introduced the double window decoding scheme, an online decoding strategy tailored to decoder switching systems. This scheme dramatically suppresses the instability of backlog trajectories, overcoming the limitations of naive online decoding and simple sliding window approaches. Our theoretical analyses also clarified that satisfying specific conditions given in Theorem~\ref{thm: sufficient condition} is sufficient to prevent the exponential growth of backlog and ensure stable logical operations under decoder switching.

Importantly, decoder switching enables us to develop high-accuracy decoders and high-speed soft-output decoders separately. For high-accuracy decoders, we can seek more complex decoding algorithms or neural network structures by freeing ourselves from strong constraints related to hardware implementation and decoding time. Moreover, we might not need to implement such costly devices in the TeraQuop region, where the switching rate is significantly suppressed and a post-selection-based decoder, i.e., the exclusive decoder~\cite{Smith2024mitigation}, might work well.
For high-speed decoders, it will be possible to improve their scalability and latency, for example, by discretizing or compressing the information of decoding graphs, since we no longer need to pursue high accuracy for these decoders.
We believe that these perspectives will open up new directions for developing future real-time decoders.

%Finally, we address the remaining issues that have not been covered in this paper. To apply our proposal to realistic real-time decoding systems, it is necessary to integrate the framework of decoder switching with online decoding schemes. This challenge involves determining how to estimate soft output, like complementary gap, in an online setting and how to optimize the size of the decoding window under decoder switching. These aspects are somewhat complex beyond our analyses and could serve as intriguing future directions for our proposal.

\section{Acknowledgement}

We are grateful to thank Yugo Takada, Yutaka Hirano, Tomohiro Itogawa, Takumi Akiyama, Yutaro Akahoshi, Moeto Mishima, Shinichiro Yamano, Mitsuki Katsuda, and Hoiki Liu for fruitful discussions. K. F. is supported by MEXT Quantum Leap Flagship Program (MEXT Q-LEAP)
Grant No. JPMXS0120319794, JST COI-NEXT Grant No. JPMJPF2014, JST
Moonshot R\&D Grant No. JPMJMS2061, and JST CREST JPMJCR24I3.

{\bf Author contributions}: R. T., K. K., and K. F. conceived the basic idea of decoder switching concurrently. Then, R. T. formulated the theoretical details, performed the numerical simulations for all figures, analyzed the data, and wrote the original draft of this paper. 
K. K. contributed to the theoretical formulation, provided source codes for the calculation of the cluster gap, and optimized the codes to accelerate the simulations.
J. F., H. O., and S. S. provided overall supervision, environments, and resources for this work and guided the research direction.
K. F. provided technical supervision for this work and contributed to the conceptualization and the interpretation of the numerical results.
All authors participated in the discussion of the results and the review of the manuscript.

\appendix

{\renewcommand{\arraystretch}{1.5}
\begin{table*}
    \centering
    \caption{Our notations used in this paper.}
    \begin{tabular}{p{2.5cm}p{14cm}}
    \hline \hline
    Notation\hspace{5mm}  & Meaning\\
    \hline
    $d$ & Code distance of the QEC code.\\
    $p_{\text{ph}}$ & Physical error rate in the uniform circuit-level noise model.\\
    $g$ & Value of the soft output generated from the soft-output decoder (weak decoder).\\
    $g_{\text{comp}}$ & Value of an unsigned complementary gap. \\
    $\tilde{g}_{\text{comp}}$ & Value of a signed complementary gap. \\
    $g_{\text{cluster}}$ & Value of a cluster gap. \\
    $g_{\text{th}}$ & Threshold value of soft output, which determines whether switching to the strong decoder is needed or not.\\
    $p(g)$ & Probability distribution of the soft output $g$ obtained within a window of $d$-rounds.\\
    $P_{L}$ & Logical error rate per shot (i.e. $d$-rounds) of the QEC code when using a specific decoder.\\
    $P(e|g)$ & $g$-conditional logical error rate per shot of the soft-output decoder.\\
    $P_{\text{weak}}(e|g)$ & $g$-conditional logical error rate per shot of the weak decoder. which is equivalent to $P(e|g)$. \\
    $P_{\text{strong}}(e|g)$ & Logical error rate per shot of the strong decoder under the condition that the weak decoder outputs $g$. \\
    $P_{L,\text{switch}}(g_{\text{th}})$ & Logical error rate per shot of the QEC code when using the decoder switching system with the threshold of $g_{\text{th}}$.\\
    $P_{L,\text{th}}(g_{\text{th}})$ &  Thresholded logical error rate per shot of the weak decoder, given the threshold of $g_{\text{th}}$.\\
    $\gamma_{\text{switch}}(g_{\text{th}})$ & Probability that we encounter a soft output smaller than $g_{\text{th}}$ within a window of $d$-rounds.\\
    $\tau_{\text{gen}}$ & Syndrome generation time per round.\\
    $\tau_{\text{dec}}$ & Decoding time per round of the decoder in a single-decoder approach.\\
    $\tau_{\text{dec}}^{\text{weak}}$ & Decoding time per round of the weak decoder.\\
    $\tau_{\text{dec}}^{\text{strong}}$ & Decoding time per round of the strong decoder.\\
    $T_{\text{comm}}$ & Communication latency for transmitting the syndrome data for each round and the decoding output between the system controller and the decoder in a single-decoder approach.\\
    $T_{\text{comm}}^{\text{weak}}$ & Communication latency for transmitting the syndrome data for each round and the decoding output between the system controller and the weak decoder.\\
    $T_{\text{comm}}^{\text{strong}}$ & Communication latency for transmitting the syndrome data for each round and the decoding output between the system controller and the strong decoder.\\
    $r_i$ & Number of rounds accumulated between destructive measurements to determine the feedback operations for the previous and the next non-Clifford gates. \\
    $r_{op,i}$ & Number of rounds required for executing the quantum circuit of the $i$-th non-Clifford gate. \\
    $r_{\text{com}}$ & Number of rounds in the commit region of a decoding window. \\
    $r_{\text{buf}}$ & Number of rounds in the buffer region of a decoding window. \\
    $r_{\text{strong}}$ & Number of rounds assigned to the strong decoder once a switching event arises \\
    \hline \hline
    \end{tabular}
    \label{tab:notations}
\end{table*}
}

\section{Notations in this paper}
\label{Appendix:notations}

In Table.~\ref{tab:notations}, we list the notations frequently used in this paper.

\section{Error threshold of belief-matching decoder}
\label{appendix:belief-matching}

\begin{figure}
    \centering
    \includegraphics[width=0.9\linewidth]{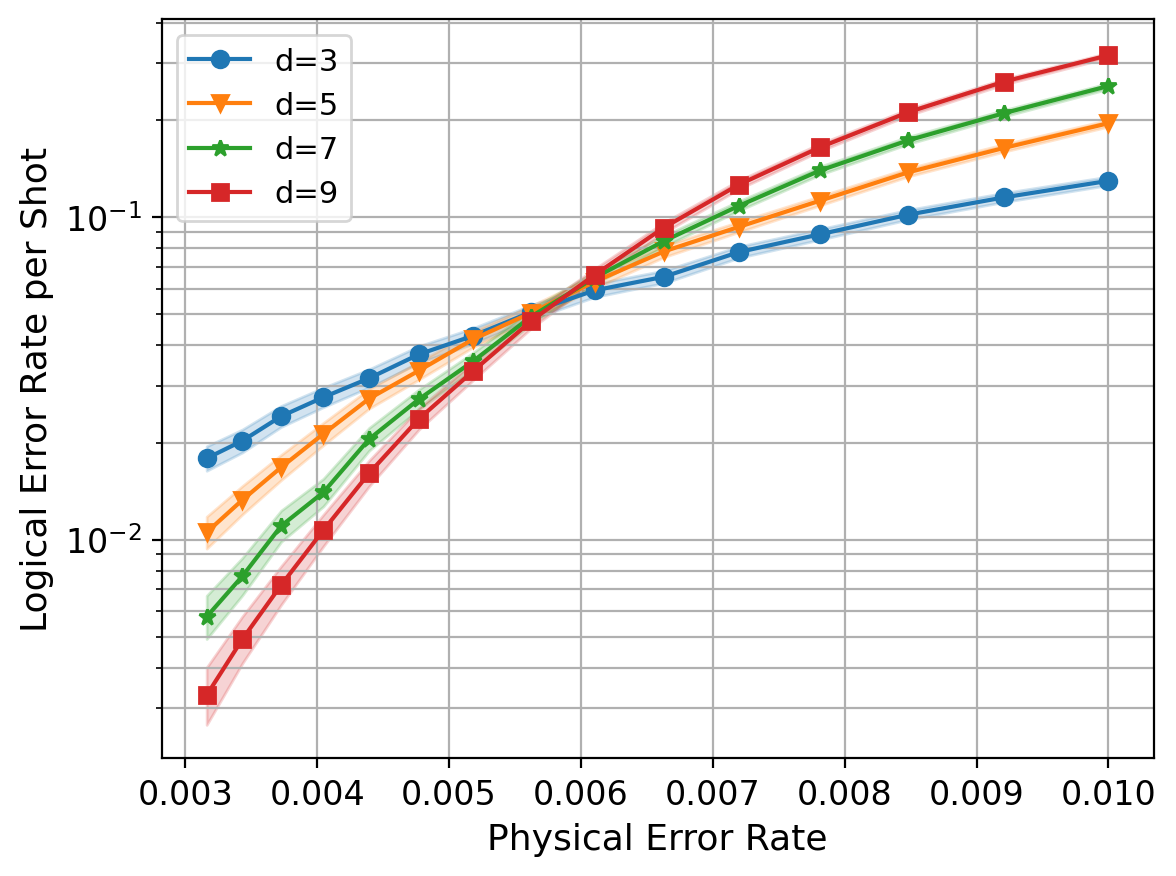}
    \caption{Logical error rate around the error threshold when utilizing the belief-matching decoder~\cite{Higgott2023belief_matching}. Here we set parameters to the same values as in Fig.~\ref{fig:250822_error_rate_w_complementary_and_threshold=20.0_N=1000000}.}
    \label{fig:error_rate_w_BP_MWPM}
\end{figure}

In this section, we show the error threshold of the belief-matching decoder~\cite{Higgott2023belief_matching} for the rotated surface code under our setup of decoding problems. We calculated it with the Python library ``BeliefMatching" and plotted the results in Fig.~\ref{fig:error_rate_w_BP_MWPM}.
We find that the error threshold is located at almost the same value as that of the decoder switching systems in Fig.~\ref{fig:250822_error_rate_w_complementary_and_threshold=20.0_N=1000000}.

\begin{figure*}[h]
    \centering
    \includegraphics[width=0.8\linewidth]{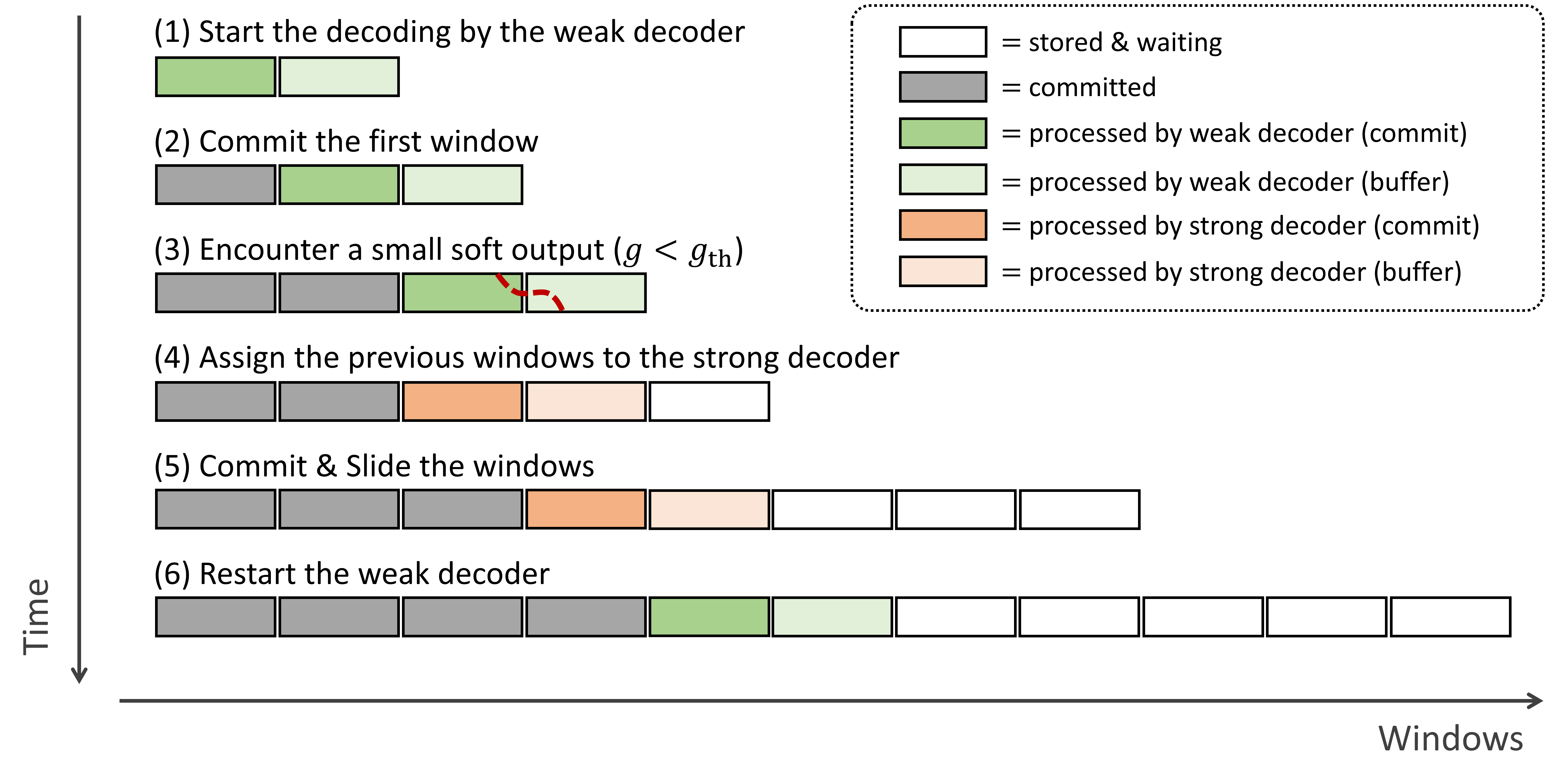}
    \caption{Naive application of the sliding window decoding scheme to a decoder switching system. Here, for simplicity, we assume the case where both the commit and buffer sizes are equal to $d$ rounds: $r_{\text{com}}=r_{\text{buf}}=d$. Each block represents stored syndrome data for $d$-rounds syndrome measurement. In this case, during the decoding process by the strong decoder, the weak decoder has to await the outcome from the strong decoder and is not assigned a new task.}
    \label{fig: sliding window decoding}
\end{figure*}

\begin{figure}[h]
    \centering
    \includegraphics[width=0.95\linewidth]{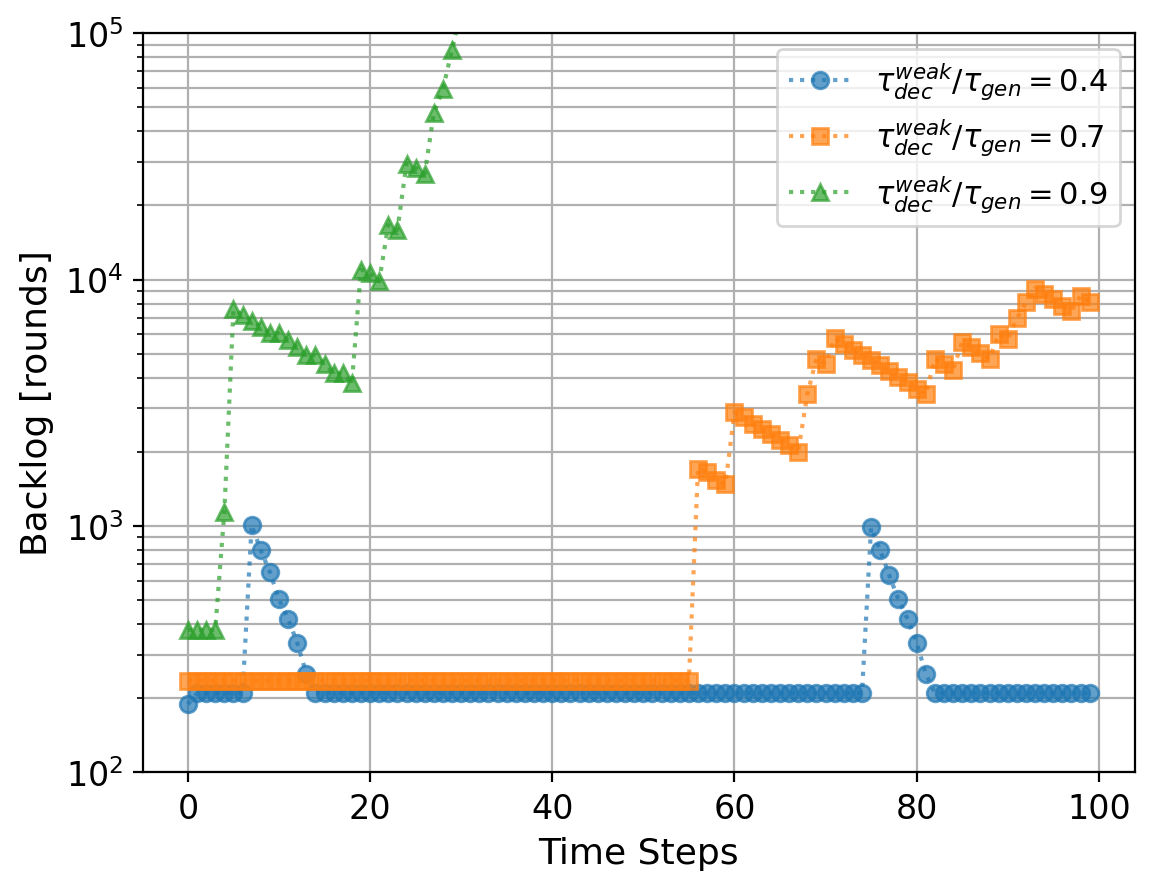}
    \caption{Sampled trajectories of backlog $r_i$ in the sliding decoding scheme with a decoder switching system. 
    Here we vary the decoding time of weak decoder as $\tau_{\text{dec}}^{\text{weak}}/\tau_{\text{gen}}=0.4, 0.7, 0.9$ and fix the other parameters with $d=21$, $r_{op,i}=9d$, $\gamma_{\text{switch}}=10^{-3}$, $T_{\text{comm}}^{\text{weak}}=\tau_{\text{gen}}$ and $T_{\text{comm}}^{\text{strong}}=\tau_{\text{dec}}^{\text{strong}}=10\tau_{\text{gen}}$.}
    \label{fig:backlog_w_sliding_decoding}
\end{figure}

\begin{figure*}[h]
    \centering
  \begin{tabular}{lll}
{\normalsize (a)} & \hspace{0.3cm} & {\normalsize (b)} \\
   \includegraphics[width=8.2cm]{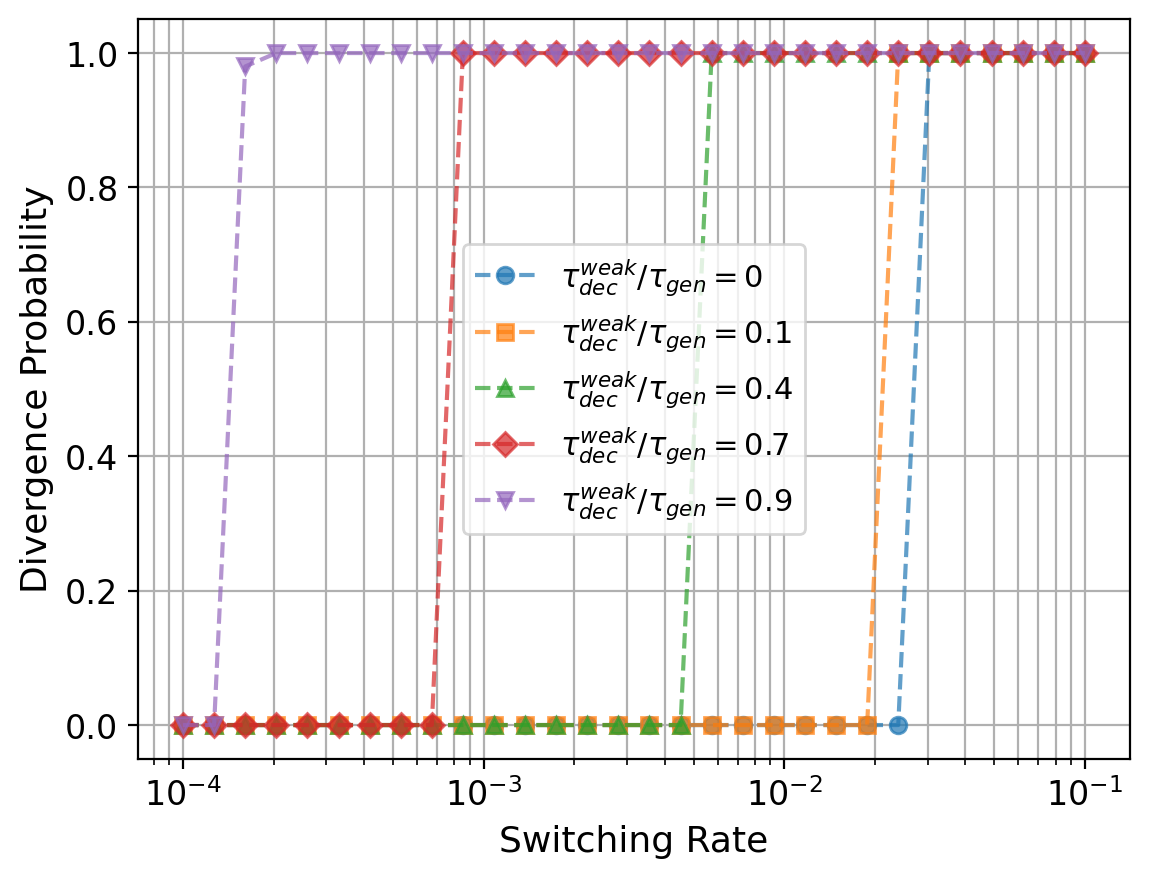}& &
   \includegraphics[width=8.6cm]{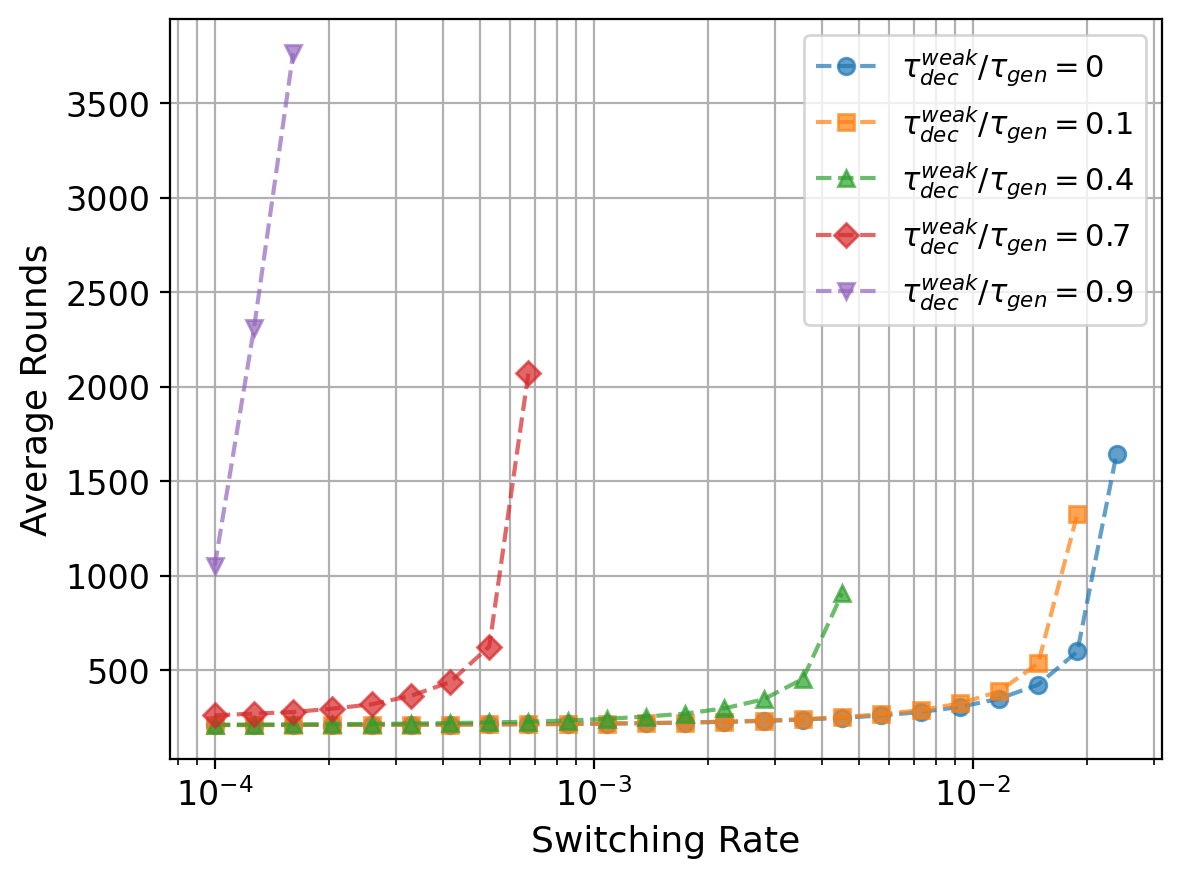}\\
  \end{tabular}
    \caption{(a) The divergence probability and (b) the average size of backlogs in the sliding window decoding scheme with a decoder switching system. We sample $10^3$ backlog trajectories, and each sample is judged to be diverging when the backlog $r_i$ exceeds $10^6$ rounds even once during $N_{\text{gate}}=10^4$ time steps. The average size of backlogs is evaluated by averaging over all time steps and backlog trajectories that have not diverged. We vary the decoding time of weak decoder as $\tau_{\text{dec}}^{\text{weak}}/\tau_{\text{gen}}=0, 0.1, 0.4, 0.7, 0.9$ and  fix the other parameters with $d=21$, $r_{op,i}=9d$, $T_{\text{comm}}^{\text{weak}}=\tau_{\text{gen}}$ and $T_{\text{comm}}^{\text{strong}}=\tau_{\text{dec}}^{\text{strong}}=10\tau_{\text{gen}}$.}
    \label{fig:divergence_prob_w_sliding_window}
\end{figure*}

\section{Backlog dynamics in sliding window decoding scheme}
\label{appendix:sliding window}

This appendix analyzes the backlog dynamics within the conventional sliding window decoding scheme~\cite{Dennis2002} for a decoder
switching system. 

Fig.~\ref{fig: sliding window decoding} illustrates the specific steps of this scheme. The decoding system 
sequentially processes the syndrome data within a decoding window, which consists of the commit and buffer regions. In Fig.~\ref{fig: sliding window decoding}, we fix the size of these regions as $r_{\text{com}}=r_{\text{buf}}=d$. Each window is first processed by the weak decoder, and the reliability of the decoding result is assessed with a soft output $g$. If we encounter a small soft output, the corresponding window is assigned to the strong decoder. During the decoding process by the strong decoder, new syndrome data keeps being added continuously.
Once the strong decoder completes the assigned task, we process the next decoding window with the weak decoder.

Generally, as discussed in Theorem~\ref{thm: sufficient condition}, the buffer size $r_{\text{com}}$ has to satisfy
Eq.~\eqref{eq:condition for commit size} to guarantee that the weak decoder can keep up with the generation of new syndrome data as long as no switching event occurs. However, it is important to note that merely imposing Eq.~\eqref{eq:condition for commit size} does not uniquely determine the buffer size. If it is set too small, the weak decoder may not be able to make up for decoding delays. Conversely, if the buffer size is too large, the size of rounds assigned to the strong decoder becomes large.
In our numerical simulations below, we set the buffer size as 
\begin{equation}
    r_{\text{com}} = \max \left(d, \left\lceil \frac{\tau_{\rm{dec}}^{\rm{weak}}}{0.95\tau_{\rm{gen}}-\tau_{\rm{dec}}^{\rm{weak}}}r_{\rm{buf}} \right\rceil \right).
\end{equation}
Although it might be possible to improve the performance of the sliding window decoding by fine-tuning this value, such tuning is beyond the purpose of our work.

In Fig.~\ref{fig:backlog_w_sliding_decoding}, we plot illustrative examples of backlog trajectories $\{r_i\}_{i=0,1,\cdots}$ for several values of $\tau_{\text{dec}}^{\text{weak}}$, assuming $\gamma_{\text{switch}}=10^{-3}$. Moreover, Fig.~\ref{fig:divergence_prob_w_sliding_window} displays the divergence probability and the average size of backlogs. These plots look similar to those in Fig.~\ref{fig:divergence_prob_w_double_window.png}, rather than Fig.~\ref{fig:divergence}. However, the threshold values of the switching rate are fairly smaller than those in Fig.~\ref{fig:divergence_prob_w_double_window.png}. In particular, this gap becomes more prominent for smaller values of $\tau_{\text{dec}}^{\text{weak}}$, since the accumulation of the weak decoder's backlog leads to a more pronounced detrimental effect.

\newpage

\bibliography{citation}

\begin{thebibliography}{92}%
\makeatletter
\providecommand \@ifxundefined [1]{%
 \@ifx{#1\undefined}
}%
\providecommand \@ifnum [1]{%
 \ifnum #1\expandafter \@firstoftwo
 \else \expandafter \@secondoftwo
 \fi
}%
\providecommand \@ifx [1]{%
 \ifx #1\expandafter \@firstoftwo
 \else \expandafter \@secondoftwo
 \fi
}%
\providecommand \natexlab [1]{#1}%
\providecommand \enquote  [1]{``#1''}%
\providecommand \bibnamefont  [1]{#1}%
\providecommand \bibfnamefont [1]{#1}%
\providecommand \citenamefont [1]{#1}%
\providecommand \href@noop [0]{\@secondoftwo}%
\providecommand \href [0]{\begingroup \@sanitize@url \@href}%
\providecommand \@href[1]{\@@startlink{#1}\@@href}%
\providecommand \@@href[1]{\endgroup#1\@@endlink}%
\providecommand \@sanitize@url [0]{\catcode `\\12\catcode `\$12\catcode `\&12\catcode `\#12\catcode `\^12\catcode `\_12\catcode `\%12\relax}%
\providecommand \@@startlink[1]{}%
\providecommand \@@endlink[0]{}%
\providecommand \url  [0]{\begingroup\@sanitize@url \@url }%
\providecommand \@url [1]{\endgroup\@href {#1}{\urlprefix }}%
\providecommand \urlprefix  [0]{URL }%
\providecommand \Eprint [0]{\href }%
\providecommand \doibase [0]{https://doi.org/}%
\providecommand \selectlanguage [0]{\@gobble}%
\providecommand \bibinfo  [0]{\@secondoftwo}%
\providecommand \bibfield  [0]{\@secondoftwo}%
\providecommand \translation [1]{[#1]}%
\providecommand \BibitemOpen [0]{}%
\providecommand \bibitemStop [0]{}%
\providecommand \bibitemNoStop [0]{.\EOS\space}%
\providecommand \EOS [0]{\spacefactor3000\relax}%
\providecommand \BibitemShut  [1]{\csname bibitem#1\endcsname}%
\let\auto@bib@innerbib\@empty
%</preamble>
\bibitem [{\citenamefont {Shor}(1994)}]{Shor1994}%
  \BibitemOpen
  \bibfield  {author} {\bibinfo {author} {\bibfnamefont {P.}~\bibnamefont {Shor}},\ }\bibfield  {title} {\bibinfo {title} {Algorithms for quantum computation: discrete logarithms and factoring},\ }in\ \href {https://doi.org/10.1109/SFCS.1994.365700} {\emph {\bibinfo {booktitle} {Proceedings 35th Annual Symposium on Foundations of Computer Science}}}\ (\bibinfo {year} {1994})\ pp.\ \bibinfo {pages} {124--134}\BibitemShut {NoStop}%
\bibitem [{\citenamefont {Shor}(1999)}]{Shor1999}%
  \BibitemOpen
  \bibfield  {author} {\bibinfo {author} {\bibfnamefont {P.~W.}\ \bibnamefont {Shor}},\ }\bibfield  {title} {\bibinfo {title} {Polynomial-time algorithms for prime factorization and discrete logarithms on a quantum computer},\ }\href {https://doi.org/10.1137/S0036144598347011} {\bibfield  {journal} {\bibinfo  {journal} {SIAM Review}\ }\textbf {\bibinfo {volume} {41}},\ \bibinfo {pages} {303} (\bibinfo {year} {1999})},\ \Eprint {https://arxiv.org/abs/https://doi.org/10.1137/S0036144598347011} {https://doi.org/10.1137/S0036144598347011} \BibitemShut {NoStop}%
\bibitem [{\citenamefont {Gidney}\ and\ \citenamefont {Eker{\aa{}}}(2021)}]{Gidney2021RSA}%
  \BibitemOpen
  \bibfield  {author} {\bibinfo {author} {\bibfnamefont {C.}~\bibnamefont {Gidney}}\ and\ \bibinfo {author} {\bibfnamefont {M.}~\bibnamefont {Eker{\aa{}}}},\ }\bibfield  {title} {\bibinfo {title} {How to factor 2048 bit {RSA} integers in 8 hours using 20 million noisy qubits},\ }\href {https://doi.org/10.22331/q-2021-04-15-433} {\bibfield  {journal} {\bibinfo  {journal} {{Quantum}}\ }\textbf {\bibinfo {volume} {5}},\ \bibinfo {pages} {433} (\bibinfo {year} {2021})}\BibitemShut {NoStop}%
\bibitem [{\citenamefont {Gidney}(2025)}]{Gidney2025RSA}%
  \BibitemOpen
  \bibfield  {author} {\bibinfo {author} {\bibfnamefont {C.}~\bibnamefont {Gidney}},\ }\href {https://arxiv.org/abs/2505.15917} {\bibinfo {title} {How to factor 2048 bit rsa integers with less than a million noisy qubits}} (\bibinfo {year} {2025}),\ \Eprint {https://arxiv.org/abs/2505.15917} {arXiv:2505.15917 [quant-ph]} \BibitemShut {NoStop}%
\bibitem [{\citenamefont {Lloyd}(1996)}]{Lloyd1996}%
  \BibitemOpen
  \bibfield  {author} {\bibinfo {author} {\bibfnamefont {S.}~\bibnamefont {Lloyd}},\ }\bibfield  {title} {\bibinfo {title} {Universal quantum simulators},\ }\href {https://doi.org/10.1126/science.273.5278.1073} {\bibfield  {journal} {\bibinfo  {journal} {Science}\ }\textbf {\bibinfo {volume} {273}},\ \bibinfo {pages} {1073} (\bibinfo {year} {1996})},\ \Eprint {https://arxiv.org/abs/https://www.science.org/doi/pdf/10.1126/science.273.5278.1073} {https://www.science.org/doi/pdf/10.1126/science.273.5278.1073} \BibitemShut {NoStop}%
\bibitem [{\citenamefont {Abrams}\ and\ \citenamefont {Lloyd}(1999)}]{Abrams1999}%
  \BibitemOpen
  \bibfield  {author} {\bibinfo {author} {\bibfnamefont {D.~S.}\ \bibnamefont {Abrams}}\ and\ \bibinfo {author} {\bibfnamefont {S.}~\bibnamefont {Lloyd}},\ }\bibfield  {title} {\bibinfo {title} {Quantum algorithm providing exponential speed increase for finding eigenvalues and eigenvectors},\ }\href {https://doi.org/10.1103/PhysRevLett.83.5162} {\bibfield  {journal} {\bibinfo  {journal} {Phys. Rev. Lett.}\ }\textbf {\bibinfo {volume} {83}},\ \bibinfo {pages} {5162} (\bibinfo {year} {1999})}\BibitemShut {NoStop}%
\bibitem [{\citenamefont {Aspuru-Guzik}\ \emph {et~al.}(2005)\citenamefont {Aspuru-Guzik}, \citenamefont {Dutoi}, \citenamefont {Love},\ and\ \citenamefont {Head-Gordon}}]{Aspuru-Guzik2005}%
  \BibitemOpen
  \bibfield  {author} {\bibinfo {author} {\bibfnamefont {A.}~\bibnamefont {Aspuru-Guzik}}, \bibinfo {author} {\bibfnamefont {A.~D.}\ \bibnamefont {Dutoi}}, \bibinfo {author} {\bibfnamefont {P.~J.}\ \bibnamefont {Love}},\ and\ \bibinfo {author} {\bibfnamefont {M.}~\bibnamefont {Head-Gordon}},\ }\bibfield  {title} {\bibinfo {title} {Simulated quantum computation of molecular energies},\ }\href {https://doi.org/10.1126/science.1113479} {\bibfield  {journal} {\bibinfo  {journal} {Science}\ }\textbf {\bibinfo {volume} {309}},\ \bibinfo {pages} {1704} (\bibinfo {year} {2005})},\ \Eprint {https://arxiv.org/abs/https://www.science.org/doi/pdf/10.1126/science.1113479} {https://www.science.org/doi/pdf/10.1126/science.1113479} \BibitemShut {NoStop}%
\bibitem [{\citenamefont {Lee}\ \emph {et~al.}(2021)\citenamefont {Lee}, \citenamefont {Berry}, \citenamefont {Gidney}, \citenamefont {Huggins}, \citenamefont {McClean}, \citenamefont {Wiebe},\ and\ \citenamefont {Babbush}}]{Lee2021}%
  \BibitemOpen
  \bibfield  {author} {\bibinfo {author} {\bibfnamefont {J.}~\bibnamefont {Lee}}, \bibinfo {author} {\bibfnamefont {D.~W.}\ \bibnamefont {Berry}}, \bibinfo {author} {\bibfnamefont {C.}~\bibnamefont {Gidney}}, \bibinfo {author} {\bibfnamefont {W.~J.}\ \bibnamefont {Huggins}}, \bibinfo {author} {\bibfnamefont {J.~R.}\ \bibnamefont {McClean}}, \bibinfo {author} {\bibfnamefont {N.}~\bibnamefont {Wiebe}},\ and\ \bibinfo {author} {\bibfnamefont {R.}~\bibnamefont {Babbush}},\ }\bibfield  {title} {\bibinfo {title} {Even more efficient quantum computations of chemistry through tensor hypercontraction},\ }\href {https://doi.org/10.1103/PRXQuantum.2.030305} {\bibfield  {journal} {\bibinfo  {journal} {PRX Quantum}\ }\textbf {\bibinfo {volume} {2}},\ \bibinfo {pages} {030305} (\bibinfo {year} {2021})}\BibitemShut {NoStop}%
\bibitem [{\citenamefont {Low}\ \emph {et~al.}(2025)\citenamefont {Low}, \citenamefont {King}, \citenamefont {Berry}, \citenamefont {Han}, \citenamefont {III}, \citenamefont {White}, \citenamefont {Babbush}, \citenamefont {Somma},\ and\ \citenamefont {Rubin}}]{Low2025}%
  \BibitemOpen
  \bibfield  {author} {\bibinfo {author} {\bibfnamefont {G.~H.}\ \bibnamefont {Low}}, \bibinfo {author} {\bibfnamefont {R.}~\bibnamefont {King}}, \bibinfo {author} {\bibfnamefont {D.~W.}\ \bibnamefont {Berry}}, \bibinfo {author} {\bibfnamefont {Q.}~\bibnamefont {Han}}, \bibinfo {author} {\bibfnamefont {A.~E.~D.}\ \bibnamefont {III}}, \bibinfo {author} {\bibfnamefont {A.}~\bibnamefont {White}}, \bibinfo {author} {\bibfnamefont {R.}~\bibnamefont {Babbush}}, \bibinfo {author} {\bibfnamefont {R.~D.}\ \bibnamefont {Somma}},\ and\ \bibinfo {author} {\bibfnamefont {N.~C.}\ \bibnamefont {Rubin}},\ }\href {https://arxiv.org/abs/2502.15882} {\bibinfo {title} {Fast quantum simulation of electronic structure by spectrum amplification}} (\bibinfo {year} {2025}),\ \Eprint {https://arxiv.org/abs/2502.15882} {arXiv:2502.15882 [quant-ph]} \BibitemShut {NoStop}%
\bibitem [{\citenamefont {Babbush}\ \emph {et~al.}(2018)\citenamefont {Babbush}, \citenamefont {Gidney}, \citenamefont {Berry}, \citenamefont {Wiebe}, \citenamefont {McClean}, \citenamefont {Paler}, \citenamefont {Fowler},\ and\ \citenamefont {Neven}}]{Babbush2018qubitization}%
  \BibitemOpen
  \bibfield  {author} {\bibinfo {author} {\bibfnamefont {R.}~\bibnamefont {Babbush}}, \bibinfo {author} {\bibfnamefont {C.}~\bibnamefont {Gidney}}, \bibinfo {author} {\bibfnamefont {D.~W.}\ \bibnamefont {Berry}}, \bibinfo {author} {\bibfnamefont {N.}~\bibnamefont {Wiebe}}, \bibinfo {author} {\bibfnamefont {J.}~\bibnamefont {McClean}}, \bibinfo {author} {\bibfnamefont {A.}~\bibnamefont {Paler}}, \bibinfo {author} {\bibfnamefont {A.}~\bibnamefont {Fowler}},\ and\ \bibinfo {author} {\bibfnamefont {H.}~\bibnamefont {Neven}},\ }\bibfield  {title} {\bibinfo {title} {Encoding electronic spectra in quantum circuits with linear t complexity},\ }\href {https://doi.org/10.1103/PhysRevX.8.041015} {\bibfield  {journal} {\bibinfo  {journal} {Phys. Rev. X}\ }\textbf {\bibinfo {volume} {8}},\ \bibinfo {pages} {041015} (\bibinfo {year} {2018})}\BibitemShut {NoStop}%
\bibitem [{\citenamefont {Kivlichan}\ \emph {et~al.}(2020)\citenamefont {Kivlichan}, \citenamefont {Gidney}, \citenamefont {Berry}, \citenamefont {Wiebe}, \citenamefont {McClean}, \citenamefont {Sun}, \citenamefont {Jiang}, \citenamefont {Rubin}, \citenamefont {Fowler}, \citenamefont {Aspuru-Guzik} \emph {et~al.}}]{Kivlichan2020improved}%
  \BibitemOpen
  \bibfield  {author} {\bibinfo {author} {\bibfnamefont {I.~D.}\ \bibnamefont {Kivlichan}}, \bibinfo {author} {\bibfnamefont {C.}~\bibnamefont {Gidney}}, \bibinfo {author} {\bibfnamefont {D.~W.}\ \bibnamefont {Berry}}, \bibinfo {author} {\bibfnamefont {N.}~\bibnamefont {Wiebe}}, \bibinfo {author} {\bibfnamefont {J.}~\bibnamefont {McClean}}, \bibinfo {author} {\bibfnamefont {W.}~\bibnamefont {Sun}}, \bibinfo {author} {\bibfnamefont {Z.}~\bibnamefont {Jiang}}, \bibinfo {author} {\bibfnamefont {N.}~\bibnamefont {Rubin}}, \bibinfo {author} {\bibfnamefont {A.}~\bibnamefont {Fowler}}, \bibinfo {author} {\bibfnamefont {A.}~\bibnamefont {Aspuru-Guzik}}, \emph {et~al.},\ }\bibfield  {title} {\bibinfo {title} {Improved fault-tolerant quantum simulation of condensed-phase correlated electrons via trotterization},\ }\href@noop {} {\bibfield  {journal} {\bibinfo  {journal} {Quantum}\ }\textbf {\bibinfo {volume} {4}},\ \bibinfo {pages} {296} (\bibinfo {year} {2020})}\BibitemShut {NoStop}%
\bibitem [{\citenamefont {Yoshioka}\ \emph {et~al.}(2022)\citenamefont {Yoshioka}, \citenamefont {Okubo}, \citenamefont {Suzuki}, \citenamefont {Koizumi},\ and\ \citenamefont {Mizukami}}]{Yoshioka2022hunting}%
  \BibitemOpen
  \bibfield  {author} {\bibinfo {author} {\bibfnamefont {N.}~\bibnamefont {Yoshioka}}, \bibinfo {author} {\bibfnamefont {T.}~\bibnamefont {Okubo}}, \bibinfo {author} {\bibfnamefont {Y.}~\bibnamefont {Suzuki}}, \bibinfo {author} {\bibfnamefont {Y.}~\bibnamefont {Koizumi}},\ and\ \bibinfo {author} {\bibfnamefont {W.}~\bibnamefont {Mizukami}},\ }\href@noop {} {\bibinfo {title} {Hunting for quantum-classical crossover in condensed matter problems}} (\bibinfo {year} {2022}),\ \Eprint {https://arxiv.org/abs/arXiv:2210.14109} {arXiv:2210.14109} \BibitemShut {NoStop}%
\bibitem [{\citenamefont {Toshio}\ \emph {et~al.}(2025)\citenamefont {Toshio}, \citenamefont {Akahoshi}, \citenamefont {Fujisaki}, \citenamefont {Oshima}, \citenamefont {Sato},\ and\ \citenamefont {Fujii}}]{Toshio2024}%
  \BibitemOpen
  \bibfield  {author} {\bibinfo {author} {\bibfnamefont {R.}~\bibnamefont {Toshio}}, \bibinfo {author} {\bibfnamefont {Y.}~\bibnamefont {Akahoshi}}, \bibinfo {author} {\bibfnamefont {J.}~\bibnamefont {Fujisaki}}, \bibinfo {author} {\bibfnamefont {H.}~\bibnamefont {Oshima}}, \bibinfo {author} {\bibfnamefont {S.}~\bibnamefont {Sato}},\ and\ \bibinfo {author} {\bibfnamefont {K.}~\bibnamefont {Fujii}},\ }\bibfield  {title} {\bibinfo {title} {Practical quantum advantage on partially fault-tolerant quantum computer},\ }\href {https://doi.org/10.1103/PhysRevX.15.021057} {\bibfield  {journal} {\bibinfo  {journal} {Phys. Rev. X}\ }\textbf {\bibinfo {volume} {15}},\ \bibinfo {pages} {021057} (\bibinfo {year} {2025})}\BibitemShut {NoStop}%
\bibitem [{\citenamefont {Akahoshi}\ \emph {et~al.}(2024)\citenamefont {Akahoshi}, \citenamefont {Toshio}, \citenamefont {Fujisaki}, \citenamefont {Oshima}, \citenamefont {Sato},\ and\ \citenamefont {Fujii}}]{Akahoshi2024}%
  \BibitemOpen
  \bibfield  {author} {\bibinfo {author} {\bibfnamefont {Y.}~\bibnamefont {Akahoshi}}, \bibinfo {author} {\bibfnamefont {R.}~\bibnamefont {Toshio}}, \bibinfo {author} {\bibfnamefont {J.}~\bibnamefont {Fujisaki}}, \bibinfo {author} {\bibfnamefont {H.}~\bibnamefont {Oshima}}, \bibinfo {author} {\bibfnamefont {S.}~\bibnamefont {Sato}},\ and\ \bibinfo {author} {\bibfnamefont {K.}~\bibnamefont {Fujii}},\ }\href@noop {} {\bibinfo {title} {Compilation of trotter-based time evolution for partially fault-tolerant quantum computing architecture}} (\bibinfo {year} {2024}),\ \Eprint {https://arxiv.org/abs/arXiv:2408.14929} {arXiv:2408.14929} \BibitemShut {NoStop}%
\bibitem [{\citenamefont {Harrow}\ \emph {et~al.}(2009)\citenamefont {Harrow}, \citenamefont {Hassidim},\ and\ \citenamefont {Lloyd}}]{Harrow2009}%
  \BibitemOpen
  \bibfield  {author} {\bibinfo {author} {\bibfnamefont {A.~W.}\ \bibnamefont {Harrow}}, \bibinfo {author} {\bibfnamefont {A.}~\bibnamefont {Hassidim}},\ and\ \bibinfo {author} {\bibfnamefont {S.}~\bibnamefont {Lloyd}},\ }\bibfield  {title} {\bibinfo {title} {Quantum algorithm for linear systems of equations},\ }\href {https://doi.org/10.1103/PhysRevLett.103.150502} {\bibfield  {journal} {\bibinfo  {journal} {Phys. Rev. Lett.}\ }\textbf {\bibinfo {volume} {103}},\ \bibinfo {pages} {150502} (\bibinfo {year} {2009})}\BibitemShut {NoStop}%
\bibitem [{\citenamefont {Shor}(1996)}]{Shor1996}%
  \BibitemOpen
  \bibfield  {author} {\bibinfo {author} {\bibfnamefont {P.}~\bibnamefont {Shor}},\ }\bibfield  {title} {\bibinfo {title} {Fault-tolerant quantum computation},\ }in\ \href {https://doi.org/10.1109/SFCS.1996.548464} {\emph {\bibinfo {booktitle} {Proceedings of 37th Conference on Foundations of Computer Science}}}\ (\bibinfo {year} {1996})\ pp.\ \bibinfo {pages} {56--65}\BibitemShut {NoStop}%
\bibitem [{\citenamefont {Aharonov}\ and\ \citenamefont {Ben-Or}(1997)}]{Aharonov1997}%
  \BibitemOpen
  \bibfield  {author} {\bibinfo {author} {\bibfnamefont {D.}~\bibnamefont {Aharonov}}\ and\ \bibinfo {author} {\bibfnamefont {M.}~\bibnamefont {Ben-Or}},\ }\bibfield  {title} {\bibinfo {title} {Fault-tolerant quantum computation with constant error},\ }in\ \href {https://doi.org/10.1145/258533.258579} {\emph {\bibinfo {booktitle} {Proceedings of the Twenty-Ninth Annual ACM Symposium on Theory of Computing}}},\ \bibinfo {series and number} {STOC '97}\ (\bibinfo  {publisher} {Association for Computing Machinery},\ \bibinfo {address} {New York, NY, USA},\ \bibinfo {year} {1997})\ p.\ \bibinfo {pages} {176–188}\BibitemShut {NoStop}%
\bibitem [{\citenamefont {Kitaev}(1997)}]{Kitaev1997}%
  \BibitemOpen
  \bibfield  {author} {\bibinfo {author} {\bibfnamefont {A.~Y.}\ \bibnamefont {Kitaev}},\ }\bibinfo {title} {Quantum error correction with imperfect gates},\ in\ \href {https://doi.org/10.1007/978-1-4615-5923-8_19} {\emph {\bibinfo {booktitle} {Quantum Communication, Computing, and Measurement}}},\ \bibinfo {editor} {edited by\ \bibinfo {editor} {\bibfnamefont {O.}~\bibnamefont {Hirota}}, \bibinfo {editor} {\bibfnamefont {A.~S.}\ \bibnamefont {Holevo}},\ and\ \bibinfo {editor} {\bibfnamefont {C.~M.}\ \bibnamefont {Caves}}}\ (\bibinfo  {publisher} {Springer US},\ \bibinfo {address} {Boston, MA},\ \bibinfo {year} {1997})\ pp.\ \bibinfo {pages} {181--188}\BibitemShut {NoStop}%
\bibitem [{\citenamefont {Nielsen}\ and\ \citenamefont {Chuang}(2000)}]{Nielsen2000}%
  \BibitemOpen
  \bibfield  {author} {\bibinfo {author} {\bibfnamefont {M.~A.}\ \bibnamefont {Nielsen}}\ and\ \bibinfo {author} {\bibfnamefont {I.~L.}\ \bibnamefont {Chuang}},\ }\href@noop {} {\emph {\bibinfo {title} {Quantum Computation and Quantum Information}}}\ (\bibinfo  {publisher} {Cambridge University Press},\ \bibinfo {year} {2000})\BibitemShut {NoStop}%
\bibitem [{\citenamefont {Battistel}\ \emph {et~al.}(2023)\citenamefont {Battistel}, \citenamefont {Chamberland}, \citenamefont {Johar}, \citenamefont {Overwater}, \citenamefont {Sebastiano}, \citenamefont {Skoric}, \citenamefont {Ueno},\ and\ \citenamefont {Usman}}]{Battistel2023review}%
  \BibitemOpen
  \bibfield  {author} {\bibinfo {author} {\bibfnamefont {F.}~\bibnamefont {Battistel}}, \bibinfo {author} {\bibfnamefont {C.}~\bibnamefont {Chamberland}}, \bibinfo {author} {\bibfnamefont {K.}~\bibnamefont {Johar}}, \bibinfo {author} {\bibfnamefont {R.~W.~J.}\ \bibnamefont {Overwater}}, \bibinfo {author} {\bibfnamefont {F.}~\bibnamefont {Sebastiano}}, \bibinfo {author} {\bibfnamefont {L.}~\bibnamefont {Skoric}}, \bibinfo {author} {\bibfnamefont {Y.}~\bibnamefont {Ueno}},\ and\ \bibinfo {author} {\bibfnamefont {M.}~\bibnamefont {Usman}},\ }\bibfield  {title} {\bibinfo {title} {Real-time decoding for fault-tolerant quantum computing: progress, challenges and outlook},\ }\href {https://doi.org/10.1088/2399-1984/aceba6} {\bibfield  {journal} {\bibinfo  {journal} {Nano Futures}\ }\textbf {\bibinfo {volume} {7}},\ \bibinfo {pages} {032003} (\bibinfo {year} {2023})}\BibitemShut {NoStop}%
\bibitem [{\citenamefont {deMarti iOlius}\ \emph {et~al.}(2024)\citenamefont {deMarti iOlius}, \citenamefont {Fuentes}, \citenamefont {Or{\'{u}}s}, \citenamefont {Crespo},\ and\ \citenamefont {Etxezarreta~Martinez}}]{iOlius2024review}%
  \BibitemOpen
  \bibfield  {author} {\bibinfo {author} {\bibfnamefont {A.}~\bibnamefont {deMarti iOlius}}, \bibinfo {author} {\bibfnamefont {P.}~\bibnamefont {Fuentes}}, \bibinfo {author} {\bibfnamefont {R.}~\bibnamefont {Or{\'{u}}s}}, \bibinfo {author} {\bibfnamefont {P.~M.}\ \bibnamefont {Crespo}},\ and\ \bibinfo {author} {\bibfnamefont {J.}~\bibnamefont {Etxezarreta~Martinez}},\ }\bibfield  {title} {\bibinfo {title} {Decoding algorithms for surface codes},\ }\href {https://doi.org/10.22331/q-2024-10-10-1498} {\bibfield  {journal} {\bibinfo  {journal} {{Quantum}}\ }\textbf {\bibinfo {volume} {8}},\ \bibinfo {pages} {1498} (\bibinfo {year} {2024})}\BibitemShut {NoStop}%
\bibitem [{\citenamefont {Terhal}(2015)}]{Terhal2015Review}%
  \BibitemOpen
  \bibfield  {author} {\bibinfo {author} {\bibfnamefont {B.~M.}\ \bibnamefont {Terhal}},\ }\bibfield  {title} {\bibinfo {title} {Quantum error correction for quantum memories},\ }\href {https://doi.org/10.1103/RevModPhys.87.307} {\bibfield  {journal} {\bibinfo  {journal} {Rev. Mod. Phys.}\ }\textbf {\bibinfo {volume} {87}},\ \bibinfo {pages} {307} (\bibinfo {year} {2015})}\BibitemShut {NoStop}%
\bibitem [{\citenamefont {Skoric}\ \emph {et~al.}(2023)\citenamefont {Skoric}, \citenamefont {Browne}, \citenamefont {Barnes}, \citenamefont {Gillespie},\ and\ \citenamefont {Campbell}}]{Skoric2023}%
  \BibitemOpen
  \bibfield  {author} {\bibinfo {author} {\bibfnamefont {L.}~\bibnamefont {Skoric}}, \bibinfo {author} {\bibfnamefont {D.~E.}\ \bibnamefont {Browne}}, \bibinfo {author} {\bibfnamefont {K.~M.}\ \bibnamefont {Barnes}}, \bibinfo {author} {\bibfnamefont {N.~I.}\ \bibnamefont {Gillespie}},\ and\ \bibinfo {author} {\bibfnamefont {E.~T.}\ \bibnamefont {Campbell}},\ }\bibfield  {title} {\bibinfo {title} {Parallel window decoding enables scalable fault tolerant quantum computation},\ }\href {https://doi.org/10.1038/s41467-023-42482-1} {\bibfield  {journal} {\bibinfo  {journal} {Nature Communications}\ }\textbf {\bibinfo {volume} {14}},\ \bibinfo {pages} {7040} (\bibinfo {year} {2023})}\BibitemShut {NoStop}%
\bibitem [{\citenamefont {Bausch}\ \emph {et~al.}(2024)\citenamefont {Bausch}, \citenamefont {Senior}, \citenamefont {Heras}, \citenamefont {Edlich}, \citenamefont {Davies}, \citenamefont {Newman}, \citenamefont {Jones}, \citenamefont {Satzinger}, \citenamefont {Niu}, \citenamefont {Blackwell}, \citenamefont {Holland}, \citenamefont {Kafri}, \citenamefont {Atalaya}, \citenamefont {Gidney}, \citenamefont {Hassabis}, \citenamefont {Boixo}, \citenamefont {Neven},\ and\ \citenamefont {Kohli}}]{Bausch2024alphaqubit}%
  \BibitemOpen
  \bibfield  {author} {\bibinfo {author} {\bibfnamefont {J.}~\bibnamefont {Bausch}}, \bibinfo {author} {\bibfnamefont {A.~W.}\ \bibnamefont {Senior}}, \bibinfo {author} {\bibfnamefont {F.~J.~H.}\ \bibnamefont {Heras}}, \bibinfo {author} {\bibfnamefont {T.}~\bibnamefont {Edlich}}, \bibinfo {author} {\bibfnamefont {A.}~\bibnamefont {Davies}}, \bibinfo {author} {\bibfnamefont {M.}~\bibnamefont {Newman}}, \bibinfo {author} {\bibfnamefont {C.}~\bibnamefont {Jones}}, \bibinfo {author} {\bibfnamefont {K.}~\bibnamefont {Satzinger}}, \bibinfo {author} {\bibfnamefont {M.~Y.}\ \bibnamefont {Niu}}, \bibinfo {author} {\bibfnamefont {S.}~\bibnamefont {Blackwell}}, \bibinfo {author} {\bibfnamefont {G.}~\bibnamefont {Holland}}, \bibinfo {author} {\bibfnamefont {D.}~\bibnamefont {Kafri}}, \bibinfo {author} {\bibfnamefont {J.}~\bibnamefont {Atalaya}}, \bibinfo {author} {\bibfnamefont {C.}~\bibnamefont {Gidney}}, \bibinfo {author} {\bibfnamefont {D.}~\bibnamefont {Hassabis}}, \bibinfo {author} {\bibfnamefont {S.}~\bibnamefont
  {Boixo}}, \bibinfo {author} {\bibfnamefont {H.}~\bibnamefont {Neven}},\ and\ \bibinfo {author} {\bibfnamefont {P.}~\bibnamefont {Kohli}},\ }\bibfield  {title} {\bibinfo {title} {Learning high-accuracy error decoding for quantum processors},\ }\href {https://doi.org/10.1038/s41586-024-08148-8} {\bibfield  {journal} {\bibinfo  {journal} {Nature}\ }\textbf {\bibinfo {volume} {635}},\ \bibinfo {pages} {834} (\bibinfo {year} {2024})}\BibitemShut {NoStop}%
\bibitem [{\citenamefont {Dennis}\ \emph {et~al.}(2002)\citenamefont {Dennis}, \citenamefont {Kitaev}, \citenamefont {Landahl},\ and\ \citenamefont {Preskill}}]{Dennis2002}%
  \BibitemOpen
  \bibfield  {author} {\bibinfo {author} {\bibfnamefont {E.}~\bibnamefont {Dennis}}, \bibinfo {author} {\bibfnamefont {A.}~\bibnamefont {Kitaev}}, \bibinfo {author} {\bibfnamefont {A.}~\bibnamefont {Landahl}},\ and\ \bibinfo {author} {\bibfnamefont {J.}~\bibnamefont {Preskill}},\ }\bibfield  {title} {\bibinfo {title} {Topological quantum memory},\ }\href {https://doi.org/10.1063/1.1499754} {\bibfield  {journal} {\bibinfo  {journal} {Journal of Mathematical Physics}\ }\textbf {\bibinfo {volume} {43}},\ \bibinfo {pages} {4452} (\bibinfo {year} {2002})},\ \Eprint {https://arxiv.org/abs/https://doi.org/10.1063/1.1499754} {https://doi.org/10.1063/1.1499754} \BibitemShut {NoStop}%
\bibitem [{\citenamefont {Fowler}(2015)}]{Fowler2015mwpm}%
  \BibitemOpen
  \bibfield  {author} {\bibinfo {author} {\bibfnamefont {A.~G.}\ \bibnamefont {Fowler}},\ }\bibfield  {title} {\bibinfo {title} {Minimum weight perfect matching of fault-tolerant topological quantum error correction in average o(1) parallel time},\ }\href@noop {} {\bibfield  {journal} {\bibinfo  {journal} {Quantum Info. Comput.}\ }\textbf {\bibinfo {volume} {15}},\ \bibinfo {pages} {145–158} (\bibinfo {year} {2015})}\BibitemShut {NoStop}%
\bibitem [{\citenamefont {Higgott}\ and\ \citenamefont {Gidney}(2023)}]{Higgott2023sparse}%
  \BibitemOpen
  \bibfield  {author} {\bibinfo {author} {\bibfnamefont {O.}~\bibnamefont {Higgott}}\ and\ \bibinfo {author} {\bibfnamefont {C.}~\bibnamefont {Gidney}},\ }\bibfield  {title} {\bibinfo {title} {Sparse blossom: correcting a million errors per core second with minimum-weight matching},\ }\href@noop {} {\bibfield  {journal} {\bibinfo  {journal} {arXiv preprint arXiv:2303.15933}\ } (\bibinfo {year} {2023})}\BibitemShut {NoStop}%
\bibitem [{\citenamefont {Wu}\ and\ \citenamefont {Zhong}(2023)}]{Wu2023fusion}%
  \BibitemOpen
  \bibfield  {author} {\bibinfo {author} {\bibfnamefont {Y.}~\bibnamefont {Wu}}\ and\ \bibinfo {author} {\bibfnamefont {L.}~\bibnamefont {Zhong}},\ }\bibfield  {title} {\bibinfo {title} {Fusion blossom: Fast mwpm decoders for qec},\ }in\ \href@noop {} {\emph {\bibinfo {booktitle} {2023 IEEE International Conference on Quantum Computing and Engineering (QCE)}}},\ Vol.~\bibinfo {volume} {1}\ (\bibinfo {organization} {IEEE},\ \bibinfo {year} {2023})\ pp.\ \bibinfo {pages} {928--938}\BibitemShut {NoStop}%
\bibitem [{\citenamefont {Ferris}\ and\ \citenamefont {Poulin}(2014)}]{Ferris2014}%
  \BibitemOpen
  \bibfield  {author} {\bibinfo {author} {\bibfnamefont {A.~J.}\ \bibnamefont {Ferris}}\ and\ \bibinfo {author} {\bibfnamefont {D.}~\bibnamefont {Poulin}},\ }\bibfield  {title} {\bibinfo {title} {Tensor networks and quantum error correction},\ }\href {https://doi.org/10.1103/PhysRevLett.113.030501} {\bibfield  {journal} {\bibinfo  {journal} {Phys. Rev. Lett.}\ }\textbf {\bibinfo {volume} {113}},\ \bibinfo {pages} {030501} (\bibinfo {year} {2014})}\BibitemShut {NoStop}%
\bibitem [{\citenamefont {Bravyi}\ \emph {et~al.}(2014)\citenamefont {Bravyi}, \citenamefont {Suchara},\ and\ \citenamefont {Vargo}}]{Bravyi2014}%
  \BibitemOpen
  \bibfield  {author} {\bibinfo {author} {\bibfnamefont {S.}~\bibnamefont {Bravyi}}, \bibinfo {author} {\bibfnamefont {M.}~\bibnamefont {Suchara}},\ and\ \bibinfo {author} {\bibfnamefont {A.}~\bibnamefont {Vargo}},\ }\bibfield  {title} {\bibinfo {title} {Efficient algorithms for maximum likelihood decoding in the surface code},\ }\href {https://doi.org/10.1103/PhysRevA.90.032326} {\bibfield  {journal} {\bibinfo  {journal} {Phys. Rev. A}\ }\textbf {\bibinfo {volume} {90}},\ \bibinfo {pages} {032326} (\bibinfo {year} {2014})}\BibitemShut {NoStop}%
\bibitem [{\citenamefont {Acharya}\ \emph {et~al.}(2025)\citenamefont {Acharya}, \citenamefont {Abanin}, \citenamefont {Aghababaie-Beni}, \citenamefont {Aleiner}, \citenamefont {Andersen}, \citenamefont {Ansmann}, \citenamefont {Arute}, \citenamefont {Arya}, \citenamefont {Asfaw}, \citenamefont {Astrakhantsev}, \citenamefont {Atalaya}, \citenamefont {Babbush}, \citenamefont {Bacon}, \citenamefont {Ballard}, \citenamefont {Bardin}, \citenamefont {Bausch}, \citenamefont {Bengtsson}, \citenamefont {Bilmes}, \citenamefont {Blackwell}, \citenamefont {Boixo}, \citenamefont {Bortoli}, \citenamefont {Bourassa}, \citenamefont {Bovaird}, \citenamefont {Brill}, \citenamefont {Broughton}, \citenamefont {Browne}, \citenamefont {Buchea}, \citenamefont {Buckley}, \citenamefont {Buell}, \citenamefont {Burger}, \citenamefont {Burkett}, \citenamefont {Bushnell}, \citenamefont {Cabrera}, \citenamefont {Campero}, \citenamefont {Chang}, \citenamefont {Chen}, \citenamefont {Chen}, \citenamefont {Chiaro}, \citenamefont {Chik},
  \citenamefont {Chou}, \citenamefont {Claes}, \citenamefont {Cleland}, \citenamefont {Cogan}, \citenamefont {Collins}, \citenamefont {Conner}, \citenamefont {Courtney}, \citenamefont {Crook}, \citenamefont {Curtin}, \citenamefont {Das}, \citenamefont {Davies}, \citenamefont {De~Lorenzo}, \citenamefont {Debroy}, \citenamefont {Demura}, \citenamefont {Devoret}, \citenamefont {Di~Paolo}, \citenamefont {Donohoe}, \citenamefont {Drozdov}, \citenamefont {Dunsworth}, \citenamefont {Earle}, \citenamefont {Edlich}, \citenamefont {Eickbusch}, \citenamefont {Elbag}, \citenamefont {Elzouka}, \citenamefont {Erickson}, \citenamefont {Faoro}, \citenamefont {Farhi}, \citenamefont {Ferreira}, \citenamefont {Burgos}, \citenamefont {Forati}, \citenamefont {Fowler}, \citenamefont {Foxen}, \citenamefont {Ganjam}, \citenamefont {Garcia}, \citenamefont {Gasca}, \citenamefont {Genois}, \citenamefont {Giang}, \citenamefont {Gidney}, \citenamefont {Gilboa}, \citenamefont {Gosula}, \citenamefont {Dau}, \citenamefont {Graumann},
  \citenamefont {Greene}, \citenamefont {Gross}, \citenamefont {Habegger}, \citenamefont {Hall}, \citenamefont {Hamilton}, \citenamefont {Hansen}, \citenamefont {Harrigan}, \citenamefont {Harrington}, \citenamefont {Heras}, \citenamefont {Heslin}, \citenamefont {Heu}, \citenamefont {Higgott}, \citenamefont {Hill}, \citenamefont {Hilton}, \citenamefont {Holland}, \citenamefont {Hong}, \citenamefont {Huang}, \citenamefont {Huff}, \citenamefont {Huggins}, \citenamefont {Ioffe}, \citenamefont {Isakov}, \citenamefont {Iveland}, \citenamefont {Jeffrey}, \citenamefont {Jiang}, \citenamefont {Jones}, \citenamefont {Jordan}, \citenamefont {Joshi}, \citenamefont {Juhas}, \citenamefont {Kafri}, \citenamefont {Kang}, \citenamefont {Karamlou}, \citenamefont {Kechedzhi}, \citenamefont {Kelly}, \citenamefont {Khaire}, \citenamefont {Khattar}, \citenamefont {Khezri}, \citenamefont {Kim}, \citenamefont {Klimov}, \citenamefont {Klots}, \citenamefont {Kobrin}, \citenamefont {Kohli}, \citenamefont {Korotkov}, \citenamefont
  {Kostritsa}, \citenamefont {Kothari}, \citenamefont {Kozlovskii}, \citenamefont {Kreikebaum}, \citenamefont {Kurilovich}, \citenamefont {Lacroix}, \citenamefont {Landhuis}, \citenamefont {Lange-Dei}, \citenamefont {Langley}, \citenamefont {Laptev}, \citenamefont {Lau}, \citenamefont {Le~Guevel}, \citenamefont {Ledford}, \citenamefont {Lee}, \citenamefont {Lee}, \citenamefont {Lensky}, \citenamefont {Leon}, \citenamefont {Lester}, \citenamefont {Li}, \citenamefont {Li}, \citenamefont {Lill}, \citenamefont {Liu}, \citenamefont {Livingston}, \citenamefont {Locharla}, \citenamefont {Lucero}, \citenamefont {Lundahl}, \citenamefont {Lunt}, \citenamefont {Madhuk}, \citenamefont {Malone}, \citenamefont {Maloney}, \citenamefont {Mandr{\`a}}, \citenamefont {Manyika}, \citenamefont {Martin}, \citenamefont {Martin}, \citenamefont {Martin}, \citenamefont {Maxfield}, \citenamefont {McClean}, \citenamefont {McEwen}, \citenamefont {Meeks}, \citenamefont {Megrant}, \citenamefont {Mi}, \citenamefont {Miao}, \citenamefont
  {Mieszala}, \citenamefont {Molavi}, \citenamefont {Molina}, \citenamefont {Montazeri}, \citenamefont {Morvan}, \citenamefont {Movassagh}, \citenamefont {Mruczkiewicz}, \citenamefont {Naaman}, \citenamefont {Neeley}, \citenamefont {Neill}, \citenamefont {Nersisyan}, \citenamefont {Neven}, \citenamefont {Newman}, \citenamefont {Ng}, \citenamefont {Nguyen}, \citenamefont {Nguyen}, \citenamefont {Ni}, \citenamefont {Niu}, \citenamefont {O'Brien}, \citenamefont {Oliver}, \citenamefont {Opremcak}, \citenamefont {Ottosson}, \citenamefont {Petukhov}, \citenamefont {Pizzuto}, \citenamefont {Platt}, \citenamefont {Potter}, \citenamefont {Pritchard}, \citenamefont {Pryadko}, \citenamefont {Quintana}, \citenamefont {Ramachandran}, \citenamefont {Reagor}, \citenamefont {Redding}, \citenamefont {Rhodes}, \citenamefont {Roberts}, \citenamefont {Rosenberg}, \citenamefont {Rosenfeld}, \citenamefont {Roushan}, \citenamefont {Rubin}, \citenamefont {Saei}, \citenamefont {Sank}, \citenamefont {Sankaragomathi}, \citenamefont
  {Satzinger}, \citenamefont {Schurkus}, \citenamefont {Schuster}, \citenamefont {Senior}, \citenamefont {Shearn}, \citenamefont {Shorter}, \citenamefont {Shutty}, \citenamefont {Shvarts}, \citenamefont {Singh}, \citenamefont {Sivak}, \citenamefont {Skruzny}, \citenamefont {Small}, \citenamefont {Smelyanskiy}, \citenamefont {Smith}, \citenamefont {Somma}, \citenamefont {Springer}, \citenamefont {Sterling}, \citenamefont {Strain}, \citenamefont {Suchard}, \citenamefont {Szasz}, \citenamefont {Sztein}, \citenamefont {Thor}, \citenamefont {Torres}, \citenamefont {Torunbalci}, \citenamefont {Vaishnav}, \citenamefont {Vargas}, \citenamefont {Vdovichev}, \citenamefont {Vidal}, \citenamefont {Villalonga}, \citenamefont {Heidweiller}, \citenamefont {Waltman}, \citenamefont {Wang}, \citenamefont {Ware}, \citenamefont {Weber}, \citenamefont {Weidel}, \citenamefont {White}, \citenamefont {Wong}, \citenamefont {Woo}, \citenamefont {Xing}, \citenamefont {Yao}, \citenamefont {Yeh}, \citenamefont {Ying}, \citenamefont
  {Yoo}, \citenamefont {Yosri}, \citenamefont {Young}, \citenamefont {Zalcman}, \citenamefont {Zhang}, \citenamefont {Zhu}, \citenamefont {Zobrist}, \citenamefont {AI},\ and\ \citenamefont {{Collaborators}}}]{Acharya2025google}%
  \BibitemOpen
  \bibfield  {author} {\bibinfo {author} {\bibfnamefont {R.}~\bibnamefont {Acharya}}, \bibinfo {author} {\bibfnamefont {D.~A.}\ \bibnamefont {Abanin}}, \bibinfo {author} {\bibfnamefont {L.}~\bibnamefont {Aghababaie-Beni}}, \bibinfo {author} {\bibfnamefont {I.}~\bibnamefont {Aleiner}}, \bibinfo {author} {\bibfnamefont {T.~I.}\ \bibnamefont {Andersen}}, \bibinfo {author} {\bibfnamefont {M.}~\bibnamefont {Ansmann}}, \bibinfo {author} {\bibfnamefont {F.}~\bibnamefont {Arute}}, \bibinfo {author} {\bibfnamefont {K.}~\bibnamefont {Arya}}, \bibinfo {author} {\bibfnamefont {A.}~\bibnamefont {Asfaw}}, \bibinfo {author} {\bibfnamefont {N.}~\bibnamefont {Astrakhantsev}}, \bibinfo {author} {\bibfnamefont {J.}~\bibnamefont {Atalaya}}, \bibinfo {author} {\bibfnamefont {R.}~\bibnamefont {Babbush}}, \bibinfo {author} {\bibfnamefont {D.}~\bibnamefont {Bacon}}, \bibinfo {author} {\bibfnamefont {B.}~\bibnamefont {Ballard}}, \bibinfo {author} {\bibfnamefont {J.~C.}\ \bibnamefont {Bardin}}, \bibinfo {author} {\bibfnamefont
  {J.}~\bibnamefont {Bausch}}, \bibinfo {author} {\bibfnamefont {A.}~\bibnamefont {Bengtsson}}, \bibinfo {author} {\bibfnamefont {A.}~\bibnamefont {Bilmes}}, \bibinfo {author} {\bibfnamefont {S.}~\bibnamefont {Blackwell}}, \bibinfo {author} {\bibfnamefont {S.}~\bibnamefont {Boixo}}, \bibinfo {author} {\bibfnamefont {G.}~\bibnamefont {Bortoli}}, \bibinfo {author} {\bibfnamefont {A.}~\bibnamefont {Bourassa}}, \bibinfo {author} {\bibfnamefont {J.}~\bibnamefont {Bovaird}}, \bibinfo {author} {\bibfnamefont {L.}~\bibnamefont {Brill}}, \bibinfo {author} {\bibfnamefont {M.}~\bibnamefont {Broughton}}, \bibinfo {author} {\bibfnamefont {D.~A.}\ \bibnamefont {Browne}}, \bibinfo {author} {\bibfnamefont {B.}~\bibnamefont {Buchea}}, \bibinfo {author} {\bibfnamefont {B.~B.}\ \bibnamefont {Buckley}}, \bibinfo {author} {\bibfnamefont {D.~A.}\ \bibnamefont {Buell}}, \bibinfo {author} {\bibfnamefont {T.}~\bibnamefont {Burger}}, \bibinfo {author} {\bibfnamefont {B.}~\bibnamefont {Burkett}}, \bibinfo {author} {\bibfnamefont
  {N.}~\bibnamefont {Bushnell}}, \bibinfo {author} {\bibfnamefont {A.}~\bibnamefont {Cabrera}}, \bibinfo {author} {\bibfnamefont {J.}~\bibnamefont {Campero}}, \bibinfo {author} {\bibfnamefont {H.-S.}\ \bibnamefont {Chang}}, \bibinfo {author} {\bibfnamefont {Y.}~\bibnamefont {Chen}}, \bibinfo {author} {\bibfnamefont {Z.}~\bibnamefont {Chen}}, \bibinfo {author} {\bibfnamefont {B.}~\bibnamefont {Chiaro}}, \bibinfo {author} {\bibfnamefont {D.}~\bibnamefont {Chik}}, \bibinfo {author} {\bibfnamefont {C.}~\bibnamefont {Chou}}, \bibinfo {author} {\bibfnamefont {J.}~\bibnamefont {Claes}}, \bibinfo {author} {\bibfnamefont {A.~Y.}\ \bibnamefont {Cleland}}, \bibinfo {author} {\bibfnamefont {J.}~\bibnamefont {Cogan}}, \bibinfo {author} {\bibfnamefont {R.}~\bibnamefont {Collins}}, \bibinfo {author} {\bibfnamefont {P.}~\bibnamefont {Conner}}, \bibinfo {author} {\bibfnamefont {W.}~\bibnamefont {Courtney}}, \bibinfo {author} {\bibfnamefont {A.~L.}\ \bibnamefont {Crook}}, \bibinfo {author} {\bibfnamefont {B.}~\bibnamefont
  {Curtin}}, \bibinfo {author} {\bibfnamefont {S.}~\bibnamefont {Das}}, \bibinfo {author} {\bibfnamefont {A.}~\bibnamefont {Davies}}, \bibinfo {author} {\bibfnamefont {L.}~\bibnamefont {De~Lorenzo}}, \bibinfo {author} {\bibfnamefont {D.~M.}\ \bibnamefont {Debroy}}, \bibinfo {author} {\bibfnamefont {S.}~\bibnamefont {Demura}}, \bibinfo {author} {\bibfnamefont {M.}~\bibnamefont {Devoret}}, \bibinfo {author} {\bibfnamefont {A.}~\bibnamefont {Di~Paolo}}, \bibinfo {author} {\bibfnamefont {P.}~\bibnamefont {Donohoe}}, \bibinfo {author} {\bibfnamefont {I.}~\bibnamefont {Drozdov}}, \bibinfo {author} {\bibfnamefont {A.}~\bibnamefont {Dunsworth}}, \bibinfo {author} {\bibfnamefont {C.}~\bibnamefont {Earle}}, \bibinfo {author} {\bibfnamefont {T.}~\bibnamefont {Edlich}}, \bibinfo {author} {\bibfnamefont {A.}~\bibnamefont {Eickbusch}}, \bibinfo {author} {\bibfnamefont {A.~M.}\ \bibnamefont {Elbag}}, \bibinfo {author} {\bibfnamefont {M.}~\bibnamefont {Elzouka}}, \bibinfo {author} {\bibfnamefont {C.}~\bibnamefont
  {Erickson}}, \bibinfo {author} {\bibfnamefont {L.}~\bibnamefont {Faoro}}, \bibinfo {author} {\bibfnamefont {E.}~\bibnamefont {Farhi}}, \bibinfo {author} {\bibfnamefont {V.~S.}\ \bibnamefont {Ferreira}}, \bibinfo {author} {\bibfnamefont {L.~F.}\ \bibnamefont {Burgos}}, \bibinfo {author} {\bibfnamefont {E.}~\bibnamefont {Forati}}, \bibinfo {author} {\bibfnamefont {A.~G.}\ \bibnamefont {Fowler}}, \bibinfo {author} {\bibfnamefont {B.}~\bibnamefont {Foxen}}, \bibinfo {author} {\bibfnamefont {S.}~\bibnamefont {Ganjam}}, \bibinfo {author} {\bibfnamefont {G.}~\bibnamefont {Garcia}}, \bibinfo {author} {\bibfnamefont {R.}~\bibnamefont {Gasca}}, \bibinfo {author} {\bibfnamefont {{\'E}.}~\bibnamefont {Genois}}, \bibinfo {author} {\bibfnamefont {W.}~\bibnamefont {Giang}}, \bibinfo {author} {\bibfnamefont {C.}~\bibnamefont {Gidney}}, \bibinfo {author} {\bibfnamefont {D.}~\bibnamefont {Gilboa}}, \bibinfo {author} {\bibfnamefont {R.}~\bibnamefont {Gosula}}, \bibinfo {author} {\bibfnamefont {A.~G.}\ \bibnamefont {Dau}},
  \bibinfo {author} {\bibfnamefont {D.}~\bibnamefont {Graumann}}, \bibinfo {author} {\bibfnamefont {A.}~\bibnamefont {Greene}}, \bibinfo {author} {\bibfnamefont {J.~A.}\ \bibnamefont {Gross}}, \bibinfo {author} {\bibfnamefont {S.}~\bibnamefont {Habegger}}, \bibinfo {author} {\bibfnamefont {J.}~\bibnamefont {Hall}}, \bibinfo {author} {\bibfnamefont {M.~C.}\ \bibnamefont {Hamilton}}, \bibinfo {author} {\bibfnamefont {M.}~\bibnamefont {Hansen}}, \bibinfo {author} {\bibfnamefont {M.~P.}\ \bibnamefont {Harrigan}}, \bibinfo {author} {\bibfnamefont {S.~D.}\ \bibnamefont {Harrington}}, \bibinfo {author} {\bibfnamefont {F.~J.~H.}\ \bibnamefont {Heras}}, \bibinfo {author} {\bibfnamefont {S.}~\bibnamefont {Heslin}}, \bibinfo {author} {\bibfnamefont {P.}~\bibnamefont {Heu}}, \bibinfo {author} {\bibfnamefont {O.}~\bibnamefont {Higgott}}, \bibinfo {author} {\bibfnamefont {G.}~\bibnamefont {Hill}}, \bibinfo {author} {\bibfnamefont {J.}~\bibnamefont {Hilton}}, \bibinfo {author} {\bibfnamefont {G.}~\bibnamefont {Holland}},
  \bibinfo {author} {\bibfnamefont {S.}~\bibnamefont {Hong}}, \bibinfo {author} {\bibfnamefont {H.-Y.}\ \bibnamefont {Huang}}, \bibinfo {author} {\bibfnamefont {A.}~\bibnamefont {Huff}}, \bibinfo {author} {\bibfnamefont {W.~J.}\ \bibnamefont {Huggins}}, \bibinfo {author} {\bibfnamefont {L.~B.}\ \bibnamefont {Ioffe}}, \bibinfo {author} {\bibfnamefont {S.~V.}\ \bibnamefont {Isakov}}, \bibinfo {author} {\bibfnamefont {J.}~\bibnamefont {Iveland}}, \bibinfo {author} {\bibfnamefont {E.}~\bibnamefont {Jeffrey}}, \bibinfo {author} {\bibfnamefont {Z.}~\bibnamefont {Jiang}}, \bibinfo {author} {\bibfnamefont {C.}~\bibnamefont {Jones}}, \bibinfo {author} {\bibfnamefont {S.}~\bibnamefont {Jordan}}, \bibinfo {author} {\bibfnamefont {C.}~\bibnamefont {Joshi}}, \bibinfo {author} {\bibfnamefont {P.}~\bibnamefont {Juhas}}, \bibinfo {author} {\bibfnamefont {D.}~\bibnamefont {Kafri}}, \bibinfo {author} {\bibfnamefont {H.}~\bibnamefont {Kang}}, \bibinfo {author} {\bibfnamefont {A.~H.}\ \bibnamefont {Karamlou}}, \bibinfo {author}
  {\bibfnamefont {K.}~\bibnamefont {Kechedzhi}}, \bibinfo {author} {\bibfnamefont {J.}~\bibnamefont {Kelly}}, \bibinfo {author} {\bibfnamefont {T.}~\bibnamefont {Khaire}}, \bibinfo {author} {\bibfnamefont {T.}~\bibnamefont {Khattar}}, \bibinfo {author} {\bibfnamefont {M.}~\bibnamefont {Khezri}}, \bibinfo {author} {\bibfnamefont {S.}~\bibnamefont {Kim}}, \bibinfo {author} {\bibfnamefont {P.~V.}\ \bibnamefont {Klimov}}, \bibinfo {author} {\bibfnamefont {A.~R.}\ \bibnamefont {Klots}}, \bibinfo {author} {\bibfnamefont {B.}~\bibnamefont {Kobrin}}, \bibinfo {author} {\bibfnamefont {P.}~\bibnamefont {Kohli}}, \bibinfo {author} {\bibfnamefont {A.~N.}\ \bibnamefont {Korotkov}}, \bibinfo {author} {\bibfnamefont {F.}~\bibnamefont {Kostritsa}}, \bibinfo {author} {\bibfnamefont {R.}~\bibnamefont {Kothari}}, \bibinfo {author} {\bibfnamefont {B.}~\bibnamefont {Kozlovskii}}, \bibinfo {author} {\bibfnamefont {J.~M.}\ \bibnamefont {Kreikebaum}}, \bibinfo {author} {\bibfnamefont {V.~D.}\ \bibnamefont {Kurilovich}}, \bibinfo
  {author} {\bibfnamefont {N.}~\bibnamefont {Lacroix}}, \bibinfo {author} {\bibfnamefont {D.}~\bibnamefont {Landhuis}}, \bibinfo {author} {\bibfnamefont {T.}~\bibnamefont {Lange-Dei}}, \bibinfo {author} {\bibfnamefont {B.~W.}\ \bibnamefont {Langley}}, \bibinfo {author} {\bibfnamefont {P.}~\bibnamefont {Laptev}}, \bibinfo {author} {\bibfnamefont {K.-M.}\ \bibnamefont {Lau}}, \bibinfo {author} {\bibfnamefont {L.}~\bibnamefont {Le~Guevel}}, \bibinfo {author} {\bibfnamefont {J.}~\bibnamefont {Ledford}}, \bibinfo {author} {\bibfnamefont {J.}~\bibnamefont {Lee}}, \bibinfo {author} {\bibfnamefont {K.}~\bibnamefont {Lee}}, \bibinfo {author} {\bibfnamefont {Y.~D.}\ \bibnamefont {Lensky}}, \bibinfo {author} {\bibfnamefont {S.}~\bibnamefont {Leon}}, \bibinfo {author} {\bibfnamefont {B.~J.}\ \bibnamefont {Lester}}, \bibinfo {author} {\bibfnamefont {W.~Y.}\ \bibnamefont {Li}}, \bibinfo {author} {\bibfnamefont {Y.}~\bibnamefont {Li}}, \bibinfo {author} {\bibfnamefont {A.~T.}\ \bibnamefont {Lill}}, \bibinfo {author}
  {\bibfnamefont {W.}~\bibnamefont {Liu}}, \bibinfo {author} {\bibfnamefont {W.~P.}\ \bibnamefont {Livingston}}, \bibinfo {author} {\bibfnamefont {A.}~\bibnamefont {Locharla}}, \bibinfo {author} {\bibfnamefont {E.}~\bibnamefont {Lucero}}, \bibinfo {author} {\bibfnamefont {D.}~\bibnamefont {Lundahl}}, \bibinfo {author} {\bibfnamefont {A.}~\bibnamefont {Lunt}}, \bibinfo {author} {\bibfnamefont {S.}~\bibnamefont {Madhuk}}, \bibinfo {author} {\bibfnamefont {F.~D.}\ \bibnamefont {Malone}}, \bibinfo {author} {\bibfnamefont {A.}~\bibnamefont {Maloney}}, \bibinfo {author} {\bibfnamefont {S.}~\bibnamefont {Mandr{\`a}}}, \bibinfo {author} {\bibfnamefont {J.}~\bibnamefont {Manyika}}, \bibinfo {author} {\bibfnamefont {L.~S.}\ \bibnamefont {Martin}}, \bibinfo {author} {\bibfnamefont {O.}~\bibnamefont {Martin}}, \bibinfo {author} {\bibfnamefont {S.}~\bibnamefont {Martin}}, \bibinfo {author} {\bibfnamefont {C.}~\bibnamefont {Maxfield}}, \bibinfo {author} {\bibfnamefont {J.~R.}\ \bibnamefont {McClean}}, \bibinfo {author}
  {\bibfnamefont {M.}~\bibnamefont {McEwen}}, \bibinfo {author} {\bibfnamefont {S.}~\bibnamefont {Meeks}}, \bibinfo {author} {\bibfnamefont {A.}~\bibnamefont {Megrant}}, \bibinfo {author} {\bibfnamefont {X.}~\bibnamefont {Mi}}, \bibinfo {author} {\bibfnamefont {K.~C.}\ \bibnamefont {Miao}}, \bibinfo {author} {\bibfnamefont {A.}~\bibnamefont {Mieszala}}, \bibinfo {author} {\bibfnamefont {R.}~\bibnamefont {Molavi}}, \bibinfo {author} {\bibfnamefont {S.}~\bibnamefont {Molina}}, \bibinfo {author} {\bibfnamefont {S.}~\bibnamefont {Montazeri}}, \bibinfo {author} {\bibfnamefont {A.}~\bibnamefont {Morvan}}, \bibinfo {author} {\bibfnamefont {R.}~\bibnamefont {Movassagh}}, \bibinfo {author} {\bibfnamefont {W.}~\bibnamefont {Mruczkiewicz}}, \bibinfo {author} {\bibfnamefont {O.}~\bibnamefont {Naaman}}, \bibinfo {author} {\bibfnamefont {M.}~\bibnamefont {Neeley}}, \bibinfo {author} {\bibfnamefont {C.}~\bibnamefont {Neill}}, \bibinfo {author} {\bibfnamefont {A.}~\bibnamefont {Nersisyan}}, \bibinfo {author} {\bibfnamefont
  {H.}~\bibnamefont {Neven}}, \bibinfo {author} {\bibfnamefont {M.}~\bibnamefont {Newman}}, \bibinfo {author} {\bibfnamefont {J.~H.}\ \bibnamefont {Ng}}, \bibinfo {author} {\bibfnamefont {A.}~\bibnamefont {Nguyen}}, \bibinfo {author} {\bibfnamefont {M.}~\bibnamefont {Nguyen}}, \bibinfo {author} {\bibfnamefont {C.-H.}\ \bibnamefont {Ni}}, \bibinfo {author} {\bibfnamefont {M.~Y.}\ \bibnamefont {Niu}}, \bibinfo {author} {\bibfnamefont {T.~E.}\ \bibnamefont {O'Brien}}, \bibinfo {author} {\bibfnamefont {W.~D.}\ \bibnamefont {Oliver}}, \bibinfo {author} {\bibfnamefont {A.}~\bibnamefont {Opremcak}}, \bibinfo {author} {\bibfnamefont {K.}~\bibnamefont {Ottosson}}, \bibinfo {author} {\bibfnamefont {A.}~\bibnamefont {Petukhov}}, \bibinfo {author} {\bibfnamefont {A.}~\bibnamefont {Pizzuto}}, \bibinfo {author} {\bibfnamefont {J.}~\bibnamefont {Platt}}, \bibinfo {author} {\bibfnamefont {R.}~\bibnamefont {Potter}}, \bibinfo {author} {\bibfnamefont {O.}~\bibnamefont {Pritchard}}, \bibinfo {author} {\bibfnamefont {L.~P.}\
  \bibnamefont {Pryadko}}, \bibinfo {author} {\bibfnamefont {C.}~\bibnamefont {Quintana}}, \bibinfo {author} {\bibfnamefont {G.}~\bibnamefont {Ramachandran}}, \bibinfo {author} {\bibfnamefont {M.~J.}\ \bibnamefont {Reagor}}, \bibinfo {author} {\bibfnamefont {J.}~\bibnamefont {Redding}}, \bibinfo {author} {\bibfnamefont {D.~M.}\ \bibnamefont {Rhodes}}, \bibinfo {author} {\bibfnamefont {G.}~\bibnamefont {Roberts}}, \bibinfo {author} {\bibfnamefont {E.}~\bibnamefont {Rosenberg}}, \bibinfo {author} {\bibfnamefont {E.}~\bibnamefont {Rosenfeld}}, \bibinfo {author} {\bibfnamefont {P.}~\bibnamefont {Roushan}}, \bibinfo {author} {\bibfnamefont {N.~C.}\ \bibnamefont {Rubin}}, \bibinfo {author} {\bibfnamefont {N.}~\bibnamefont {Saei}}, \bibinfo {author} {\bibfnamefont {D.}~\bibnamefont {Sank}}, \bibinfo {author} {\bibfnamefont {K.}~\bibnamefont {Sankaragomathi}}, \bibinfo {author} {\bibfnamefont {K.~J.}\ \bibnamefont {Satzinger}}, \bibinfo {author} {\bibfnamefont {H.~F.}\ \bibnamefont {Schurkus}}, \bibinfo {author}
  {\bibfnamefont {C.}~\bibnamefont {Schuster}}, \bibinfo {author} {\bibfnamefont {A.~W.}\ \bibnamefont {Senior}}, \bibinfo {author} {\bibfnamefont {M.~J.}\ \bibnamefont {Shearn}}, \bibinfo {author} {\bibfnamefont {A.}~\bibnamefont {Shorter}}, \bibinfo {author} {\bibfnamefont {N.}~\bibnamefont {Shutty}}, \bibinfo {author} {\bibfnamefont {V.}~\bibnamefont {Shvarts}}, \bibinfo {author} {\bibfnamefont {S.}~\bibnamefont {Singh}}, \bibinfo {author} {\bibfnamefont {V.}~\bibnamefont {Sivak}}, \bibinfo {author} {\bibfnamefont {J.}~\bibnamefont {Skruzny}}, \bibinfo {author} {\bibfnamefont {S.}~\bibnamefont {Small}}, \bibinfo {author} {\bibfnamefont {V.}~\bibnamefont {Smelyanskiy}}, \bibinfo {author} {\bibfnamefont {W.~C.}\ \bibnamefont {Smith}}, \bibinfo {author} {\bibfnamefont {R.~D.}\ \bibnamefont {Somma}}, \bibinfo {author} {\bibfnamefont {S.}~\bibnamefont {Springer}}, \bibinfo {author} {\bibfnamefont {G.}~\bibnamefont {Sterling}}, \bibinfo {author} {\bibfnamefont {D.}~\bibnamefont {Strain}}, \bibinfo {author}
  {\bibfnamefont {J.}~\bibnamefont {Suchard}}, \bibinfo {author} {\bibfnamefont {A.}~\bibnamefont {Szasz}}, \bibinfo {author} {\bibfnamefont {A.}~\bibnamefont {Sztein}}, \bibinfo {author} {\bibfnamefont {D.}~\bibnamefont {Thor}}, \bibinfo {author} {\bibfnamefont {A.}~\bibnamefont {Torres}}, \bibinfo {author} {\bibfnamefont {M.~M.}\ \bibnamefont {Torunbalci}}, \bibinfo {author} {\bibfnamefont {A.}~\bibnamefont {Vaishnav}}, \bibinfo {author} {\bibfnamefont {J.}~\bibnamefont {Vargas}}, \bibinfo {author} {\bibfnamefont {S.}~\bibnamefont {Vdovichev}}, \bibinfo {author} {\bibfnamefont {G.}~\bibnamefont {Vidal}}, \bibinfo {author} {\bibfnamefont {B.}~\bibnamefont {Villalonga}}, \bibinfo {author} {\bibfnamefont {C.~V.}\ \bibnamefont {Heidweiller}}, \bibinfo {author} {\bibfnamefont {S.}~\bibnamefont {Waltman}}, \bibinfo {author} {\bibfnamefont {S.~X.}\ \bibnamefont {Wang}}, \bibinfo {author} {\bibfnamefont {B.}~\bibnamefont {Ware}}, \bibinfo {author} {\bibfnamefont {K.}~\bibnamefont {Weber}}, \bibinfo {author}
  {\bibfnamefont {T.}~\bibnamefont {Weidel}}, \bibinfo {author} {\bibfnamefont {T.}~\bibnamefont {White}}, \bibinfo {author} {\bibfnamefont {K.}~\bibnamefont {Wong}}, \bibinfo {author} {\bibfnamefont {B.~W.~K.}\ \bibnamefont {Woo}}, \bibinfo {author} {\bibfnamefont {C.}~\bibnamefont {Xing}}, \bibinfo {author} {\bibfnamefont {Z.~J.}\ \bibnamefont {Yao}}, \bibinfo {author} {\bibfnamefont {P.}~\bibnamefont {Yeh}}, \bibinfo {author} {\bibfnamefont {B.}~\bibnamefont {Ying}}, \bibinfo {author} {\bibfnamefont {J.}~\bibnamefont {Yoo}}, \bibinfo {author} {\bibfnamefont {N.}~\bibnamefont {Yosri}}, \bibinfo {author} {\bibfnamefont {G.}~\bibnamefont {Young}}, \bibinfo {author} {\bibfnamefont {A.}~\bibnamefont {Zalcman}}, \bibinfo {author} {\bibfnamefont {Y.}~\bibnamefont {Zhang}}, \bibinfo {author} {\bibfnamefont {N.}~\bibnamefont {Zhu}}, \bibinfo {author} {\bibfnamefont {N.}~\bibnamefont {Zobrist}}, \bibinfo {author} {\bibfnamefont {G.~Q.}\ \bibnamefont {AI}},\ and\ \bibinfo {author} {\bibnamefont {{Collaborators}}},\
  }\bibfield  {title} {\bibinfo {title} {Quantum error correction below the surface code threshold},\ }\href {https://doi.org/10.1038/s41586-024-08449-y} {\bibfield  {journal} {\bibinfo  {journal} {Nature}\ }\textbf {\bibinfo {volume} {638}},\ \bibinfo {pages} {920} (\bibinfo {year} {2025})}\BibitemShut {NoStop}%
\bibitem [{\citenamefont {Jeffrey}\ \emph {et~al.}(2014)\citenamefont {Jeffrey}, \citenamefont {Sank}, \citenamefont {Mutus}, \citenamefont {White}, \citenamefont {Kelly}, \citenamefont {Barends}, \citenamefont {Chen}, \citenamefont {Chen}, \citenamefont {Chiaro}, \citenamefont {Dunsworth}, \citenamefont {Megrant}, \citenamefont {O'Malley}, \citenamefont {Neill}, \citenamefont {Roushan}, \citenamefont {Vainsencher}, \citenamefont {Wenner}, \citenamefont {Cleland},\ and\ \citenamefont {Martinis}}]{Jeffrey2014}%
  \BibitemOpen
  \bibfield  {author} {\bibinfo {author} {\bibfnamefont {E.}~\bibnamefont {Jeffrey}}, \bibinfo {author} {\bibfnamefont {D.}~\bibnamefont {Sank}}, \bibinfo {author} {\bibfnamefont {J.~Y.}\ \bibnamefont {Mutus}}, \bibinfo {author} {\bibfnamefont {T.~C.}\ \bibnamefont {White}}, \bibinfo {author} {\bibfnamefont {J.}~\bibnamefont {Kelly}}, \bibinfo {author} {\bibfnamefont {R.}~\bibnamefont {Barends}}, \bibinfo {author} {\bibfnamefont {Y.}~\bibnamefont {Chen}}, \bibinfo {author} {\bibfnamefont {Z.}~\bibnamefont {Chen}}, \bibinfo {author} {\bibfnamefont {B.}~\bibnamefont {Chiaro}}, \bibinfo {author} {\bibfnamefont {A.}~\bibnamefont {Dunsworth}}, \bibinfo {author} {\bibfnamefont {A.}~\bibnamefont {Megrant}}, \bibinfo {author} {\bibfnamefont {P.~J.~J.}\ \bibnamefont {O'Malley}}, \bibinfo {author} {\bibfnamefont {C.}~\bibnamefont {Neill}}, \bibinfo {author} {\bibfnamefont {P.}~\bibnamefont {Roushan}}, \bibinfo {author} {\bibfnamefont {A.}~\bibnamefont {Vainsencher}}, \bibinfo {author} {\bibfnamefont {J.}~\bibnamefont
  {Wenner}}, \bibinfo {author} {\bibfnamefont {A.~N.}\ \bibnamefont {Cleland}},\ and\ \bibinfo {author} {\bibfnamefont {J.~M.}\ \bibnamefont {Martinis}},\ }\bibfield  {title} {\bibinfo {title} {Fast accurate state measurement with superconducting qubits},\ }\href {https://doi.org/10.1103/PhysRevLett.112.190504} {\bibfield  {journal} {\bibinfo  {journal} {Phys. Rev. Lett.}\ }\textbf {\bibinfo {volume} {112}},\ \bibinfo {pages} {190504} (\bibinfo {year} {2014})}\BibitemShut {NoStop}%
\bibitem [{\citenamefont {Arute}\ \emph {et~al.}(2019)\citenamefont {Arute}, \citenamefont {Arya}, \citenamefont {Babbush}, \citenamefont {Bacon}, \citenamefont {Bardin}, \citenamefont {Barends}, \citenamefont {Biswas}, \citenamefont {Boixo}, \citenamefont {Brandao}, \citenamefont {Buell}, \citenamefont {Burkett}, \citenamefont {Chen}, \citenamefont {Chen}, \citenamefont {Chiaro}, \citenamefont {Collins}, \citenamefont {Courtney}, \citenamefont {Dunsworth}, \citenamefont {Farhi}, \citenamefont {Foxen}, \citenamefont {Fowler}, \citenamefont {Gidney}, \citenamefont {Giustina}, \citenamefont {Graff}, \citenamefont {Guerin}, \citenamefont {Habegger}, \citenamefont {Harrigan}, \citenamefont {Hartmann}, \citenamefont {Ho}, \citenamefont {Hoffmann}, \citenamefont {Huang}, \citenamefont {Humble}, \citenamefont {Isakov}, \citenamefont {Jeffrey}, \citenamefont {Jiang}, \citenamefont {Kafri}, \citenamefont {Kechedzhi}, \citenamefont {Kelly}, \citenamefont {Klimov}, \citenamefont {Knysh}, \citenamefont {Korotkov},
  \citenamefont {Kostritsa}, \citenamefont {Landhuis}, \citenamefont {Lindmark}, \citenamefont {Lucero}, \citenamefont {Lyakh}, \citenamefont {Mandr{\`a}}, \citenamefont {McClean}, \citenamefont {McEwen}, \citenamefont {Megrant}, \citenamefont {Mi}, \citenamefont {Michielsen}, \citenamefont {Mohseni}, \citenamefont {Mutus}, \citenamefont {Naaman}, \citenamefont {Neeley}, \citenamefont {Neill}, \citenamefont {Niu}, \citenamefont {Ostby}, \citenamefont {Petukhov}, \citenamefont {Platt}, \citenamefont {Quintana}, \citenamefont {Rieffel}, \citenamefont {Roushan}, \citenamefont {Rubin}, \citenamefont {Sank}, \citenamefont {Satzinger}, \citenamefont {Smelyanskiy}, \citenamefont {Sung}, \citenamefont {Trevithick}, \citenamefont {Vainsencher}, \citenamefont {Villalonga}, \citenamefont {White}, \citenamefont {Yao}, \citenamefont {Yeh}, \citenamefont {Zalcman}, \citenamefont {Neven},\ and\ \citenamefont {Martinis}}]{Arute2019}%
  \BibitemOpen
  \bibfield  {author} {\bibinfo {author} {\bibfnamefont {F.}~\bibnamefont {Arute}}, \bibinfo {author} {\bibfnamefont {K.}~\bibnamefont {Arya}}, \bibinfo {author} {\bibfnamefont {R.}~\bibnamefont {Babbush}}, \bibinfo {author} {\bibfnamefont {D.}~\bibnamefont {Bacon}}, \bibinfo {author} {\bibfnamefont {J.~C.}\ \bibnamefont {Bardin}}, \bibinfo {author} {\bibfnamefont {R.}~\bibnamefont {Barends}}, \bibinfo {author} {\bibfnamefont {R.}~\bibnamefont {Biswas}}, \bibinfo {author} {\bibfnamefont {S.}~\bibnamefont {Boixo}}, \bibinfo {author} {\bibfnamefont {F.~G. S.~L.}\ \bibnamefont {Brandao}}, \bibinfo {author} {\bibfnamefont {D.~A.}\ \bibnamefont {Buell}}, \bibinfo {author} {\bibfnamefont {B.}~\bibnamefont {Burkett}}, \bibinfo {author} {\bibfnamefont {Y.}~\bibnamefont {Chen}}, \bibinfo {author} {\bibfnamefont {Z.}~\bibnamefont {Chen}}, \bibinfo {author} {\bibfnamefont {B.}~\bibnamefont {Chiaro}}, \bibinfo {author} {\bibfnamefont {R.}~\bibnamefont {Collins}}, \bibinfo {author} {\bibfnamefont {W.}~\bibnamefont
  {Courtney}}, \bibinfo {author} {\bibfnamefont {A.}~\bibnamefont {Dunsworth}}, \bibinfo {author} {\bibfnamefont {E.}~\bibnamefont {Farhi}}, \bibinfo {author} {\bibfnamefont {B.}~\bibnamefont {Foxen}}, \bibinfo {author} {\bibfnamefont {A.}~\bibnamefont {Fowler}}, \bibinfo {author} {\bibfnamefont {C.}~\bibnamefont {Gidney}}, \bibinfo {author} {\bibfnamefont {M.}~\bibnamefont {Giustina}}, \bibinfo {author} {\bibfnamefont {R.}~\bibnamefont {Graff}}, \bibinfo {author} {\bibfnamefont {K.}~\bibnamefont {Guerin}}, \bibinfo {author} {\bibfnamefont {S.}~\bibnamefont {Habegger}}, \bibinfo {author} {\bibfnamefont {M.~P.}\ \bibnamefont {Harrigan}}, \bibinfo {author} {\bibfnamefont {M.~J.}\ \bibnamefont {Hartmann}}, \bibinfo {author} {\bibfnamefont {A.}~\bibnamefont {Ho}}, \bibinfo {author} {\bibfnamefont {M.}~\bibnamefont {Hoffmann}}, \bibinfo {author} {\bibfnamefont {T.}~\bibnamefont {Huang}}, \bibinfo {author} {\bibfnamefont {T.~S.}\ \bibnamefont {Humble}}, \bibinfo {author} {\bibfnamefont {S.~V.}\ \bibnamefont
  {Isakov}}, \bibinfo {author} {\bibfnamefont {E.}~\bibnamefont {Jeffrey}}, \bibinfo {author} {\bibfnamefont {Z.}~\bibnamefont {Jiang}}, \bibinfo {author} {\bibfnamefont {D.}~\bibnamefont {Kafri}}, \bibinfo {author} {\bibfnamefont {K.}~\bibnamefont {Kechedzhi}}, \bibinfo {author} {\bibfnamefont {J.}~\bibnamefont {Kelly}}, \bibinfo {author} {\bibfnamefont {P.~V.}\ \bibnamefont {Klimov}}, \bibinfo {author} {\bibfnamefont {S.}~\bibnamefont {Knysh}}, \bibinfo {author} {\bibfnamefont {A.}~\bibnamefont {Korotkov}}, \bibinfo {author} {\bibfnamefont {F.}~\bibnamefont {Kostritsa}}, \bibinfo {author} {\bibfnamefont {D.}~\bibnamefont {Landhuis}}, \bibinfo {author} {\bibfnamefont {M.}~\bibnamefont {Lindmark}}, \bibinfo {author} {\bibfnamefont {E.}~\bibnamefont {Lucero}}, \bibinfo {author} {\bibfnamefont {D.}~\bibnamefont {Lyakh}}, \bibinfo {author} {\bibfnamefont {S.}~\bibnamefont {Mandr{\`a}}}, \bibinfo {author} {\bibfnamefont {J.~R.}\ \bibnamefont {McClean}}, \bibinfo {author} {\bibfnamefont {M.}~\bibnamefont
  {McEwen}}, \bibinfo {author} {\bibfnamefont {A.}~\bibnamefont {Megrant}}, \bibinfo {author} {\bibfnamefont {X.}~\bibnamefont {Mi}}, \bibinfo {author} {\bibfnamefont {K.}~\bibnamefont {Michielsen}}, \bibinfo {author} {\bibfnamefont {M.}~\bibnamefont {Mohseni}}, \bibinfo {author} {\bibfnamefont {J.}~\bibnamefont {Mutus}}, \bibinfo {author} {\bibfnamefont {O.}~\bibnamefont {Naaman}}, \bibinfo {author} {\bibfnamefont {M.}~\bibnamefont {Neeley}}, \bibinfo {author} {\bibfnamefont {C.}~\bibnamefont {Neill}}, \bibinfo {author} {\bibfnamefont {M.~Y.}\ \bibnamefont {Niu}}, \bibinfo {author} {\bibfnamefont {E.}~\bibnamefont {Ostby}}, \bibinfo {author} {\bibfnamefont {A.}~\bibnamefont {Petukhov}}, \bibinfo {author} {\bibfnamefont {J.~C.}\ \bibnamefont {Platt}}, \bibinfo {author} {\bibfnamefont {C.}~\bibnamefont {Quintana}}, \bibinfo {author} {\bibfnamefont {E.~G.}\ \bibnamefont {Rieffel}}, \bibinfo {author} {\bibfnamefont {P.}~\bibnamefont {Roushan}}, \bibinfo {author} {\bibfnamefont {N.~C.}\ \bibnamefont {Rubin}},
  \bibinfo {author} {\bibfnamefont {D.}~\bibnamefont {Sank}}, \bibinfo {author} {\bibfnamefont {K.~J.}\ \bibnamefont {Satzinger}}, \bibinfo {author} {\bibfnamefont {V.}~\bibnamefont {Smelyanskiy}}, \bibinfo {author} {\bibfnamefont {K.~J.}\ \bibnamefont {Sung}}, \bibinfo {author} {\bibfnamefont {M.~D.}\ \bibnamefont {Trevithick}}, \bibinfo {author} {\bibfnamefont {A.}~\bibnamefont {Vainsencher}}, \bibinfo {author} {\bibfnamefont {B.}~\bibnamefont {Villalonga}}, \bibinfo {author} {\bibfnamefont {T.}~\bibnamefont {White}}, \bibinfo {author} {\bibfnamefont {Z.~J.}\ \bibnamefont {Yao}}, \bibinfo {author} {\bibfnamefont {P.}~\bibnamefont {Yeh}}, \bibinfo {author} {\bibfnamefont {A.}~\bibnamefont {Zalcman}}, \bibinfo {author} {\bibfnamefont {H.}~\bibnamefont {Neven}},\ and\ \bibinfo {author} {\bibfnamefont {J.~M.}\ \bibnamefont {Martinis}},\ }\bibfield  {title} {\bibinfo {title} {Quantum supremacy using a programmable superconducting processor},\ }\href {https://doi.org/10.1038/s41586-019-1666-5} {\bibfield
  {journal} {\bibinfo  {journal} {Nature}\ }\textbf {\bibinfo {volume} {574}},\ \bibinfo {pages} {505} (\bibinfo {year} {2019})}\BibitemShut {NoStop}%
\bibitem [{Goo(2023)}]{Google2023suppressing}%
  \BibitemOpen
  \bibfield  {title} {\bibinfo {title} {Suppressing quantum errors by scaling a surface code logical qubit},\ }\href@noop {} {\bibfield  {journal} {\bibinfo  {journal} {Nature}\ }\textbf {\bibinfo {volume} {614}},\ \bibinfo {pages} {676} (\bibinfo {year} {2023})}\BibitemShut {NoStop}%
\bibitem [{\citenamefont {Delfosse}\ and\ \citenamefont {Nickerson}(2021)}]{Delfosse2021UF}%
  \BibitemOpen
  \bibfield  {author} {\bibinfo {author} {\bibfnamefont {N.}~\bibnamefont {Delfosse}}\ and\ \bibinfo {author} {\bibfnamefont {N.~H.}\ \bibnamefont {Nickerson}},\ }\bibfield  {title} {\bibinfo {title} {Almost-linear time decoding algorithm for topological codes},\ }\href {https://doi.org/10.22331/q-2021-12-02-595} {\bibfield  {journal} {\bibinfo  {journal} {{Quantum}}\ }\textbf {\bibinfo {volume} {5}},\ \bibinfo {pages} {595} (\bibinfo {year} {2021})}\BibitemShut {NoStop}%
\bibitem [{\citenamefont {Heer}\ \emph {et~al.}(2023)\citenamefont {Heer}, \citenamefont {Sozzo}, \citenamefont {Fujii},\ and\ \citenamefont {Sano}}]{Heer2023}%
  \BibitemOpen
  \bibfield  {author} {\bibinfo {author} {\bibfnamefont {M.~J.}\ \bibnamefont {Heer}}, \bibinfo {author} {\bibfnamefont {E.~D.}\ \bibnamefont {Sozzo}}, \bibinfo {author} {\bibfnamefont {K.}~\bibnamefont {Fujii}},\ and\ \bibinfo {author} {\bibfnamefont {K.}~\bibnamefont {Sano}},\ }\bibfield  {title} {\bibinfo {title} {Novel union-find-based decoders for scalable quantum error correction on systolic arrays},\ }in\ \href {https://doi.org/10.1109/IPDPSW59300.2023.00092} {\emph {\bibinfo {booktitle} {2023 IEEE International Parallel and Distributed Processing Symposium Workshops (IPDPSW)}}}\ (\bibinfo {year} {2023})\ pp.\ \bibinfo {pages} {524--533}\BibitemShut {NoStop}%
\bibitem [{\citenamefont {Chan}\ and\ \citenamefont {Benjamin}(2023)}]{Chan2023actis}%
  \BibitemOpen
  \bibfield  {author} {\bibinfo {author} {\bibfnamefont {T.}~\bibnamefont {Chan}}\ and\ \bibinfo {author} {\bibfnamefont {S.~C.}\ \bibnamefont {Benjamin}},\ }\bibfield  {title} {\bibinfo {title} {Actis: {A} {S}trictly {L}ocal {U}nion–{F}ind {D}ecoder},\ }\href {https://doi.org/10.22331/q-2023-11-14-1183} {\bibfield  {journal} {\bibinfo  {journal} {{Quantum}}\ }\textbf {\bibinfo {volume} {7}},\ \bibinfo {pages} {1183} (\bibinfo {year} {2023})}\BibitemShut {NoStop}%
\bibitem [{\citenamefont {Liyanage}\ \emph {et~al.}(2024)\citenamefont {Liyanage}, \citenamefont {Wu}, \citenamefont {Tagare},\ and\ \citenamefont {Zhong}}]{Liyanage2024FPGA}%
  \BibitemOpen
  \bibfield  {author} {\bibinfo {author} {\bibfnamefont {N.}~\bibnamefont {Liyanage}}, \bibinfo {author} {\bibfnamefont {Y.}~\bibnamefont {Wu}}, \bibinfo {author} {\bibfnamefont {S.}~\bibnamefont {Tagare}},\ and\ \bibinfo {author} {\bibfnamefont {L.}~\bibnamefont {Zhong}},\ }\bibfield  {title} {\bibinfo {title} {Fpga-based distributed union-find decoder for surface codes},\ }\href {https://doi.org/10.1109/TQE.2024.3467271} {\bibfield  {journal} {\bibinfo  {journal} {IEEE Transactions on Quantum Engineering}\ }\textbf {\bibinfo {volume} {5}},\ \bibinfo {pages} {1} (\bibinfo {year} {2024})}\BibitemShut {NoStop}%
\bibitem [{\citenamefont {Liyanage}\ \emph {et~al.}(2025)\citenamefont {Liyanage}, \citenamefont {Wu}, \citenamefont {Houghton},\ and\ \citenamefont {Zhong}}]{Liyanage2025}%
  \BibitemOpen
  \bibfield  {author} {\bibinfo {author} {\bibfnamefont {N.}~\bibnamefont {Liyanage}}, \bibinfo {author} {\bibfnamefont {Y.}~\bibnamefont {Wu}}, \bibinfo {author} {\bibfnamefont {E.}~\bibnamefont {Houghton}},\ and\ \bibinfo {author} {\bibfnamefont {L.}~\bibnamefont {Zhong}},\ }\href {https://arxiv.org/abs/2504.11805} {\bibinfo {title} {Network-integrated decoding system for real-time quantum error correction with lattice surgery}} (\bibinfo {year} {2025}),\ \Eprint {https://arxiv.org/abs/2504.11805} {arXiv:2504.11805 [quant-ph]} \BibitemShut {NoStop}%
\bibitem [{\citenamefont {Barber}\ \emph {et~al.}(2025)\citenamefont {Barber}, \citenamefont {Barnes}, \citenamefont {Bialas}, \citenamefont {Bu{\u{g}}dayc{\i}}, \citenamefont {Campbell}, \citenamefont {Gillespie}, \citenamefont {Johar}, \citenamefont {Rajan}, \citenamefont {Richardson}, \citenamefont {Skoric}, \citenamefont {Topal}, \citenamefont {Turner},\ and\ \citenamefont {Ziad}}]{Barber2025}%
  \BibitemOpen
  \bibfield  {author} {\bibinfo {author} {\bibfnamefont {B.}~\bibnamefont {Barber}}, \bibinfo {author} {\bibfnamefont {K.~M.}\ \bibnamefont {Barnes}}, \bibinfo {author} {\bibfnamefont {T.}~\bibnamefont {Bialas}}, \bibinfo {author} {\bibfnamefont {O.}~\bibnamefont {Bu{\u{g}}dayc{\i}}}, \bibinfo {author} {\bibfnamefont {E.~T.}\ \bibnamefont {Campbell}}, \bibinfo {author} {\bibfnamefont {N.~I.}\ \bibnamefont {Gillespie}}, \bibinfo {author} {\bibfnamefont {K.}~\bibnamefont {Johar}}, \bibinfo {author} {\bibfnamefont {R.}~\bibnamefont {Rajan}}, \bibinfo {author} {\bibfnamefont {A.~W.}\ \bibnamefont {Richardson}}, \bibinfo {author} {\bibfnamefont {L.}~\bibnamefont {Skoric}}, \bibinfo {author} {\bibfnamefont {C.}~\bibnamefont {Topal}}, \bibinfo {author} {\bibfnamefont {M.~L.}\ \bibnamefont {Turner}},\ and\ \bibinfo {author} {\bibfnamefont {A.~B.}\ \bibnamefont {Ziad}},\ }\bibfield  {title} {\bibinfo {title} {A real-time, scalable, fast and resource-efficient decoder for a quantum computer},\ }\href
  {https://doi.org/10.1038/s41928-024-01319-5} {\bibfield  {journal} {\bibinfo  {journal} {Nature Electronics}\ }\textbf {\bibinfo {volume} {8}},\ \bibinfo {pages} {84} (\bibinfo {year} {2025})}\BibitemShut {NoStop}%
\bibitem [{\citenamefont {Huang}\ \emph {et~al.}(2020)\citenamefont {Huang}, \citenamefont {Newman},\ and\ \citenamefont {Brown}}]{Huang2020weightedUF}%
  \BibitemOpen
  \bibfield  {author} {\bibinfo {author} {\bibfnamefont {S.}~\bibnamefont {Huang}}, \bibinfo {author} {\bibfnamefont {M.}~\bibnamefont {Newman}},\ and\ \bibinfo {author} {\bibfnamefont {K.~R.}\ \bibnamefont {Brown}},\ }\bibfield  {title} {\bibinfo {title} {Fault-tolerant weighted union-find decoding on the toric code},\ }\href {https://doi.org/10.1103/PhysRevA.102.012419} {\bibfield  {journal} {\bibinfo  {journal} {Phys. Rev. A}\ }\textbf {\bibinfo {volume} {102}},\ \bibinfo {pages} {012419} (\bibinfo {year} {2020})}\BibitemShut {NoStop}%
\bibitem [{\citenamefont {Overwater}\ \emph {et~al.}(2022)\citenamefont {Overwater}, \citenamefont {Babaie},\ and\ \citenamefont {Sebastiano}}]{Overwater2022}%
  \BibitemOpen
  \bibfield  {author} {\bibinfo {author} {\bibfnamefont {R.~W.~J.}\ \bibnamefont {Overwater}}, \bibinfo {author} {\bibfnamefont {M.}~\bibnamefont {Babaie}},\ and\ \bibinfo {author} {\bibfnamefont {F.}~\bibnamefont {Sebastiano}},\ }\bibfield  {title} {\bibinfo {title} {Neural-network decoders for quantum error correction using surface codes: A space exploration of the hardware cost-performance tradeoffs},\ }\href {https://doi.org/10.1109/TQE.2022.3174017} {\bibfield  {journal} {\bibinfo  {journal} {IEEE Transactions on Quantum Engineering}\ }\textbf {\bibinfo {volume} {3}},\ \bibinfo {pages} {1} (\bibinfo {year} {2022})}\BibitemShut {NoStop}%
\bibitem [{\citenamefont {Liao}\ \emph {et~al.}(2023)\citenamefont {Liao}, \citenamefont {Suzuki}, \citenamefont {Tanimoto}, \citenamefont {Ueno},\ and\ \citenamefont {Tokunaga}}]{Liao2023}%
  \BibitemOpen
  \bibfield  {author} {\bibinfo {author} {\bibfnamefont {W.}~\bibnamefont {Liao}}, \bibinfo {author} {\bibfnamefont {Y.}~\bibnamefont {Suzuki}}, \bibinfo {author} {\bibfnamefont {T.}~\bibnamefont {Tanimoto}}, \bibinfo {author} {\bibfnamefont {Y.}~\bibnamefont {Ueno}},\ and\ \bibinfo {author} {\bibfnamefont {Y.}~\bibnamefont {Tokunaga}},\ }\bibfield  {title} {\bibinfo {title} {Wit-greedy: Hardware system design of weighted iterative greedy decoder for surface code},\ }in\ \href {https://doi.org/10.1145/3566097.3567933} {\emph {\bibinfo {booktitle} {Proceedings of the 28th Asia and South Pacific Design Automation Conference}}},\ \bibinfo {series and number} {ASPDAC '23}\ (\bibinfo  {publisher} {Association for Computing Machinery},\ \bibinfo {address} {New York, NY, USA},\ \bibinfo {year} {2023})\ p.\ \bibinfo {pages} {209–215}\BibitemShut {NoStop}%
\bibitem [{\citenamefont {Das}\ \emph {et~al.}(2022)\citenamefont {Das}, \citenamefont {Locharla},\ and\ \citenamefont {Jones}}]{Das2022}%
  \BibitemOpen
  \bibfield  {author} {\bibinfo {author} {\bibfnamefont {P.}~\bibnamefont {Das}}, \bibinfo {author} {\bibfnamefont {A.}~\bibnamefont {Locharla}},\ and\ \bibinfo {author} {\bibfnamefont {C.}~\bibnamefont {Jones}},\ }\bibfield  {title} {\bibinfo {title} {Lilliput: a lightweight low-latency lookup-table decoder for near-term quantum error correction},\ }in\ \href {https://doi.org/10.1145/3503222.3507707} {\emph {\bibinfo {booktitle} {Proceedings of the 27th ACM International Conference on Architectural Support for Programming Languages and Operating Systems}}},\ \bibinfo {series and number} {ASPLOS '22}\ (\bibinfo  {publisher} {Association for Computing Machinery},\ \bibinfo {address} {New York, NY, USA},\ \bibinfo {year} {2022})\ p.\ \bibinfo {pages} {541–553}\BibitemShut {NoStop}%
\bibitem [{\citenamefont {Vittal}\ \emph {et~al.}(2023)\citenamefont {Vittal}, \citenamefont {Das},\ and\ \citenamefont {Qureshi}}]{Vittal2023Astrea}%
  \BibitemOpen
  \bibfield  {author} {\bibinfo {author} {\bibfnamefont {S.}~\bibnamefont {Vittal}}, \bibinfo {author} {\bibfnamefont {P.}~\bibnamefont {Das}},\ and\ \bibinfo {author} {\bibfnamefont {M.}~\bibnamefont {Qureshi}},\ }\bibfield  {title} {\bibinfo {title} {Astrea: Accurate quantum error-decoding via practical minimum-weight perfect-matching},\ }in\ \href {https://doi.org/10.1145/3579371.3589037} {\emph {\bibinfo {booktitle} {Proceedings of the 50th Annual International Symposium on Computer Architecture}}},\ \bibinfo {series and number} {ISCA '23}\ (\bibinfo  {publisher} {Association for Computing Machinery},\ \bibinfo {address} {New York, NY, USA},\ \bibinfo {year} {2023})\BibitemShut {NoStop}%
\bibitem [{\citenamefont {Gidney}\ \emph {et~al.}(2025)\citenamefont {Gidney}, \citenamefont {Newman}, \citenamefont {Brooks},\ and\ \citenamefont {Jones}}]{Gidney2025yoked}%
  \BibitemOpen
  \bibfield  {author} {\bibinfo {author} {\bibfnamefont {C.}~\bibnamefont {Gidney}}, \bibinfo {author} {\bibfnamefont {M.}~\bibnamefont {Newman}}, \bibinfo {author} {\bibfnamefont {P.}~\bibnamefont {Brooks}},\ and\ \bibinfo {author} {\bibfnamefont {C.}~\bibnamefont {Jones}},\ }\bibfield  {title} {\bibinfo {title} {Yoked surface codes},\ }\href {https://doi.org/10.1038/s41467-025-59714-1} {\bibfield  {journal} {\bibinfo  {journal} {Nature Communications}\ }\textbf {\bibinfo {volume} {16}},\ \bibinfo {pages} {4498} (\bibinfo {year} {2025})}\BibitemShut {NoStop}%
\bibitem [{\citenamefont {Meister}\ \emph {et~al.}(2024)\citenamefont {Meister}, \citenamefont {Pattison},\ and\ \citenamefont {Preskill}}]{Meister2024}%
  \BibitemOpen
  \bibfield  {author} {\bibinfo {author} {\bibfnamefont {N.}~\bibnamefont {Meister}}, \bibinfo {author} {\bibfnamefont {C.~A.}\ \bibnamefont {Pattison}},\ and\ \bibinfo {author} {\bibfnamefont {J.}~\bibnamefont {Preskill}},\ }\href {https://arxiv.org/abs/2405.07433} {\bibinfo {title} {Efficient soft-output decoders for the surface code}} (\bibinfo {year} {2024}),\ \Eprint {https://arxiv.org/abs/2405.07433} {arXiv:2405.07433 [quant-ph]} \BibitemShut {NoStop}%
\bibitem [{\citenamefont {Lee}\ \emph {et~al.}(2025)\citenamefont {Lee}, \citenamefont {English},\ and\ \citenamefont {Bartlett}}]{Lee2025soft}%
  \BibitemOpen
  \bibfield  {author} {\bibinfo {author} {\bibfnamefont {S.-H.}\ \bibnamefont {Lee}}, \bibinfo {author} {\bibfnamefont {L.}~\bibnamefont {English}},\ and\ \bibinfo {author} {\bibfnamefont {S.~D.}\ \bibnamefont {Bartlett}},\ }\href {https://arxiv.org/abs/2510.05795} {\bibinfo {title} {Efficient post-selection for general quantum ldpc codes}} (\bibinfo {year} {2025}),\ \Eprint {https://arxiv.org/abs/2510.05795} {arXiv:2510.05795 [quant-ph]} \BibitemShut {NoStop}%
\bibitem [{Kis()}]{Kishi2025}%
  \BibitemOpen
  \href@noop {} {}\bibinfo {note} {K. Kishi, R. Toshio and J. Fujisaki and H. Oshima and S. Sato and K. Fujii, ``Early Stopping for Fast Soft-Output Calculation in Cluster-Based Decoders". To appear.}\BibitemShut {Stop}%
\bibitem [{\citenamefont {Hutter}\ \emph {et~al.}(2014)\citenamefont {Hutter}, \citenamefont {Wootton},\ and\ \citenamefont {Loss}}]{Hutter2014}%
  \BibitemOpen
  \bibfield  {author} {\bibinfo {author} {\bibfnamefont {A.}~\bibnamefont {Hutter}}, \bibinfo {author} {\bibfnamefont {J.~R.}\ \bibnamefont {Wootton}},\ and\ \bibinfo {author} {\bibfnamefont {D.}~\bibnamefont {Loss}},\ }\bibfield  {title} {\bibinfo {title} {Efficient markov chain monte carlo algorithm for the surface code},\ }\href {https://doi.org/10.1103/PhysRevA.89.022326} {\bibfield  {journal} {\bibinfo  {journal} {Phys. Rev. A}\ }\textbf {\bibinfo {volume} {89}},\ \bibinfo {pages} {022326} (\bibinfo {year} {2014})}\BibitemShut {NoStop}%
\bibitem [{\citenamefont {Bomb\'{\i}n}\ \emph {et~al.}(2024)\citenamefont {Bomb\'{\i}n}, \citenamefont {Pant}, \citenamefont {Roberts},\ and\ \citenamefont {Seetharam}}]{Bombin2024}%
  \BibitemOpen
  \bibfield  {author} {\bibinfo {author} {\bibfnamefont {H.}~\bibnamefont {Bomb\'{\i}n}}, \bibinfo {author} {\bibfnamefont {M.}~\bibnamefont {Pant}}, \bibinfo {author} {\bibfnamefont {S.}~\bibnamefont {Roberts}},\ and\ \bibinfo {author} {\bibfnamefont {K.~I.}\ \bibnamefont {Seetharam}},\ }\bibfield  {title} {\bibinfo {title} {Fault-tolerant postselection for low-overhead magic state preparation},\ }\href {https://doi.org/10.1103/PRXQuantum.5.010302} {\bibfield  {journal} {\bibinfo  {journal} {PRX Quantum}\ }\textbf {\bibinfo {volume} {5}},\ \bibinfo {pages} {010302} (\bibinfo {year} {2024})}\BibitemShut {NoStop}%
\bibitem [{\citenamefont {Gidney}\ \emph {et~al.}(2024)\citenamefont {Gidney}, \citenamefont {Shutty},\ and\ \citenamefont {Jones}}]{Gidney2024cultivation}%
  \BibitemOpen
  \bibfield  {author} {\bibinfo {author} {\bibfnamefont {C.}~\bibnamefont {Gidney}}, \bibinfo {author} {\bibfnamefont {N.}~\bibnamefont {Shutty}},\ and\ \bibinfo {author} {\bibfnamefont {C.}~\bibnamefont {Jones}},\ }\href {https://arxiv.org/abs/2409.17595} {\bibinfo {title} {Magic state cultivation: growing t states as cheap as cnot gates}} (\bibinfo {year} {2024}),\ \Eprint {https://arxiv.org/abs/2409.17595} {arXiv:2409.17595 [quant-ph]} \BibitemShut {NoStop}%
\bibitem [{\citenamefont {Smith}\ \emph {et~al.}(2024)\citenamefont {Smith}, \citenamefont {Brown},\ and\ \citenamefont {Bartlett}}]{Smith2024mitigation}%
  \BibitemOpen
  \bibfield  {author} {\bibinfo {author} {\bibfnamefont {S.~C.}\ \bibnamefont {Smith}}, \bibinfo {author} {\bibfnamefont {B.~J.}\ \bibnamefont {Brown}},\ and\ \bibinfo {author} {\bibfnamefont {S.~D.}\ \bibnamefont {Bartlett}},\ }\bibfield  {title} {\bibinfo {title} {Mitigating errors in logical qubits},\ }\href {https://doi.org/10.1038/s42005-024-01883-4} {\bibfield  {journal} {\bibinfo  {journal} {Communications Physics}\ }\textbf {\bibinfo {volume} {7}},\ \bibinfo {pages} {386} (\bibinfo {year} {2024})}\BibitemShut {NoStop}%
\bibitem [{\citenamefont {Varbanov}\ \emph {et~al.}(2025)\citenamefont {Varbanov}, \citenamefont {Serra-Peralta}, \citenamefont {Byfield},\ and\ \citenamefont {Terhal}}]{Varbanov2025}%
  \BibitemOpen
  \bibfield  {author} {\bibinfo {author} {\bibfnamefont {B.~M.}\ \bibnamefont {Varbanov}}, \bibinfo {author} {\bibfnamefont {M.}~\bibnamefont {Serra-Peralta}}, \bibinfo {author} {\bibfnamefont {D.}~\bibnamefont {Byfield}},\ and\ \bibinfo {author} {\bibfnamefont {B.~M.}\ \bibnamefont {Terhal}},\ }\bibfield  {title} {\bibinfo {title} {Neural network decoder for near-term surface-code experiments},\ }\href {https://doi.org/10.1103/PhysRevResearch.7.013029} {\bibfield  {journal} {\bibinfo  {journal} {Phys. Rev. Res.}\ }\textbf {\bibinfo {volume} {7}},\ \bibinfo {pages} {013029} (\bibinfo {year} {2025})}\BibitemShut {NoStop}%
\bibitem [{\citenamefont {Hu}\ \emph {et~al.}(2025)\citenamefont {Hu}, \citenamefont {Ouyang}, \citenamefont {Lu}, \citenamefont {Lin},\ and\ \citenamefont {Zhong}}]{Hu2025}%
  \BibitemOpen
  \bibfield  {author} {\bibinfo {author} {\bibfnamefont {G.}~\bibnamefont {Hu}}, \bibinfo {author} {\bibfnamefont {W.}~\bibnamefont {Ouyang}}, \bibinfo {author} {\bibfnamefont {C.-Y.}\ \bibnamefont {Lu}}, \bibinfo {author} {\bibfnamefont {C.}~\bibnamefont {Lin}},\ and\ \bibinfo {author} {\bibfnamefont {H.-S.}\ \bibnamefont {Zhong}},\ }\href {https://arxiv.org/abs/2502.19971} {\bibinfo {title} {Efficient and universal neural-network decoder for stabilizer-based quantum error correction}} (\bibinfo {year} {2025}),\ \Eprint {https://arxiv.org/abs/2502.19971} {arXiv:2502.19971 [quant-ph]} \BibitemShut {NoStop}%
\bibitem [{\citenamefont {Blue}\ \emph {et~al.}(2025)\citenamefont {Blue}, \citenamefont {Avlani}, \citenamefont {He}, \citenamefont {Ziyin},\ and\ \citenamefont {Chuang}}]{Blue2025}%
  \BibitemOpen
  \bibfield  {author} {\bibinfo {author} {\bibfnamefont {J.}~\bibnamefont {Blue}}, \bibinfo {author} {\bibfnamefont {H.}~\bibnamefont {Avlani}}, \bibinfo {author} {\bibfnamefont {Z.}~\bibnamefont {He}}, \bibinfo {author} {\bibfnamefont {L.}~\bibnamefont {Ziyin}},\ and\ \bibinfo {author} {\bibfnamefont {I.~L.}\ \bibnamefont {Chuang}},\ }\href {https://arxiv.org/abs/2504.13043} {\bibinfo {title} {Machine learning decoding of circuit-level noise for bivariate bicycle codes}} (\bibinfo {year} {2025}),\ \Eprint {https://arxiv.org/abs/2504.13043} {arXiv:2504.13043 [quant-ph]} \BibitemShut {NoStop}%
\bibitem [{\citenamefont {Higgott}\ \emph {et~al.}(2023)\citenamefont {Higgott}, \citenamefont {Bohdanowicz}, \citenamefont {Kubica}, \citenamefont {Flammia},\ and\ \citenamefont {Campbell}}]{Higgott2023belief_matching}%
  \BibitemOpen
  \bibfield  {author} {\bibinfo {author} {\bibfnamefont {O.}~\bibnamefont {Higgott}}, \bibinfo {author} {\bibfnamefont {T.~C.}\ \bibnamefont {Bohdanowicz}}, \bibinfo {author} {\bibfnamefont {A.}~\bibnamefont {Kubica}}, \bibinfo {author} {\bibfnamefont {S.~T.}\ \bibnamefont {Flammia}},\ and\ \bibinfo {author} {\bibfnamefont {E.~T.}\ \bibnamefont {Campbell}},\ }\bibfield  {title} {\bibinfo {title} {Improved decoding of circuit noise and fragile boundaries of tailored surface codes},\ }\href {https://doi.org/10.1103/PhysRevX.13.031007} {\bibfield  {journal} {\bibinfo  {journal} {Phys. Rev. X}\ }\textbf {\bibinfo {volume} {13}},\ \bibinfo {pages} {031007} (\bibinfo {year} {2023})}\BibitemShut {NoStop}%
\bibitem [{\citenamefont {Iyer}\ and\ \citenamefont {Poulin}(2015)}]{Iyer2015}%
  \BibitemOpen
  \bibfield  {author} {\bibinfo {author} {\bibfnamefont {P.}~\bibnamefont {Iyer}}\ and\ \bibinfo {author} {\bibfnamefont {D.}~\bibnamefont {Poulin}},\ }\bibfield  {title} {\bibinfo {title} {Hardness of decoding quantum stabilizer codes},\ }\href {https://doi.org/10.1109/TIT.2015.2422294} {\bibfield  {journal} {\bibinfo  {journal} {IEEE Transactions on Information Theory}\ }\textbf {\bibinfo {volume} {61}},\ \bibinfo {pages} {5209} (\bibinfo {year} {2015})}\BibitemShut {NoStop}%
\bibitem [{\citenamefont {Poulin}(2006)}]{Poulin2006}%
  \BibitemOpen
  \bibfield  {author} {\bibinfo {author} {\bibfnamefont {D.}~\bibnamefont {Poulin}},\ }\bibfield  {title} {\bibinfo {title} {Optimal and efficient decoding of concatenated quantum block codes},\ }\href {https://doi.org/10.1103/PhysRevA.74.052333} {\bibfield  {journal} {\bibinfo  {journal} {Phys. Rev. A}\ }\textbf {\bibinfo {volume} {74}},\ \bibinfo {pages} {052333} (\bibinfo {year} {2006})}\BibitemShut {NoStop}%
\bibitem [{\citenamefont {Edmonds}(1965{\natexlab{a}})}]{Edmonds1965_1}%
  \BibitemOpen
  \bibfield  {author} {\bibinfo {author} {\bibfnamefont {J.}~\bibnamefont {Edmonds}},\ }\bibfield  {title} {\bibinfo {title} {Paths, trees, and flowers},\ }\href {https://doi.org/10.4153/CJM-1965-045-4} {\bibfield  {journal} {\bibinfo  {journal} {Canadian Journal of Mathematics}\ }\textbf {\bibinfo {volume} {17}},\ \bibinfo {pages} {449–467} (\bibinfo {year} {1965}{\natexlab{a}})}\BibitemShut {NoStop}%
\bibitem [{\citenamefont {Edmonds}(1965{\natexlab{b}})}]{Edmonds1965_2}%
  \BibitemOpen
  \bibfield  {author} {\bibinfo {author} {\bibfnamefont {J.}~\bibnamefont {Edmonds}},\ }\bibfield  {title} {\bibinfo {title} {Maximum matching and a polyhedron with 0,1-vertices},\ }\href {https://api.semanticscholar.org/CorpusID:15379135} {\bibfield  {journal} {\bibinfo  {journal} {Journal of Research of the National Bureau of Standards Section B Mathematics and Mathematical Physics}\ ,\ \bibinfo {pages} {125}} (\bibinfo {year} {1965}{\natexlab{b}})}\BibitemShut {NoStop}%
\bibitem [{\citenamefont {Riesebos}\ \emph {et~al.}(2017)\citenamefont {Riesebos}, \citenamefont {Fu}, \citenamefont {Varsamopoulos}, \citenamefont {Almudever},\ and\ \citenamefont {Bertels}}]{Riesebos2017}%
  \BibitemOpen
  \bibfield  {author} {\bibinfo {author} {\bibfnamefont {L.}~\bibnamefont {Riesebos}}, \bibinfo {author} {\bibfnamefont {X.}~\bibnamefont {Fu}}, \bibinfo {author} {\bibfnamefont {S.}~\bibnamefont {Varsamopoulos}}, \bibinfo {author} {\bibfnamefont {C.~G.}\ \bibnamefont {Almudever}},\ and\ \bibinfo {author} {\bibfnamefont {K.}~\bibnamefont {Bertels}},\ }\bibfield  {title} {\bibinfo {title} {Pauli frames for quantum computer architectures},\ }in\ \href {https://doi.org/10.1145/3061639.3062300} {\emph {\bibinfo {booktitle} {Proceedings of the 54th Annual Design Automation Conference 2017}}},\ \bibinfo {series and number} {DAC '17}\ (\bibinfo  {publisher} {Association for Computing Machinery},\ \bibinfo {address} {New York, NY, USA},\ \bibinfo {year} {2017})\BibitemShut {NoStop}%
\bibitem [{\citenamefont {Delfosse}\ \emph {et~al.}(2023)\citenamefont {Delfosse}, \citenamefont {Paz}, \citenamefont {Vaschillo},\ and\ \citenamefont {Svore}}]{Delfosse2023tradeoff}%
  \BibitemOpen
  \bibfield  {author} {\bibinfo {author} {\bibfnamefont {N.}~\bibnamefont {Delfosse}}, \bibinfo {author} {\bibfnamefont {A.}~\bibnamefont {Paz}}, \bibinfo {author} {\bibfnamefont {A.}~\bibnamefont {Vaschillo}},\ and\ \bibinfo {author} {\bibfnamefont {K.~M.}\ \bibnamefont {Svore}},\ }\href {https://arxiv.org/abs/2310.15313} {\bibinfo {title} {How to choose a decoder for a fault-tolerant quantum computer? the speed vs accuracy trade-off}} (\bibinfo {year} {2023}),\ \Eprint {https://arxiv.org/abs/2310.15313} {arXiv:2310.15313 [quant-ph]} \BibitemShut {NoStop}%
\bibitem [{\citenamefont {Tan}\ \emph {et~al.}(2023)\citenamefont {Tan}, \citenamefont {Zhang}, \citenamefont {Chao}, \citenamefont {Shi},\ and\ \citenamefont {Chen}}]{Tan2023}%
  \BibitemOpen
  \bibfield  {author} {\bibinfo {author} {\bibfnamefont {X.}~\bibnamefont {Tan}}, \bibinfo {author} {\bibfnamefont {F.}~\bibnamefont {Zhang}}, \bibinfo {author} {\bibfnamefont {R.}~\bibnamefont {Chao}}, \bibinfo {author} {\bibfnamefont {Y.}~\bibnamefont {Shi}},\ and\ \bibinfo {author} {\bibfnamefont {J.}~\bibnamefont {Chen}},\ }\bibfield  {title} {\bibinfo {title} {Scalable surface-code decoders with parallelization in time},\ }\href {https://doi.org/10.1103/PRXQuantum.4.040344} {\bibfield  {journal} {\bibinfo  {journal} {PRX Quantum}\ }\textbf {\bibinfo {volume} {4}},\ \bibinfo {pages} {040344} (\bibinfo {year} {2023})}\BibitemShut {NoStop}%
\bibitem [{\citenamefont {Viszlai}\ \emph {et~al.}(2024)\citenamefont {Viszlai}, \citenamefont {Chadwick}, \citenamefont {Joshi}, \citenamefont {Ravi}, \citenamefont {Li},\ and\ \citenamefont {Chong}}]{Viszlai2024}%
  \BibitemOpen
  \bibfield  {author} {\bibinfo {author} {\bibfnamefont {J.}~\bibnamefont {Viszlai}}, \bibinfo {author} {\bibfnamefont {J.~D.}\ \bibnamefont {Chadwick}}, \bibinfo {author} {\bibfnamefont {S.}~\bibnamefont {Joshi}}, \bibinfo {author} {\bibfnamefont {G.~S.}\ \bibnamefont {Ravi}}, \bibinfo {author} {\bibfnamefont {Y.}~\bibnamefont {Li}},\ and\ \bibinfo {author} {\bibfnamefont {F.~T.}\ \bibnamefont {Chong}},\ }\href {https://arxiv.org/abs/2412.05115} {\bibinfo {title} {Predictive window decoding for fault-tolerant quantum programs}} (\bibinfo {year} {2024}),\ \Eprint {https://arxiv.org/abs/2412.05115} {arXiv:2412.05115 [quant-ph]} \BibitemShut {NoStop}%
\bibitem [{\citenamefont {Pattison}\ \emph {et~al.}(2021)\citenamefont {Pattison}, \citenamefont {Beverland}, \citenamefont {da~Silva},\ and\ \citenamefont {Delfosse}}]{Pattison2021}%
  \BibitemOpen
  \bibfield  {author} {\bibinfo {author} {\bibfnamefont {C.~A.}\ \bibnamefont {Pattison}}, \bibinfo {author} {\bibfnamefont {M.~E.}\ \bibnamefont {Beverland}}, \bibinfo {author} {\bibfnamefont {M.~P.}\ \bibnamefont {da~Silva}},\ and\ \bibinfo {author} {\bibfnamefont {N.}~\bibnamefont {Delfosse}},\ }\href {https://arxiv.org/abs/2107.13589} {\bibinfo {title} {Improved quantum error correction using soft information}} (\bibinfo {year} {2021}),\ \Eprint {https://arxiv.org/abs/2107.13589} {arXiv:2107.13589 [quant-ph]} \BibitemShut {NoStop}%
\bibitem [{\citenamefont {Hanisch}\ \emph {et~al.}(2025)\citenamefont {Hanisch}, \citenamefont {Hetényi},\ and\ \citenamefont {Wootton}}]{Hanisch2025}%
  \BibitemOpen
  \bibfield  {author} {\bibinfo {author} {\bibfnamefont {M.~D.}\ \bibnamefont {Hanisch}}, \bibinfo {author} {\bibfnamefont {B.}~\bibnamefont {Hetényi}},\ and\ \bibinfo {author} {\bibfnamefont {J.~R.}\ \bibnamefont {Wootton}},\ }\href {https://arxiv.org/abs/2411.16228} {\bibinfo {title} {Soft information decoding with superconducting qubits}} (\bibinfo {year} {2025}),\ \Eprint {https://arxiv.org/abs/2411.16228} {arXiv:2411.16228 [quant-ph]} \BibitemShut {NoStop}%
\bibitem [{\citenamefont {Majaniemi}\ and\ \citenamefont {Matekole}(2025)}]{Majaniemi2025}%
  \BibitemOpen
  \bibfield  {author} {\bibinfo {author} {\bibfnamefont {J.}~\bibnamefont {Majaniemi}}\ and\ \bibinfo {author} {\bibfnamefont {E.~S.}\ \bibnamefont {Matekole}},\ }\href {https://arxiv.org/abs/2504.03504} {\bibinfo {title} {Reducing quantum error correction overhead using soft information}} (\bibinfo {year} {2025}),\ \Eprint {https://arxiv.org/abs/2504.03504} {arXiv:2504.03504 [quant-ph]} \BibitemShut {NoStop}%
\bibitem [{\citenamefont {Pattison}\ \emph {et~al.}(2025)\citenamefont {Pattison}, \citenamefont {Krishna},\ and\ \citenamefont {Preskill}}]{Pattison2023hierarchical}%
  \BibitemOpen
  \bibfield  {author} {\bibinfo {author} {\bibfnamefont {C.~A.}\ \bibnamefont {Pattison}}, \bibinfo {author} {\bibfnamefont {A.}~\bibnamefont {Krishna}},\ and\ \bibinfo {author} {\bibfnamefont {J.}~\bibnamefont {Preskill}},\ }\bibfield  {title} {\bibinfo {title} {Hierarchical memories: {S}imulating quantum {LDPC} codes with local gates},\ }\href {https://doi.org/10.22331/q-2025-05-05-1728} {\bibfield  {journal} {\bibinfo  {journal} {{Quantum}}\ }\textbf {\bibinfo {volume} {9}},\ \bibinfo {pages} {1728} (\bibinfo {year} {2025})}\BibitemShut {NoStop}%
\bibitem [{\citenamefont {Shutty}\ \emph {et~al.}(2024)\citenamefont {Shutty}, \citenamefont {Newman},\ and\ \citenamefont {Villalonga}}]{Shutty2024}%
  \BibitemOpen
  \bibfield  {author} {\bibinfo {author} {\bibfnamefont {N.}~\bibnamefont {Shutty}}, \bibinfo {author} {\bibfnamefont {M.}~\bibnamefont {Newman}},\ and\ \bibinfo {author} {\bibfnamefont {B.}~\bibnamefont {Villalonga}},\ }\bibfield  {title} {\bibinfo {title} {Efficient near-optimal decoding of the surface code through ensembling},\ }\href@noop {} {\bibfield  {journal} {\bibinfo  {journal} {arXiv preprint arXiv:2401.12434}\ } (\bibinfo {year} {2024})}\BibitemShut {NoStop}%
\bibitem [{\citenamefont {Horsman}\ \emph {et~al.}(2012)\citenamefont {Horsman}, \citenamefont {Fowler}, \citenamefont {Devitt},\ and\ \citenamefont {Meter}}]{Horsman2012}%
  \BibitemOpen
  \bibfield  {author} {\bibinfo {author} {\bibfnamefont {C.}~\bibnamefont {Horsman}}, \bibinfo {author} {\bibfnamefont {A.~G.}\ \bibnamefont {Fowler}}, \bibinfo {author} {\bibfnamefont {S.}~\bibnamefont {Devitt}},\ and\ \bibinfo {author} {\bibfnamefont {R.~V.}\ \bibnamefont {Meter}},\ }\bibfield  {title} {\bibinfo {title} {Surface code quantum computing by lattice surgery},\ }\href {https://doi.org/10.1088/1367-2630/14/12/123011} {\bibfield  {journal} {\bibinfo  {journal} {New Journal of Physics}\ }\textbf {\bibinfo {volume} {14}},\ \bibinfo {pages} {123011} (\bibinfo {year} {2012})}\BibitemShut {NoStop}%
\bibitem [{\citenamefont {Litinski}(2019)}]{Litinski2019}%
  \BibitemOpen
  \bibfield  {author} {\bibinfo {author} {\bibfnamefont {D.}~\bibnamefont {Litinski}},\ }\bibfield  {title} {\bibinfo {title} {A {G}ame of {S}urface {C}odes: {L}arge-{S}cale {Q}uantum {C}omputing with {L}attice {S}urgery},\ }\href {https://doi.org/10.22331/q-2019-03-05-128} {\bibfield  {journal} {\bibinfo  {journal} {{Quantum}}\ }\textbf {\bibinfo {volume} {3}},\ \bibinfo {pages} {128} (\bibinfo {year} {2019})}\BibitemShut {NoStop}%
\bibitem [{\citenamefont {Dijkstra}(1959)}]{Dijkstra1959}%
  \BibitemOpen
  \bibfield  {author} {\bibinfo {author} {\bibfnamefont {E.~W.}\ \bibnamefont {Dijkstra}},\ }\bibfield  {title} {\bibinfo {title} {A note on two problems in connexion with graphs},\ }\href {https://doi.org/10.1007/BF01386390} {\bibfield  {journal} {\bibinfo  {journal} {Numer. Math.}\ }\textbf {\bibinfo {volume} {1}},\ \bibinfo {pages} {269–271} (\bibinfo {year} {1959})}\BibitemShut {NoStop}%
\bibitem [{\citenamefont {Delfosse}(2020)}]{Delfosse2020}%
  \BibitemOpen
  \bibfield  {author} {\bibinfo {author} {\bibfnamefont {N.}~\bibnamefont {Delfosse}},\ }\bibfield  {title} {\bibinfo {title} {Hierarchical decoding to reduce hardware requirements for quantum computing},\ }\href@noop {} {\bibfield  {journal} {\bibinfo  {journal} {arXiv preprint arXiv:2001.11427}\ } (\bibinfo {year} {2020})}\BibitemShut {NoStop}%
\bibitem [{\citenamefont {Smith}\ \emph {et~al.}(2023)\citenamefont {Smith}, \citenamefont {Brown},\ and\ \citenamefont {Bartlett}}]{Smith2023}%
  \BibitemOpen
  \bibfield  {author} {\bibinfo {author} {\bibfnamefont {S.~C.}\ \bibnamefont {Smith}}, \bibinfo {author} {\bibfnamefont {B.~J.}\ \bibnamefont {Brown}},\ and\ \bibinfo {author} {\bibfnamefont {S.~D.}\ \bibnamefont {Bartlett}},\ }\bibfield  {title} {\bibinfo {title} {Local predecoder to reduce the bandwidth and latency of quantum error correction},\ }\href {https://doi.org/10.1103/PhysRevApplied.19.034050} {\bibfield  {journal} {\bibinfo  {journal} {Phys. Rev. Appl.}\ }\textbf {\bibinfo {volume} {19}},\ \bibinfo {pages} {034050} (\bibinfo {year} {2023})}\BibitemShut {NoStop}%
\bibitem [{\citenamefont {Caune}\ \emph {et~al.}(2023)\citenamefont {Caune}, \citenamefont {Reid}, \citenamefont {Camps},\ and\ \citenamefont {Campbell}}]{Caune2023}%
  \BibitemOpen
  \bibfield  {author} {\bibinfo {author} {\bibfnamefont {L.}~\bibnamefont {Caune}}, \bibinfo {author} {\bibfnamefont {B.}~\bibnamefont {Reid}}, \bibinfo {author} {\bibfnamefont {J.}~\bibnamefont {Camps}},\ and\ \bibinfo {author} {\bibfnamefont {E.}~\bibnamefont {Campbell}},\ }\href {https://arxiv.org/abs/2306.17142} {\bibinfo {title} {Belief propagation as a partial decoder}} (\bibinfo {year} {2023}),\ \Eprint {https://arxiv.org/abs/2306.17142} {arXiv:2306.17142 [quant-ph]} \BibitemShut {NoStop}%
\bibitem [{\citenamefont {Jones}(2024)}]{Jones2024}%
  \BibitemOpen
  \bibfield  {author} {\bibinfo {author} {\bibfnamefont {C.}~\bibnamefont {Jones}},\ }\href {https://arxiv.org/abs/2408.12135} {\bibinfo {title} {Improved accuracy for decoding surface codes with matching synthesis}} (\bibinfo {year} {2024}),\ \Eprint {https://arxiv.org/abs/2408.12135} {arXiv:2408.12135 [quant-ph]} \BibitemShut {NoStop}%
\bibitem [{\citenamefont {Bombin}\ and\ \citenamefont {Martin-Delgado}(2007)}]{Bombin2007}%
  \BibitemOpen
  \bibfield  {author} {\bibinfo {author} {\bibfnamefont {H.}~\bibnamefont {Bombin}}\ and\ \bibinfo {author} {\bibfnamefont {M.~A.}\ \bibnamefont {Martin-Delgado}},\ }\bibfield  {title} {\bibinfo {title} {Optimal resources for topological two-dimensional stabilizer codes: Comparative study},\ }\href {https://doi.org/10.1103/PhysRevA.76.012305} {\bibfield  {journal} {\bibinfo  {journal} {Phys. Rev. A}\ }\textbf {\bibinfo {volume} {76}},\ \bibinfo {pages} {012305} (\bibinfo {year} {2007})}\BibitemShut {NoStop}%
\bibitem [{\citenamefont {Higgott}(2022)}]{Higgott2022PyMatching}%
  \BibitemOpen
  \bibfield  {author} {\bibinfo {author} {\bibfnamefont {O.}~\bibnamefont {Higgott}},\ }\bibfield  {title} {\bibinfo {title} {Pymatching: A python package for decoding quantum codes with minimum-weight perfect matching},\ }\bibfield  {journal} {\bibinfo  {journal} {ACM Transactions on Quantum Computing}\ }\textbf {\bibinfo {volume} {3}},\ \href {https://doi.org/10.1145/3505637} {10.1145/3505637} (\bibinfo {year} {2022})\BibitemShut {NoStop}%
\bibitem [{\citenamefont {Lenssen}\ and\ \citenamefont {Paler}(2025)}]{Lenssen2025fooling}%
  \BibitemOpen
  \bibfield  {author} {\bibinfo {author} {\bibfnamefont {J.}~\bibnamefont {Lenssen}}\ and\ \bibinfo {author} {\bibfnamefont {A.}~\bibnamefont {Paler}},\ }\href {https://arxiv.org/abs/2504.19651} {\bibinfo {title} {Fooling the decoder: An adversarial attack on quantum error correction}} (\bibinfo {year} {2025}),\ \Eprint {https://arxiv.org/abs/2504.19651} {arXiv:2504.19651 [quant-ph]} \BibitemShut {NoStop}%
\bibitem [{\citenamefont {Chamberland}\ \emph {et~al.}(2023)\citenamefont {Chamberland}, \citenamefont {Goncalves}, \citenamefont {Sivarajah}, \citenamefont {Peterson},\ and\ \citenamefont {Grimberg}}]{Chamberland2023NN}%
  \BibitemOpen
  \bibfield  {author} {\bibinfo {author} {\bibfnamefont {C.}~\bibnamefont {Chamberland}}, \bibinfo {author} {\bibfnamefont {L.}~\bibnamefont {Goncalves}}, \bibinfo {author} {\bibfnamefont {P.}~\bibnamefont {Sivarajah}}, \bibinfo {author} {\bibfnamefont {E.}~\bibnamefont {Peterson}},\ and\ \bibinfo {author} {\bibfnamefont {S.}~\bibnamefont {Grimberg}},\ }\bibfield  {title} {\bibinfo {title} {Techniques for combining fast local decoders with global decoders under circuit-level noise},\ }\href {https://doi.org/10.1088/2058-9565/ace64d} {\bibfield  {journal} {\bibinfo  {journal} {Quantum Science and Technology}\ }\textbf {\bibinfo {volume} {8}},\ \bibinfo {pages} {045011} (\bibinfo {year} {2023})}\BibitemShut {NoStop}%
\bibitem [{\citenamefont {Bombín}\ \emph {et~al.}(2023)\citenamefont {Bombín}, \citenamefont {Dawson}, \citenamefont {Liu}, \citenamefont {Nickerson}, \citenamefont {Pastawski},\ and\ \citenamefont {Roberts}}]{bombin2023modular}%
  \BibitemOpen
  \bibfield  {author} {\bibinfo {author} {\bibfnamefont {H.}~\bibnamefont {Bombín}}, \bibinfo {author} {\bibfnamefont {C.}~\bibnamefont {Dawson}}, \bibinfo {author} {\bibfnamefont {Y.-H.}\ \bibnamefont {Liu}}, \bibinfo {author} {\bibfnamefont {N.}~\bibnamefont {Nickerson}}, \bibinfo {author} {\bibfnamefont {F.}~\bibnamefont {Pastawski}},\ and\ \bibinfo {author} {\bibfnamefont {S.}~\bibnamefont {Roberts}},\ }\href {https://arxiv.org/abs/2303.04846} {\bibinfo {title} {Modular decoding: parallelizable real-time decoding for quantum computers}} (\bibinfo {year} {2023}),\ \Eprint {https://arxiv.org/abs/2303.04846} {arXiv:2303.04846 [quant-ph]} \BibitemShut {NoStop}%
\bibitem [{\citenamefont {Fuhui~Lin}\ \emph {et~al.}(2025)\citenamefont {Fuhui~Lin}, \citenamefont {Peterson}, \citenamefont {Sankar},\ and\ \citenamefont {Sivarajah}}]{Lin2025}%
  \BibitemOpen
  \bibfield  {author} {\bibinfo {author} {\bibfnamefont {S.}~\bibnamefont {Fuhui~Lin}}, \bibinfo {author} {\bibfnamefont {E.~C.}\ \bibnamefont {Peterson}}, \bibinfo {author} {\bibfnamefont {K.}~\bibnamefont {Sankar}},\ and\ \bibinfo {author} {\bibfnamefont {P.}~\bibnamefont {Sivarajah}},\ }\bibfield  {title} {\bibinfo {title} {Spatially parallel decoding for multi-qubit lattice surgery},\ }\href {https://doi.org/10.1088/2058-9565/adc6b6} {\bibfield  {journal} {\bibinfo  {journal} {Quantum Science and Technology}\ }\textbf {\bibinfo {volume} {10}},\ \bibinfo {pages} {035007} (\bibinfo {year} {2025})}\BibitemShut {NoStop}%
\bibitem [{\citenamefont {Zhang}\ \emph {et~al.}(2025)\citenamefont {Zhang}, \citenamefont {Xu}, \citenamefont {Zhang}, \citenamefont {Kong}, \citenamefont {Ji},\ and\ \citenamefont {Chen}}]{Zhang2025LATTE}%
  \BibitemOpen
  \bibfield  {author} {\bibinfo {author} {\bibfnamefont {K.}~\bibnamefont {Zhang}}, \bibinfo {author} {\bibfnamefont {J.}~\bibnamefont {Xu}}, \bibinfo {author} {\bibfnamefont {F.}~\bibnamefont {Zhang}}, \bibinfo {author} {\bibfnamefont {L.}~\bibnamefont {Kong}}, \bibinfo {author} {\bibfnamefont {Z.}~\bibnamefont {Ji}},\ and\ \bibinfo {author} {\bibfnamefont {J.}~\bibnamefont {Chen}},\ }\href {https://arxiv.org/abs/2509.03954} {\bibinfo {title} {Latte: A decoding architecture for quantum computing with temporal and spatial scalability}} (\bibinfo {year} {2025}),\ \Eprint {https://arxiv.org/abs/2509.03954} {arXiv:2509.03954 [quant-ph]} \BibitemShut {NoStop}%
\bibitem [{\citenamefont {Meinerz}\ \emph {et~al.}(2022)\citenamefont {Meinerz}, \citenamefont {Park},\ and\ \citenamefont {Trebst}}]{Meinerz2022}%
  \BibitemOpen
  \bibfield  {author} {\bibinfo {author} {\bibfnamefont {K.}~\bibnamefont {Meinerz}}, \bibinfo {author} {\bibfnamefont {C.-Y.}\ \bibnamefont {Park}},\ and\ \bibinfo {author} {\bibfnamefont {S.}~\bibnamefont {Trebst}},\ }\bibfield  {title} {\bibinfo {title} {Scalable neural decoder for topological surface codes},\ }\href {https://doi.org/10.1103/PhysRevLett.128.080505} {\bibfield  {journal} {\bibinfo  {journal} {Phys. Rev. Lett.}\ }\textbf {\bibinfo {volume} {128}},\ \bibinfo {pages} {080505} (\bibinfo {year} {2022})}\BibitemShut {NoStop}%
\bibitem [{\citenamefont {Ravi}\ \emph {et~al.}(2023)\citenamefont {Ravi}, \citenamefont {Baker}, \citenamefont {Fayyazi}, \citenamefont {Lin}, \citenamefont {Javadi-Abhari}, \citenamefont {Pedram},\ and\ \citenamefont {Chong}}]{Ravi2023}%
  \BibitemOpen
  \bibfield  {author} {\bibinfo {author} {\bibfnamefont {G.~S.}\ \bibnamefont {Ravi}}, \bibinfo {author} {\bibfnamefont {J.~M.}\ \bibnamefont {Baker}}, \bibinfo {author} {\bibfnamefont {A.}~\bibnamefont {Fayyazi}}, \bibinfo {author} {\bibfnamefont {S.~F.}\ \bibnamefont {Lin}}, \bibinfo {author} {\bibfnamefont {A.}~\bibnamefont {Javadi-Abhari}}, \bibinfo {author} {\bibfnamefont {M.}~\bibnamefont {Pedram}},\ and\ \bibinfo {author} {\bibfnamefont {F.~T.}\ \bibnamefont {Chong}},\ }\bibfield  {title} {\bibinfo {title} {Better than worst-case decoding for quantum error correction},\ }in\ \href {https://doi.org/10.1145/3575693.3575733} {\emph {\bibinfo {booktitle} {Proceedings of the 28th ACM International Conference on Architectural Support for Programming Languages and Operating Systems, Volume 2}}},\ \bibinfo {series and number} {ASPLOS 2023}\ (\bibinfo  {publisher} {Association for Computing Machinery},\ \bibinfo {address} {New York, NY, USA},\ \bibinfo {year} {2023})\ p.\ \bibinfo {pages} {88–102}\BibitemShut
  {NoStop}%
\bibitem [{\citenamefont {Gicev}\ \emph {et~al.}(2023)\citenamefont {Gicev}, \citenamefont {Hollenberg},\ and\ \citenamefont {Usman}}]{Gicev2023}%
  \BibitemOpen
  \bibfield  {author} {\bibinfo {author} {\bibfnamefont {S.}~\bibnamefont {Gicev}}, \bibinfo {author} {\bibfnamefont {L.~C.~L.}\ \bibnamefont {Hollenberg}},\ and\ \bibinfo {author} {\bibfnamefont {M.}~\bibnamefont {Usman}},\ }\bibfield  {title} {\bibinfo {title} {A scalable and fast artificial neural network syndrome decoder for surface codes},\ }\href {https://doi.org/10.22331/q-2023-07-12-1058} {\bibfield  {journal} {\bibinfo  {journal} {{Quantum}}\ }\textbf {\bibinfo {volume} {7}},\ \bibinfo {pages} {1058} (\bibinfo {year} {2023})}\BibitemShut {NoStop}%
\bibitem [{\citenamefont {Müller}\ \emph {et~al.}(2025)\citenamefont {Müller}, \citenamefont {Alexander}, \citenamefont {Beverland}, \citenamefont {Bühler}, \citenamefont {Johnson}, \citenamefont {Maurer},\ and\ \citenamefont {Vandeth}}]{Muller2025relayBP}%
  \BibitemOpen
  \bibfield  {author} {\bibinfo {author} {\bibfnamefont {T.}~\bibnamefont {Müller}}, \bibinfo {author} {\bibfnamefont {T.}~\bibnamefont {Alexander}}, \bibinfo {author} {\bibfnamefont {M.~E.}\ \bibnamefont {Beverland}}, \bibinfo {author} {\bibfnamefont {M.}~\bibnamefont {Bühler}}, \bibinfo {author} {\bibfnamefont {B.~R.}\ \bibnamefont {Johnson}}, \bibinfo {author} {\bibfnamefont {T.}~\bibnamefont {Maurer}},\ and\ \bibinfo {author} {\bibfnamefont {D.}~\bibnamefont {Vandeth}},\ }\href {https://arxiv.org/abs/2506.01779} {\bibinfo {title} {Improved belief propagation is sufficient for real-time decoding of quantum memory}} (\bibinfo {year} {2025}),\ \Eprint {https://arxiv.org/abs/2506.01779} {arXiv:2506.01779 [quant-ph]} \BibitemShut {NoStop}%
\bibitem [{\citenamefont {Breuckmann}\ and\ \citenamefont {Eberhardt}(2021)}]{Breuckmann2021}%
  \BibitemOpen
  \bibfield  {author} {\bibinfo {author} {\bibfnamefont {N.~P.}\ \bibnamefont {Breuckmann}}\ and\ \bibinfo {author} {\bibfnamefont {J.~N.}\ \bibnamefont {Eberhardt}},\ }\bibfield  {title} {\bibinfo {title} {Quantum low-density parity-check codes},\ }\href {https://doi.org/10.1103/PRXQuantum.2.040101} {\bibfield  {journal} {\bibinfo  {journal} {PRX Quantum}\ }\textbf {\bibinfo {volume} {2}},\ \bibinfo {pages} {040101} (\bibinfo {year} {2021})}\BibitemShut {NoStop}%
\bibitem [{\citenamefont {Bravyi}\ \emph {et~al.}(2024)\citenamefont {Bravyi}, \citenamefont {Cross}, \citenamefont {Gambetta}, \citenamefont {Maslov}, \citenamefont {Rall},\ and\ \citenamefont {Yoder}}]{Bravyi2024}%
  \BibitemOpen
  \bibfield  {author} {\bibinfo {author} {\bibfnamefont {S.}~\bibnamefont {Bravyi}}, \bibinfo {author} {\bibfnamefont {A.~W.}\ \bibnamefont {Cross}}, \bibinfo {author} {\bibfnamefont {J.~M.}\ \bibnamefont {Gambetta}}, \bibinfo {author} {\bibfnamefont {D.}~\bibnamefont {Maslov}}, \bibinfo {author} {\bibfnamefont {P.}~\bibnamefont {Rall}},\ and\ \bibinfo {author} {\bibfnamefont {T.~J.}\ \bibnamefont {Yoder}},\ }\bibfield  {title} {\bibinfo {title} {High-threshold and low-overhead fault-tolerant quantum memory},\ }\href {https://doi.org/10.1038/s41586-024-07107-7} {\bibfield  {journal} {\bibinfo  {journal} {Nature}\ }\textbf {\bibinfo {volume} {627}},\ \bibinfo {pages} {778} (\bibinfo {year} {2024})}\BibitemShut {NoStop}%
\bibitem [{\citenamefont {English}\ \emph {et~al.}(2024)\citenamefont {English}, \citenamefont {Williamson},\ and\ \citenamefont {Bartlett}}]{English2024}%
  \BibitemOpen
  \bibfield  {author} {\bibinfo {author} {\bibfnamefont {L.~H.}\ \bibnamefont {English}}, \bibinfo {author} {\bibfnamefont {D.~J.}\ \bibnamefont {Williamson}},\ and\ \bibinfo {author} {\bibfnamefont {S.~D.}\ \bibnamefont {Bartlett}},\ }\href {https://arxiv.org/abs/2410.07598} {\bibinfo {title} {Thresholds for post-selected quantum error correction from statistical mechanics}} (\bibinfo {year} {2024}),\ \Eprint {https://arxiv.org/abs/2410.07598} {arXiv:2410.07598 [quant-ph]} \BibitemShut {NoStop}%
\bibitem [{\citenamefont {Campbell}(2023)}]{Riverlane2023}%
  \BibitemOpen
  \bibfield  {author} {\bibinfo {author} {\bibfnamefont {E.}~\bibnamefont {Campbell}},\ }\href {https://www.riverlane.com/blog/what-is-a-teraquop-decoder} {\bibinfo {title} {What is a teraquop decoder?}} (\bibinfo {year} {2023})\BibitemShut {NoStop}%
\end{thebibliography}%

\end{document}